\documentclass[aps,prb,10pt,superscriptaddress,twocolumn]{revtex4-2}

\usepackage[pdftex]{graphicx}
\usepackage{dcolumn}
\usepackage{bm}
\usepackage{color}
\usepackage{amsmath}
\usepackage{nicefrac}
\usepackage{amssymb}
\usepackage{cancel}

\usepackage{braket}
\usepackage{physics}
\usepackage{bbold}
\usepackage[pdftex,colorlinks=true,linkcolor=blue,citecolor=blue,filecolor=blue]{hyperref}

\newcommand{\ff}[1]{{\boldsymbol #1}}
\newcommand{\ca}[1]{{\cal #1}}
\newcommand{\bi}{\begin{itemize}}
\newcommand{\ei}{\end{itemize}}
\newcommand{\be}{\begin{equation}}
\newcommand{\ee}{\end{equation}}
\newcommand{\ba}{\begin{eqnarray}}
\newcommand{\ea}{\end{eqnarray}}

\newcommand{\refeq}[1]{Eq.\ (\ref{eq:#1})}
\newcommand{\labeq}[1]{\label{eq:#1}}

\begin{document} 
  
\title{Microscopic theory of spin friction and dissipative spin dynamics}

\author{Nicolas Lenzing}

\affiliation{I. Institute of Theoretical Physics, Department of Physics, University of Hamburg, Notkestra{\ss}e 9-11, 22607 Hamburg, Germany}

\author{David Kr\"uger}

\affiliation{I. Institute of Theoretical Physics, Department of Physics, University of Hamburg, Notkestra{\ss}e 9-11, 22607 Hamburg, Germany}

\author{Michael Potthoff}

\affiliation{I. Institute of Theoretical Physics, Department of Physics, University of Hamburg, Notkestra{\ss}e 9-11, 22607 Hamburg, Germany}

\affiliation{The Hamburg Centre for Ultrafast Imaging, Luruper Chaussee 149, 22761 Hamburg, Germany}

\begin{abstract}
The real-time dynamics of local magnetic moments exchange coupled to a metallic system of conduction electrons is subject to dissipative friction even in the absence of spin-orbit coupling. 
Phenomenologically, this is usually described by a local Gilbert damping constant. 
Here, we use both linear response theory and adiabatic response theory to derive the spin friction microscopically for a generic single-band tight-binding model of the electronic structure. 
The resulting Gilbert damping is time-dependent and nonlocal.
For a one-dimensional model, we compare the emergent relaxation dynamics as obtained from LRT and ART against each other and against the full solution of the microscopic equations of motion and demonstrate the importance of nonlocality, while the time dependence turns out to be irrelevant.
In two dimensions and for a few magnetic moments in different geometries, it is found that the inclusion of nonlocal Gilbert damping can counterintuitively lead to longer relaxation times.
Besides the distance dependence, the directional dependence of the nonlocal Gilbert damping turns out as very important.
Our results are based on an expression relating the nonlocal Gilbert damping to the nonlocal tight-binding density of states close to the Fermi energy. 
This is exact in case of noninteracting electrons.
Effects due to electronic correlations are studied within the random-phase approximation. 
For the Hubbard model at half filling and with increasing interaction strength, we find a strong enhancement of the nonlocality of spin friction.
\end{abstract} 

\maketitle 

\section{Introduction}
\label{sec:intro}

The understanding of the relaxation dynamics of local magnetic moments on an atomistic level represents an essential step for further progress in the field of nanospintronics \cite{CFVD07,CVSH15,GL14,BMT+18}. 
A prototypical model within atomistic spin-dynamics theory \cite{TKS08,SHNE08,BMS09,EFC+14} is given by the multi-impurity Kondo model with a few localized spins, replaced by classical vectors of unit length to represent the local magnetic moments. 
This is also known as the $s$-$d$ exchange (Vonsovsky-Zener) model \cite{vz}.
Its Hamiltonian has the form
$\hat{H} = \hat{H}_{\text{el}} + \hat{H}_{\text{int}}$,
where $\hat{H}_{\text{el}}$ is a tight-binding model of the electronic structure on a $D$-dimensional lattice, and where $\hat{H}_{\text{int}}$ is a generic local exchange interaction between the classical spins $\ff S_{m}$ (with $m=1,...,M$) and the local spin-moment operators $\hat{\boldsymbol{s}}_{i_{m}}$ at the sites $i_{m}$ of the lattice. 
In its most simple form, assuming a single spin-degenerate orbital per site, 
\be
\hat{H} 
= 
\sum_{\langle ii'\rangle} \sum_{\sigma=\uparrow, \downarrow} T_{ii'} c^\dagger_{i\sigma} c_{i^\prime\sigma} + J \sum_{m} \hat{\boldsymbol{s}}_{i_{m}}\boldsymbol{S}_m
\: .
\labeq{ham}
\ee
Here $T_{ii'}$ is the hopping amplitude between sites $i$ and $i'$, and $J>0$ the local exchange-coupling strength. 

The time dependence of the electron and the classical-spin degrees of freedom follows the general rules for quantum-classical dynamics \cite{Elz12,SP15}.
As the model is quadratic in the electron annihilators and creators, $c_{i\sigma}$ and $c_{i\sigma}^{\dagger}$ ($i=1,...,L$ sites, $\sigma=\uparrow,\downarrow$ spin projection), for each spin configuration $\ff S \equiv (\ff S_{1},...,\ff S_{M})$, there is a closed system of equations of motions, for the one-particle reduced density matrix and for the classical spins. 
The dynamics of the latter is governed by classical Landau-Lifschitz equations
$\dot{\ff S}_{m} = \partial \langle \hat{H} \rangle / \partial \ff S_{m} \times \ff S_{m}$.

Here, we are interested in the relaxation dynamics of systems with a few impurity spins, $M=1,...,10$, for the single-orbital model \refeq{ham} in the thermodynamical limit $L\to \infty$.
The computational effort for solving the related coupled system of nonlinear ordinary differential equations roughly scales quadratically in the number of lattice sites $L = l^{D}$ and linearly in the propagation time $t$ \cite{SP15}.
To avoid unwanted reflections (or interferences) of propagating excitations from the system boundaries (or due to periodic boundary conditions), however, systems with a linear extension $l \sim v t$ must be considered, if $v$ is the (ballistic) propagation speed.
The Gaussian electronic part can be formally integrated out within a path-integral formalism but at the cost of a highly time-nonlocal effective action for the classical spins, which, if treated exactly, does not improve the scaling. 

For 1D systems, absorbing boundary conditions, e.g., using a generalized Lindblad master-equation approach to couple the edge sites of the conduction-electron tight-binding model to an external bath, have turned out as helpful \cite{EP20,EP21}. 
It has been demonstrated that this allows to exceed the characteristic femtosecond electronic scale set by the inverse nearest-neighbor hopping by more than five orders of magnitude.
For $D\ge 2$, however, an exact time propagation of an initial state on the pico- or even nanosecond scale appears out of reach but is quite relevant from general considerations of the time scales of magnetic processes \cite{KKR10}.
Even at thermal equilibrium and using advanced Monte-Carlo techniques, quantum-classical hybrid systems with a linear extension $l \lesssim 30$ in $D=3$ are computationally challenging \cite{Wei09}.

On the other hand, for physically relevant applications, the classical-spin dynamics takes place on a characteristic time scale $\tau_{\rm sp}$, which is more than an order of magnitude slower than the femtosecond ($\tau_{\rm el} \sim 1/T$) electron dynamics set by the nearest-neighbor hopping $T$.
This implies that the electronic quantum state follows the classical dynamics almost instantly, i.e., the electron dynamics is almost adiabatic and characterized by a typical retardation time $\tau_{\rm ret}$ much shorter than $\tau_{\rm sp}$.

This situation has motivated theoretical efforts to approximately ``integrate out'' the fast electron degrees of freedom, using the interrelated assumptions of weak $J$ and small $\tau_{\rm ret}$. 
These justify a double expansion, namely (i) a perturbative treatment of the exchange coupling $J$ followed by (ii) an expansion in the retardation time as put forward in Refs.\ \onlinecite{CG03,ON06,BNF12,UMS12}.
This results in an effective spin-only theory, given by a (generalized) Landau-Lifschitz-Gilbert (LLG) equation \cite{llg},
\be
\dot{\ff S}_{m} 
=
\sum_{m'} J_{mm'} \ff S_{m'} \times \ff S_{m} 
+
\sum_{m'} \alpha_{mm'} \ff S_{m} \times \dot{\ff S}_{m'} 
\: .
\labeq{llg}
\ee
where both, the Gilbert damping $\alpha_{mm'}$ and the indirect RKKY exchange interaction $J_{mm'}$ \cite{RK54,Kas56,Yos57} must be computed from the $J=0$ ground state of the electron system.

It has been pointed out \cite{SP15,BN19,RON24} that the Gilbert damping $\alpha_{mm'}$ is usually nonlocal, i.e., dependent on two spatial indices $m,m'$ with $\alpha_{mm'} \ne 0$ for $m \ne m'$. 
In addition and depending on the translational symmetries of the underlying system, the Gilbert damping can be inhomogenous, i.e, $\alpha_{mm}$ can be $m$ dependent or, more generally $\alpha_{mm'}$ can depend nontrivially on both $m$ and $m'$. 
Here, assuming a hopping matrix that fully respects the translational symmetries of the lattice, we consider the homogenous case only and focus on the nonlocality of the damping. 
Moreover, the purely electronic part, $\hat{H}_{\rm el}$, of the Hamiltonian (\ref{eq:ham}) is invariant under SU(2) spin rotations. 
With this choice we also disregard anisotropy effects and thus assume for $3 \times 3$ Gilbert-damping tensor that
$\alpha_{m\alpha, m'\alpha'} = \alpha_{mm'} \delta_{\alpha\alpha'}$ with $\alpha, \alpha' = x,y,z$. 

Importantly, even for the conceptually simple model Hamiltonian given by Eq.\ (\ref{eq:ham}), the local and the nonlocal elements of the Gilbert damping are nonzero and actually comparatively large. 
The physical mechanism is a retardation effect, as described in Refs.\ \onlinecite{SP15,THE15}: 
Although the impurity-spin dynamics is slow as compared to the femtosecond time scale of the electron dynamics, the quantum state does not follow the time-dependent impurity-spin configuration instantly, i.e., the electron dynamics is slightly non-adiabatic. 
Already for the ($M=1$) single-impurity case, this implies that $\ff S$ and $\langle \hat{\ff s} \rangle$ are noncollinear, which in turn produces spin damping.
This view should be contrasted with previous work, where the local damping $\alpha_{mm}$ is attributed to relativistic effects, i.e., to spin-orbit coupling
\cite{KK02,Kam07,GIS07,BTB08,HM09,GMcD09,SKB+10,Sak12,EMKK11,FI11,MKWE13,MBO16,MBNO17,GSC+19,AOT20}, see also the related discussion in Ref.\ \onlinecite{RON24}.

In the quite common continuum (as opposed to the discrete tight-binding) approach, the nonlocality of the Gilbert damping is typically accounted for by additional gradient terms $\partial_{\alpha} \ff S(\ff r,t)$ \cite{TM08,HVT08,ZZ09,KMLL12,YYXW16,MWE18,VTS18}. 
This assumes a weak spatial variation of the damping on the atomic scale and that the leading correction beyond a fully local damping suffices.
On the contrary, we will demonstrate that $\alpha_{mm'}$ is typically even more nonlocal than, e.g., the RKKY interaction $J_{mm'}$, at least for the model studied here.

With the present paper we focus the generic, non-relativistic model given by \refeq{ham} on the $D=1$ chain and on the $D=2$ square lattice and study the local and nonlocal elements of the Gilbert-damping matrix.
To this end, we derive a compact expression for $\alpha_{mm'}$ based on the local and nonlocal tight-binding density of states.

Furthermore, we discuss the impact of the spin damping on the relaxation dynamics for systems with different number $M$ of impurity spins coupled to the electron system in various geometries. 
For two classical spins ($M=2$) coupled to next-nearest-neighbor sites of a one-dimensional tight-binding chain, it has been observed \cite{EP24} that the local and the nonlocal elements of $\alpha_{mm'}$ are exactly identical, $\alpha_{11}=\alpha_{22}=\alpha_{12}=\alpha_{21}$, and that this results in a completely {\em undamped} spin dynamics. 
This counterintuitive effect actually represents a general feature of $D=1$ classical, quantum and quantum-classical bipartite multi-impurity models \cite{RON24,EP24}.
If and how this still manifests itself in higher dimensions is an obvious question to be answered.

The LLG equation (\ref{eq:llg}) is re-derived for the model given by \refeq{ham} by invoking linear response theory and, subsequently, by an expansion in the retardation time and corresponding truncation, i.e., using the two above-mentioned approximations controlled by (i) weak local exchange $J$ and (ii) short typical retardation times.
We demonstrate that the result is essentially the same when interchanging the order of the approximations, i.e., when first starting from the completely different formalism of adiabatic response theory \cite{CDH12}, adapted to the present case, and make us of a weak-$J$ approximation thereafter.
For $D=1$ the impurity-spin relaxation dynamics from both approaches will be checked against the fully numerical solution of the exact equations of motion.

Both formalisms show that $\alpha_{mm'} = \alpha_{mm'}(t)$ is actually time-dependent.
However, very different time dependencies are obtained when evaluated numerically. 
This and the impact on the long-time dynamics and the relaxation time is studied for the $D=1$ case, where the linear and the adiabatic response approach can be checked against the full solution. 

The paper is organized as follows: 
In the next section we discuss the fundamental equations of motion for the coupled spin-electron dynamics. 
The linear response and the adiabatic response approaches are introduced in Secs.\ \ref{sec:lrt} and \ref{sec:art}, respectively. 
Section \ref{sec:comp} is devoted to computational details. 
Results are presented in Sec.\ \ref{sec:results} and address the time dependence and the nonlocality of the spin friction (Secs.\ \ref{sec:alpha} and \ref{sec:nalpha}), 
the dynamics of a single spin driven by a magnetic field (Sec.\ \ref{sec:sd1}),  
and the breakdown of the effective theory if there is a van Hove singularity at the Fermi energy (Sec.\ \ref{sec:vh}). 
The anomalous dynamics of two impurity spins in $D=1$ is discussed in Sec.\ \ref{sec:sdtwo}. 
We then proceed with dimension $D=2$ and analyze the local and nonlocal spin friction in Sec.\ \ref{sec:sd2} and the distance and directional dependencies of the Gilbert damping in Sec.\ \ref{sec:dist}.
The effects of spin friction on the dynamics of two impurity spins and of impurity-spin arrays are analyzed in Secs.\ \ref{sec:non2} and \ref{sec:array}.
Electron-correlation effects, on the level of the random-phase approximation, are touched in Sec.\ \ref{sec:int}.
Our concluding remarks are given in Sec.\ \ref{sec:con}.

\section{Full spin dynamics}
\label{sec:full}

Starting from the Hamiltonian, \refeq{ham}, the exact equations of motion for the classical spins $\ff S_{m}$, with $m=1,...,M$ and $|\ff S_{m}|=1$, and of the one-particle reduced density matrix $\ff \rho(t)$ with elements
\be
\rho_{i\sigma,i'\sigma'}(t) = \langle \Psi(t) | c_{i'\sigma'}^{\dagger} c_{i\sigma} | \Psi(t) \rangle
\: , 
\ee
are readily derived (see, e.g., Ref.\ \onlinecite{SP15}). 
We find
\be
\frac{d}{dt} \ff S_{m}(t) = J \langle \ff s_{i_{m}} \rangle_{t} \times \ff S_{m}(t) - \sum_{m} \ff B_{m} \times \ff S_{m}(t)
\labeq{eoms}
\: , 
\ee
where $\langle \ff s_{i_{m}} \rangle_{t}$ is the expectation value of the local spin at site $i_{m}$ of the electron system in the $N$-electron state $|\Psi(t) \rangle$. 
If $\ff \tau$ denotes for the vector of Pauli matrices, we have $\ff s_{i} = \frac12\sum_{\sigma \sigma'} c_{i\sigma}^{\dagger} \ff \tau_{\sigma\sigma'} c_{i\sigma'}$.
The last term on the right-hand side results from local magnetic fields $\ff B_{m}$ coupling to the impurity spins $\ff S_{m}$, i.e., we have replaced the Hamiltonian in \refeq{ham} by $\hat{H} \mapsto \hat{H} - \sum_{m} \ff B_{m} \ff S_{m}$.
This will be convenient in the following.
The equation of motion for the density matrix is given by
\be
i \frac{d}{dt} \ff \rho(t) = [ \ff T^{\rm (eff)}(t) , \ff \rho(t)] 
\labeq{eomr}
\: , 
\ee
where $\ff T^{\rm (eff)}(t)$ is an effective hopping matrix with elements
\be
T_{i\sigma, i'\sigma'}^{\rm (eff)}(t)
=
\delta_{\sigma\sigma'} T_{ii'}
+
\frac{J}{2} \delta_{ii'} \sum_{m=1}^{M} \delta_{ii_{m}} \ff \tau_{\sigma\sigma'} \ff S_{m}(t)
\: .
\ee

Initially, at time $t=0$, we specify a certain start configuration $\ff S(0) = (\ff S_{1}(0), ..., \ff S_{M}(0))$ for the impurity spins.
Furthermore, we assume that the electron system, at time $t=0$, is in its ground state for $J=0$.
The corresponding ground-state one-particle reduced density matrix
\be
\ff \rho(0) = \Theta(\mu \ff 1 - \ff T) 
\: ,
\ee
is formally given in terms of the Heavyside step function $\Theta(\cdots)$ and can be computed by diagonalizing the hopping matrix $\ff T$.

In the multi-impurity-spin case ($M>1$), real-time dynamics is initiated by suddenly switching on $\hat{H}_{\rm int}$. 
For $M=1$, we suddenly switch the direction of a local magnetic field $\ff B$ that couples to the impurity spin $\ff S$.

Equations (\ref{eq:eoms}) and (\ref{eq:eomr}) form a closed set of nonlinear ordinary differential equations for the spin configuration $\ff S = (\ff S_{1}, ..., \ff S_{M})$ and $\ff \rho$. 
This can be solved numerically by standard techniques for a $D$-dimensional lattice with a finite number of sites $L=l^{D}$. 
Accessible propagation times $\tau$ are limited by the requirement $\tau \lesssim l / v$, where $v$ is the (ballistic) propagation speed.

\section{Linear response theory}
\label{sec:lrt}

The effective spin-only dynamics, determined by the LLG equation (\ref{eq:llg}), is obtained in the limit of weak $J$ and short retardation times.
In the linear response approach, we start by treating $J$ perturbatively. 
The corresponding Kubo formula reads as:
\be
\langle \ff s_{i_{m}} \rangle_{t}
= 
J \sum_{m} \int_{0}^{t} d\tau \, \chi_{mm'}(\tau) \ff S_{m'}(t-\tau)
+ \ca O(J^{2})
\: .
\labeq{kubo}
\ee
The integral kernel is given in terms of the unperturbed ($J=0$), retarded, nonlocal, and time-homogenous magnetic susceptibility
\be
\chi^{\alpha \alpha'}_{mm'}(t) 
= -i \Theta(t) e^{-\eta t} 
\left\langle [
s^{\alpha}_{i_{m}}(t) , s^{\alpha'}_{i_{m'}}(0) 
] \right\rangle^{(0)}
\labeq{chi}
\: .
\ee
Here, $\langle \cdots \rangle^{(0)}$ denotes the expectation value with respect to the unperturbed system at $J=0$. 
Furthermore, $\eta>0$ is an infinitesimal, and $\alpha = x,y,z$. 
Since $\hat{H}_{\rm el}$ is invariant under SU(2) spin rotations, the susceptibility is diagonal with respect to the directional indices and also $\alpha$ independent: $\chi^{\alpha \alpha'}_{mm'}(t) = \delta^{\alpha\alpha'}  \chi_{mm'}(t)$.

As a second approximation, we assume that the retardation time $\tau$ in \refeq{kubo} is small, i.e., that the integral kernel $\chi_{mm'}(\tau)$ is peaked at small $\tau$, on the characteristic time scale $1/T$ of the electron system, as compared to the much slower time scale on which the classical spins evolve.
This justifies a Taylor expansion $\ff S_{m'}(t-\tau) = \ff S_{m'}(t) - \tau \dot{\ff S}_{m'}(t) + \cdots$. 
Truncating the expansion after the first order and inserting into \refeq{kubo} yields:
\ba
\langle \ff s_{i_{m}} \rangle_{t}
&=& 
J \sum_{m'} \left( \int_{0}^{t} d\tau \, \chi_{mm'}(\tau) \right) \ff S_{m'}(t)
\nonumber \\
&-&
J \sum_{m'} \left( \int_{0}^{t} d\tau \, \tau \chi_{mm'}(\tau) \right) \dot{\ff S}_{m'}(t)
\: .
\ea
With \refeq{eoms} we then get
\ba
\frac{d}{dt} \ff S_{m}(t) 
& = &
\sum_{m'} J_{mm'}(t) \ff S_{m'}(t) \times \ff S_{m}(t)
\nonumber \\
& + & 
\sum_{m'} \alpha_{mm'}(t) \dot{\ff S}_{m'}(t) \times \ff S_{m}(t)
\nonumber \\
& - & 
\sum_{m} \ff B_{m} \times \ff S_{m}(t)
\labeq{eomeff}
\: , 
\ea
where we have defined the time-dependent RKKY exchange interaction
\be
J_{mm'}(t) = J^{2} \int_{0}^t d \tau \, \chi_{mm'}(\tau) 
\: , 
\labeq{trkky}
\ee
and the time-dependent Gilbert damping
\be
\alpha_{mm'}(t) = - J^{2} \int_{0}^t d \tau \, \tau \, \chi_{mm'}(\tau) 
\labeq{talpha}
\: .
\ee
Equation (\ref{eq:eomeff}) should be compared with the LLG equation \refeq{llg}. 
Apart from the additional magnetic-field term, the main difference is the time dependence of the coupling constants.

The time-independent RKKY interaction, defined as $J_{mm'} = \lim_{t\to \infty} J_{mm'}(t)$, corresponds to an effective RKKY Hamiltonian $H_{\rm RKKY} = \frac12 \sum_{mm'}^{m\ne m'} J_{mm'} \ff S_{m} \ff S_{m'}$. 
We furthermore define the time-independent Gilbert damping: $\alpha_{mm'} = \lim_{t\to \infty} \alpha_{mm'}(t)$. 
The sign in the definition (\ref{eq:talpha}) of the damping matrix is chosen such that $\alpha_{mm} > 0$. 
(This is opposite to the convention used in Refs.\ \onlinecite{SP15,EP24}).
Through Fourier transformation of the susceptibility, $\chi_{mm'}(\omega) = \int d\tau \, e^{i\omega \tau} \chi_{mm'}(\tau)$, we get the alternative representation
\be
\alpha_{mm'} 
=
i J^{2} \frac{d}{d\omega} \chi_{mm'}(\omega=0)
=
- J^{2} \frac{d}{d\omega} \mbox{Im} \chi_{mm'}(\omega=0)
\: .
\labeq{altrep}
\ee

Most convenient for the numerical simulations, however, is the representation of the damping matrix,
\be
\alpha_{mm'} 
=  
\frac{\pi}{2} J^2 A_{i_{m} i_{m'}}(\omega = 0)^{2}
\: , 
\labeq{dosform}
\ee
in terms of the local or nonlocal tight-binding density of states $A_{ii'}(\omega)$. 
In the context of a single impurity spin and local Gilbert damping ($m=m'$), a derivation of \refeq{dosform} is given in Refs.\ \cite{Sak06,KBNN06,ND08,TKD15}.
A general derivation is given in the Appendix \ref{sec:appdos}, see also Ref.\ \cite{Sak13,BNF12,SP15,STZ24}.

Assuming a translation-invariant system with lattice vectors $\ff R_{i}$ and periodic boundary conditions, we have
\be
A_{ii'}(\omega)
= 
\frac{1}{L} \sum_{\ff k} 
e^{i \ff k (\ff R_{i} - \ff R_{i'})}
\delta(\omega + \mu - \varepsilon(\ff k))
\labeq{spden}
\: ,
\ee
where $\varepsilon(\ff k) = \varepsilon(-\ff k)$ is the tight-binding dispersion and $\mu$ the chemical potential, which fixes the average particle number $\langle N \rangle$.

According to the derivation given in the Appendix \ref{sec:appdos}, the temperature dependence of the damping parameters can be obtained via
\be
\alpha_{mm'} 
=
- \frac{\pi}{2} J^{2} \int dx \, f'(x) A_{i_{m}i_{m'}}(x)^{2} 
\: .
\labeq{alphatemp}
\ee
In the low-temperature ($\beta \to \infty$) or in the wide-band limit, we can replace the one-electron spectral density by a constant, $A_{i_{m}i_{m'}}(x) \to \rho_{mm'}$. 
This yields:
\be
\alpha^{(\infty)}_{mm'} 
=
\frac{\pi}{2} J^{2} \rho_{mm'}^{2} 
\: .
\labeq{wideband}
\ee
Hence, a nontrivial temperature dependence at low temperatures is due to the variation of the spectral density near $\omega=0$.
We can make use of the Sommerfeld expansion in powers of $\beta^{-2}$ to make this explicit.
A straightforward calculation yields:
\be
\alpha_{mm'} 
=
\alpha^{(\infty)}_{mm'} 
+
\frac{\pi^{3}}{6} \frac{J^{2}}{\beta^{2}}
\left[
\left( \rho'_{mm'} \right)^{2} 
+
\rho_{mm'} \rho''_{mm'} 
\right]
+ \ca O(\beta^{-4})
\: ,
\labeq{sommer}
\ee
where we have defined
\be
\rho^{(n)}_{mm'} = \frac{d^{n}}{dx^{n}}  A_{i_{m}i_{m'}}(x) \Big|_{x=0}
\: .
\ee

\section{Adiabatic response theory}
\label{sec:art}

In the limit of weak J and short retardation times, the same effective spin-only dynamics, \refeq{llg}, is obtained when first assuming that the electron system almost adiabatically follows the spin dynamics, while the weak-$J$ approximation is done at a later stage. 
This can be seen by starting with the adiabatic response theory as outlined in Ref.\ \onlinecite{CDH12} and adapted to the case of spin dynamics.

The considered setup is that of an open quantum system, driven by external parameters $\ff S = (\ff S_{1}, ..., \ff S_{M})$ and in contact with a large thermal bath at temperature $1/\beta$.
The total Hamiltonian is 
\be
\hat{H}_{\rm total} = \hat{H} + \hat{H}_{\rm B} + \hat{H}_{\rm SB}
\labeq{htot}
\: , 
\ee
where $\hat{H}$ is the Hamiltonian given by \refeq{ham}, where $\hat{H}_{\rm B}$ denotes the Hamiltonian of the thermal bath, and where $\hat{H}_{\rm SB}$ is the system-bath interaction.
The latter terms, $\hat{H}_{\rm B}$ and $\hat{H}_{\rm SB}$, are introduced for formal reasons only and can be disregarded at the end of the consideration, assuming that the system-bath coupling is sufficiently weak. 
Finally, we will also take the zero-temperature limit $\beta \to \infty$.  

The goal is to determine the expectation value 
$\langle \ff s_{i_{m}} \rangle_{t}$
of the local spin of the electron system at site $i_{m}$ in the many-electron quantum state, for given trajectory of the classical spins $\ff S(t')$ with $t'\le t$.
This is needed for \refeq{eoms} to obtain a closed set of equations of motion for the classical spins $\ff S$ only.
To this end, we start with the thermal (canonical) equilibrium value $\langle \ff s_{i_{m}} \rangle^{\rm (eq)}_{\ff S(t)}$ for a fixed spin configuration $\ff S(t)$ at time $t$.
In the adiabatic approximation, we would have $\langle \ff s_{i_{m}} \rangle_{t} = \langle \ff s_{i_{m}} \rangle^{\rm (eq)}_{\ff S(t)}$.

Within adiabatic response theory, the difference
\be
\langle \Delta \ff s_{i_{m}}(t) \rangle 
=
\langle \ff s_{i_{m}} \rangle_{t} - \langle \ff s_{i_{m}} \rangle^{\rm (eq)}_{\ff S(t)}
\: ,
\labeq{artd}
\ee
is computed perturbatively in the deviation from a strictly adiabatic, infinitely slow spin dynamics $\ff S(t)$.
The ``small parameter'' is given by the dissipated work \cite{CDH12,Jar97}
\be
W = \int_{0}^{t} dt' \, \dot{\ff S}(t')
\left[
\nabla_{\ff S} \hat{H}(\ff S(t')) 
- 
\langle 
\nabla_{\ff S} \hat{H}(\ff S(t'))
\rangle^{\rm (eq)}_{\ff S(t')}
\right]
\: ,
\ee
where $\hat{H}(\ff S(t))$ is the total Hamiltonian (\ref{eq:ham}) for a given spin configuration. 
Note that $W$ is an operator but not an observable \cite{TLH07,CHT11}.
The dissipated work can be interpreted as the difference between the work performed on the system along a certain trajectory $\ff S(t)$ in the spin-configuration space and the work along the same path but traversed adiabatically.
For an isothermal process, the second contribution is given by the free-energy difference between the initial and the final state at $\ff S(0)$ and $\ff S(t)$, respectively. 

As detailed in Ref.\ \onlinecite{CDH12}, one finds
\be
\langle \Delta \ff s_{i_{m}}(t) \rangle = J \sum_{m'} \ff K_{mm'}(t,\ff S(t)) \, \dot{\ff S}_{m'}(t)
\labeq{arts}
\ee
up to first order in $W$.
Here, $\ff K_{mm'}$ is a $3\times 3$ matrix for each index pair $m,m'$, which depends on $t$ explicitly, and also implicitly via $\ff S(t)$.
The elements of this matrix are given by 
\ba
&& K^{\alpha\alpha'}_{mm'} (t,\ff S(t))
= 
\int_{0}^{t} dt' \int_{0}^{\beta} du \, 
\langle 
s^{\alpha}_{i_{m}}(-iu) s^{\alpha'}_{i_{m'}}(t'-t)
\rangle^{\rm (eq)}_{\ff S(t)} 
\nonumber\\
&&-
\int_{0}^{t} dt' \int_{0}^{\beta} du \,  
\langle s^{\alpha}_{i_{m}}(-iu) \rangle^{\rm (eq)}_{\ff S(t)}
\langle s^{\alpha'}_{i_{m'}}(t'-t) \rangle^{\rm (eq)}_{\ff S(t)}
\: . 
\labeq{kdef}
\ea
Here, $\alpha, \alpha' = x,y,z$ and $u$ refers to imaginary time.
Inserting $\langle \ff s_{i_{m}} \rangle_{t}$, as obtained from Eqs.\ (\ref{eq:artd}) and (\ref{eq:arts}), into \refeq{eoms}, we get:
\ba
\dot{\ff S}_{m}(t) &=& J \langle \ff s_{i_{m}} \rangle^{\rm (eq)}_{\ff S(t)} \times \ff S_{m}(t) - \sum_{m} \ff B_{m} \times \ff S_{m}(t)
\nonumber \\
&+&
J^2 \sum_{\alpha} \sum_{m'\alpha'} 
K_{mm'}^{\alpha\alpha'} (t,\ff S(t)) \dot{S}_{m'\alpha'} 
\boldsymbol{e}_{\alpha} \times \boldsymbol{S}_{m}(t)
\: .
\nonumber \\
\labeq{ins}
\ea

In the second step, we additionally assume that $J$ is weak. 
For a fixed spin configuration $\ff S(t)$, the static, equilibrium expectation value $\langle \ff s_{i_{m}} \rangle^{\rm (eq)}_{\ff S(t)}$ in the first term on the right-hand side of \refeq{ins} can be expanded in powers of $J$ as
\be
\langle \ff s_{i_{m}} \rangle^{\rm (eq)}_{\ff S(t)}
=
\langle \ff s_{i_{m}} \rangle^{\rm (eq)}_{J=0}
+
J
\sum_{m'}
\chi_{mm'} \ff S_{m'}(t)
+ \ca O(J^{2})
\: , 
\labeq{resp}
\ee
where $\chi_{mm'} \equiv \chi_{mm'}(\omega=0)$ is the static and unperturbed ($J=0$) magnetic susceptibility, i.e., the $\omega=0$ Fourier component of $\chi_{mm'}^{\alpha\alpha'}(t) = \delta^{\alpha\alpha'} \chi_{mm'}(t)$ defined in \refeq{chi}.
The first term $\langle \ff s_{i_{m}} \rangle^{\rm (eq)}_{J=0}$ vanishes, as there is no spontaneous magnetic order for a system of noninteracting conduction electrons and since the magnetic field $\ff B_{m}$ only couples to the impurity spin $\ff S_{m}$. 
With this, the first term on the right-hand side of \refeq{ins} reduces to the RKKY term $\sum_{m'} J^2 \chi_{mm'} \ff S_{m'}(t) \times \ff S_{m}(t)$, if terms of order $J^{3}$ are neglected.
Since the $K$ matrix in \refeq{ins} already carries a $J^{2}$ factor, we can then disregard its dependence on $\ff S(t)$ as this would produce terms of $\ca O(J^{3})$ as well.
With the same argument and since the model is SU(2) symmetric and does not support spontaneous magnetic order, the second term on the right-hand side of \refeq{kdef} can be disregarded.
We are left with 
\be
K^{\alpha\alpha'}_{mm'} (t)
= \delta^{\alpha\alpha'}
\int_{0}^{t} dt' \int_{0}^{\beta} du \, 
\langle 
s^{\alpha}_{i_{m}}(-iu) s^{\alpha'}_{i_{m'}}(t'-t)
\rangle^{(0)}
\: . 
\labeq{kfin}
\ee

Inserting a resolution of the identity $\ff 1 = \sum | \Psi_{n} \rangle \langle \Psi_{n} |$ with energy eigenstates between the spin operators in \refeq{kfin} and expressing the expectation value as $Z^{-1} \sum e^{-\beta E_{n}} \langle \Psi_{n'} | \cdots | \Psi_{n'} \rangle$, one can derive a Lehmann-type representation. 
This involves matrix elements $\langle \Psi_{n} | s^{\alpha}_{i} | \Psi_{n'} \rangle$ with eigenstates of the $J=0$ Hamiltonian $\hat{H}_{\rm el}$.
In the sector with even total particle number $N$, the latter is invariant under time reversal, $[\Theta, \hat{H}_{\rm el}] = 0$, where the representation of 
time reversal is given by an antiunitary operator $\Theta$ with $\Theta^{2} = \ff 1$, $\Theta \Theta^{\dagger} = \ff 1$, and $\Theta \ff s_{i} \Theta^{\dagger} = - \ff s_{i}$.
A straightforward consequence is that the matrix elements 
$\langle \Psi_{n} | s^{\alpha}_{i} | \Psi_{n'} \rangle
=
\langle \Theta \Psi_{n} | s^{\alpha}_{i} | \Theta \Psi_{n'} \rangle
=
\langle \Psi_{n} | \Theta^{\dagger} s^{\alpha}_{i} \Theta | \Psi_{n'} \rangle^{\ast}
=
- \langle \Psi_{n} | s^{\alpha}_{i} | \Psi_{n'} \rangle^{\ast} \in i \mathbb{R}$ 
are purely imaginary (see also Ref.\ \onlinecite{MP22}).
This immediately implies that $\ff K$ is symmetric $K^{\alpha\alpha'}_{mm'}(t) = K^{\alpha'\alpha}_{m'm}(t)$. 

It is important to see that there is a finite {\em antisymmetric} part for strong $J$, i.e., in the regime where the weak-$J$ approximation does not apply, e.g., for systems directly coupled to an external magnetic field, or systems where time-reversal symmetry is broken explicitly in the electronic sector, as in the case of the Haldane model \cite{Hal88}.
In the equations of motion for the classical spins, this would lead to an additional geometrical spin torque
\be
\sum_{\alpha} \sum_{m'\alpha'} 
\Omega_{mm'}^{\alpha\alpha'} (\ff S(t)) \dot{S}_{m'\alpha'} 
\boldsymbol{e}_{\alpha} \times \boldsymbol{S}_{m}(t)
\ee
resulting from the spin-Berry curvature 
\be
\Omega_{mm'}^{\alpha\alpha'}(\ff S(t))
= J^{2} \frac{1}{2} \left( 
K_{mm'}^{\alpha\alpha'}(\ff S(t)) - K_{m'm}^{\alpha'\alpha}(\ff S(t)) 
\right)
\: . 
\ee
Previously, this was derived within the framework of adiabatic spin dynamics \cite{SP17,MP22,LLP22,LKP23}.
The spin-Berry curvature represents the feedback \cite{BR93b} of the Berry-phase physics \cite{Ber84} in the electron system on the dynamics of the classical degrees of freedom. 
Here, we see that the same geometrical spin torque can be derived within the context of adiabatic response theory \cite{CDH12} as well.
Importantly, the spin-Berry curvature is a {\em geometrical} object and only depends on $t$ via $\ff S(t)$, {\em if} the explicit $t$ dependence of 
$K_{mm'}^{\alpha\alpha'}(t, \ff S(t))$ in \refeq{kdef} can be disregarded, i.e., in the long-time limit.

Let us return to our main focus, namely spin damping. 
Generally, we define
\be
\overline{K}_{mm'}^{\alpha\alpha'}(t)
= \frac{1}{2} \left( 
K_{mm'}^{\alpha\alpha'}(t,\ff S(t)) + K_{m'm}^{\alpha'\alpha}(t,\ff S(t)) 
\right)
\: . 
\ee
Specifically, for the time-reversal and spin-SU(2) symmetric model $\hat{H}_{\rm el}$ considered here and in the weak-$J$ limit, the symmetrization is superfluous and, furthermore, the $3\times 3$ tensor is actually isotropic, i.e., $\overline{K}^{\alpha\alpha'}_{mm'}(t) = \delta^{\alpha\alpha'} \overline{K}_{mm'}(t)$.
Hence, the last term on the right-hand side of \refeq{ins} reads:
\be
\sum_m \alpha_{mm'}(t) \dot{\ff S}_{m'}(t)  \times \ff S_{m}(t)
\labeq{alph}
\ee
when identifying 
\be
\alpha_{mm'}(t) = J^{2} \overline{K}_{mm'}(t)
\labeq{arta}
\: .
\ee
With the term \refeq{alph} and with \refeq{resp} inserted into \refeq{ins}, we find the same spin-only effective equation of motion (\ref{eq:eomeff}) that was derived within linear response theory and the subsequent perturbative treatment of retardation effects, but with two exceptions: 
The RKKY coupling constants are time-independent, and the expression \refeq{alph} for the damping constants differs from \refeq{talpha}.

\section{Computational details}
\label{sec:comp}

The fundamental equations of motion (\ref{eq:eoms}), (\ref{eq:eomr}) as well as the effective equations of motion (\ref{eq:llg}) can be solved using standard numerical techniques \cite{Ver10,RN17,Ste23}. 
On time scales up to $\sim 10^{9}$ in units of the inverse nearest-neighbor hopping $1/T$, the achieved numerical accuracy is sufficient, i.e., effects of numerical errors are invisible in all spin-dynamics plots shown below. 

We consider a tight-binding model of noninteracting electrons $\hat{H}_{\rm el}$ on a $D$-dimensional lattice with periodic boundary conditions such that diagonalization of the hopping matrix $\ff T = \ff U \ff \varepsilon \ff U^{\dagger}$ is achieved analytically via Fourier transformation. 
Hence, the tight-binding dispersions are given by 
\be
\varepsilon(k) = -2T \cos(k) - 2T' \cos(2k)
\labeq{1dtb}
\ee
for $D=1$, where the nearest-neighbor hopping $T=1$ fixes the energy (and with $\hbar\equiv 1$ the time scale) and where $T'$ is the next-nearest-neighbor hopping. 
The lattice constant is set to unity as well. 
For $D=2$, we have
\be
\varepsilon(\ff k) = -2T (\cos( k_x )+ \cos(k_y)) - 4T' \cos(k_x)\cos(k_y)
\: .
\labeq{tb2}
\ee

\begin{widetext}

The necessary ingredients for the effective equations (\ref{eq:llg}) or, more generally, for the effective equations of motion with time-dependent parameters, \refeq{eomeff}, can be computed numerically as follows.

We start with the RKKY interaction parameters $J_{mm'}(t)$. 
These are obtained from \refeq{trkky} by performing the $\tau$ integration numerically.
For the integrand, we use the representation
\be
\chi_{mm'}(\tau) 
=
\Theta(\tau)e^{-\eta \tau} \Im
\left[
\left(\frac{e^{i \ff T \tau}}{e^{\beta(\ff T - \mu \ff 1)} + \ff 1} \right)_{i_{m'}i_{m}} 
\left(\frac{e^{-i \ff T \tau}}{\ff 1 + e^{-\beta(\ff T - \mu \ff 1)}}\right)_{i_{m}i_{m'}}
\right]
\ee
of the ($J=0$) retarded magnetic susceptibility, where we made use of the fact that the hopping matrix is diagonal in the spin indices $\sigma, \sigma'$ and spin independent. 
With $U_{i \ff k} = \frac{1}{\sqrt{L}} e^{-i \ff k \ff R_i}$, and in the zero-temperature limit $\beta \to \infty$, one gets
\be
\chi_{mm'}(\tau) = \Theta(\tau) e^{-\eta \tau}  
\Im \left[
\frac{1}{L^2}
\sum_{\ff k}^{\rm occ}
e^{i \varepsilon(\ff k)\tau}
e^{i \ff k (\ff R_{i_{m'}} - \ff R_{i_{m}})} 
\sum_{\ff k'}^{\rm unocc} 
e^{-i \ff k'(\ff R_{i_{m}} - \ff R_{i_{m'}})}
e^{- i\varepsilon(\ff k')\tau} 
\right]
\: .
\labeq{chinum}
\ee
We see that the susceptibility is symmetric with respect to $m$, $m'$. 
As argued above, this is a consequence of the time-reversal symmetry of $\hat{H}_{\rm el}$.

Importantly, the wave-vector summations over the occupied or unoccupied points in the first Brillouin zone factorize.
This allows us to address fairly large systems and to control the thermodynamic limit. 
The $L\to \infty$ limit should be taken with a finite $\eta>0$ in the regularization factor $e^{-\eta \tau}$ which appears in the definition of the retarded susceptibility \refeq{chi} and which ensures the convergence of the limit $t \to \infty$ in \refeq{trkky}.
The latter is necessary for the computation of the time-independent RKKY coupling $J_{mm'} = \lim_{t \to \infty} J_{mm'}(t) = J^2 \chi_{mm'}(\omega=0)$.
The limit $\eta \searrow 0$ should be taken in the end. 

Within linear response theory, the time-dependent Gilbert-damping parameters $\alpha_{mm'}(t)$ are obtained from \refeq{talpha} via numerical integration. 
The time-independent parameters $\alpha_{mm'}$ are obtained from \refeq{alphados}. 

Within adiabatic response theory and for weak $J$, the identification \refeq{arta} leads us exactly to the same effective equations of motion (\ref{eq:eomeff}) that were obtained within linear response theory, albeit with a different expression for $\alpha_{mm'}(t)$ as compared to \refeq{talpha}.
For its numerical evaluation we first consider \refeq{kfin}. 
Making use of isotropy and time-reversal symmetry, $K_{mm'}^{\alpha\alpha'}(t) = \overline{K}^{\alpha\alpha'}_{mm'}(t) = \delta^{\alpha\alpha'} K_{mm'}(t)$, inserting a resolution of the identity with an orthonormal basis of eigenstates of $\hat{H}_{\rm el}$ as described below \refeq{kfin}, and evaluating the corresponding matrix elements of $s_{i_{m}}^{\alpha}$ and $s_{i_{m'}}^{\alpha'}$, yields:
\be
K_{mm'}(t) 
= 
\frac{1}{2} \int_{0}^{t} dt' \int_{0}^{\beta} du 
\left( \frac{e^{\ff T(it' + u)}}{e^{\beta(\ff T - \mu \ff 1)} + \ff 1} \right)_{m'm}  
\left( \frac{e^{-\ff T(it' + u)}}{\ff 1 + e^{-\beta(\ff T - \mu \ff 1)}} \right)_{mm'} 
\: ,
\ee
and, after diagonalization of the hopping matrix $\ff T$, 
\be
K_{mm'}(t) 
= 
\frac{1}{2} 
\frac{1}{L^2} \sum_{\ff k \ff k'} 
f(\varepsilon(\ff k) - \mu)
f(\mu - \varepsilon(\ff k'))
e^{i \ff k (\ff R_{i_m} - \ff R_{i_{m'}})} 
e^{-i \ff k' (\ff R_{i_m} - \ff R_{i_{m'}})}
\int_{0}^{t} dt' \int_{0}^{\beta} du \, 
e^{(\varepsilon(\ff k) - \varepsilon(\ff k'))(u + it^\prime)}
\: . 
\ee
Here, $f(x) = 1 / (e^{\beta x} + 1)$ is the Fermi function.
We carry out the integration over imaginary time $u$ with the case distinction (i) $\varepsilon(\ff k) \ne \varepsilon(\ff k')$ and (ii) $\varepsilon(\ff k) = \varepsilon(\ff k')$. 
Using the identity
$(e^{\beta(x-x')} - 1) f(x) f(-x') = f(-x) f(x') - f(x) f(-x')$
in addition and carrying out the $t'$ integration, we arrive at:
\be
K_{mm'}(t) 
= 
\frac{1}{L^2} \sum_{\ff k \ff k'}^{\varepsilon(\ff k) \neq \varepsilon(\ff k')}  
f(\varepsilon(\ff k) - \mu) f(\mu - \varepsilon(\ff k')) \,
\frac{
\Im[
e^{i (\ff k - \ff k')(\ff R_{i_m} - \ff R_{i_{m'}})} 
(1-e^{i (\varepsilon(\ff k) - \varepsilon(\ff k'))t} )
]
}
{(\varepsilon(\ff k) - \varepsilon(\ff k'))^2} 
+ 
K^{\rm (deg)}_{mm'}(t)
\: , 
\ee
where the first term refers to case (i) and where the second term, case (ii), is given by:
\be
K^{\rm (deg)}_{mm'}(t)
= 
\frac{1}{2} \frac{\beta t}{L^2} 
\sum_{\ff k \ff k'}^{\varepsilon(\ff k) = \varepsilon(\ff k')}  
f(\varepsilon(\ff k)-\mu)  f(\mu - \varepsilon(\ff k')) \, 
e^{i \ff k(\ff R_{i_m} - \ff R_{i_{m'}})} e^{-i \ff k' (\ff R_{i_m} - \ff R_{i_{m'}})} 
\: . 
\ee
In the thermodynamic limit $L\to \infty$, this second term vanishes, unless there is a macroscopic number of degeneracies in the tight-binding dispersion $\varepsilon(\ff k)$.
Assuming that this is not the case, taking the zero-temperature limit $\beta \to \infty$, and inserting into \refeq{arta} we find:
\be
\alpha_{mm'}(t) 
=  
J^{2}
\frac{1}{L^2} 
\sum_{\ff k}^{\rm occ} 
\sum_{\ff k'}^{\rm unocc}
\frac{
\Im[
e^{i (\ff k - \ff k')(\ff R_{i_m} - \ff R_{i_{m'}})} 
(1 - e^{i (\varepsilon(\ff k) - \varepsilon(\ff k'))t} )
]
}{
(\varepsilon(\ff k) - \varepsilon(\ff k'))^2
}
\: .
\ee
Unfortunately, the numerical evaluation is much more time-consuming as compared to linear response theory, i.e., \refeq{talpha} and \refeq{chinum}. 

\end{widetext}

\section{Results}
\label{sec:results}

\subsection{Time-dependent spin friction}
\label{sec:alpha}

\begin{figure}[t]
\includegraphics[width=0.9\linewidth]{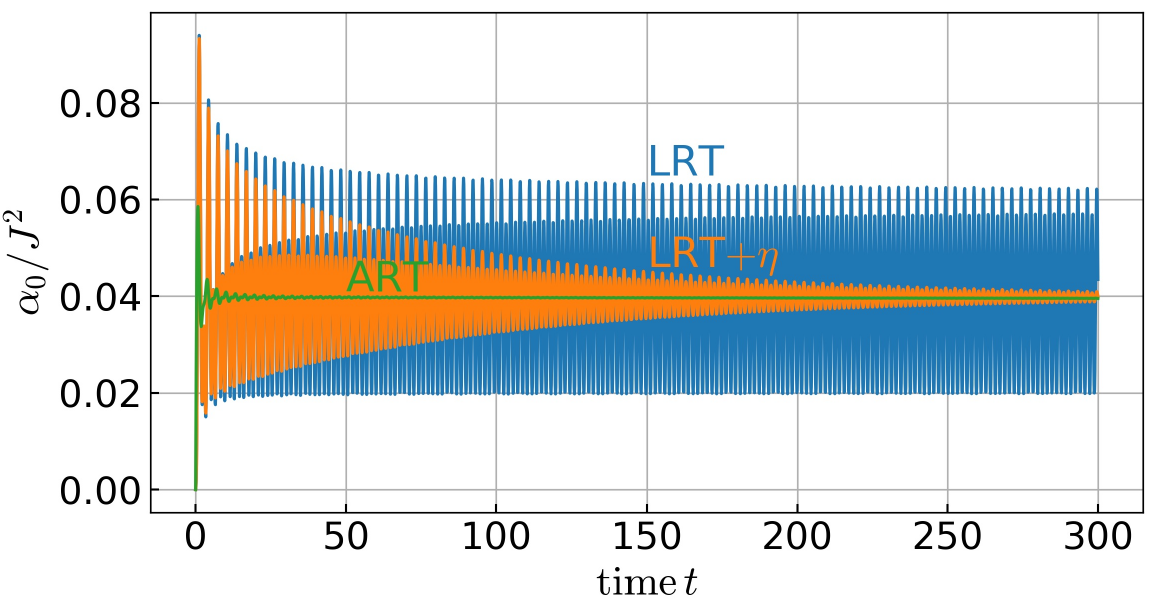}
\caption{
Time-dependent local Gilbert damping $\alpha_{0}(t) / J^{2}$ as obtained from linear response theory (blue) and from adiabatic response theory (green) for the one-dimensional system at half filling.
$T=1$ sets the energy scale (and with $\hbar=1$ the time scale).
Orange line: linear response theory with $\eta = 0.01$.
Calculations for $T'=0$, $L=10^{5}$, and periodic boundary conditions.
}
\label{fig:alpha}
\end{figure}

\begin{figure}[t]
\includegraphics[width=0.9\linewidth]{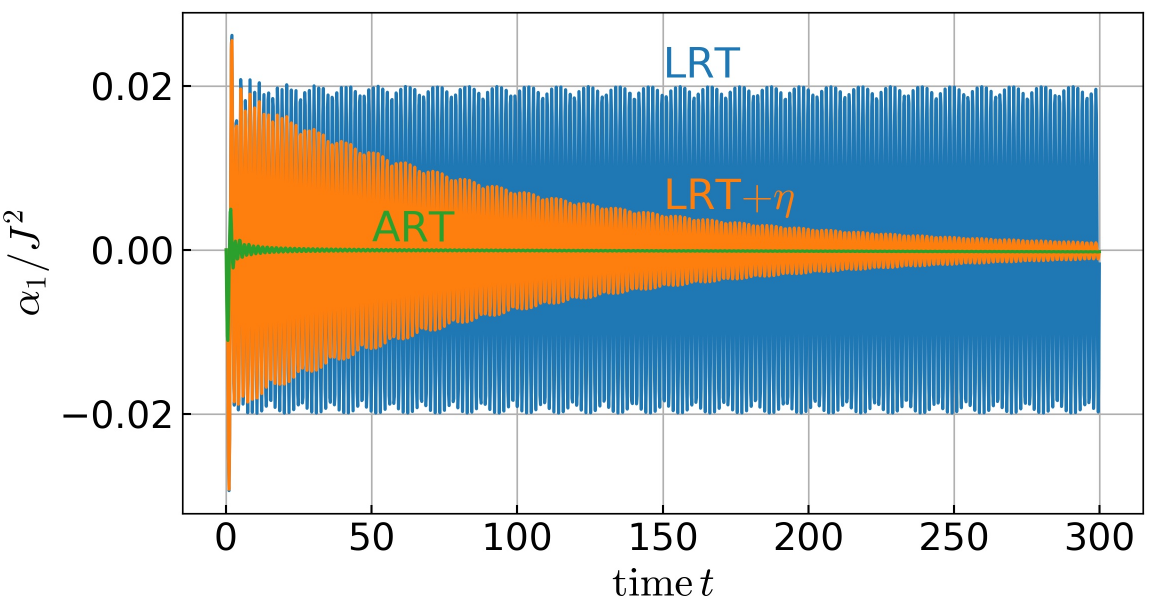}
\caption{
The same as Fig.\ \ref{fig:alpha} but for the nearest-neighbor damping $\alpha_{1}(t)/J^{2}$, i.e., $d=i_{m'} - i_{m}=1$.
}
\label{fig:alphann}
\end{figure}

In the effective spin-only theory, the relaxation towards the ground-state spin configuration is determined by the Gilbert damping $\alpha_{mm'}(t)$. 
As shown above, there are at least three different ways to calculate this quantity, namely (i) via linear response theory (LRT), \refeq{talpha}, or (ii) via adiabatic response theory (ART), \refeq{arta}. 
Finally, (iii) one usually assumes time-independent damping constants $\alpha_{mm'} = \lim_{t\to \infty} \alpha_{mm'}(t)$.
We start the discussion by a corresponding comparison.
This is done for a one-dimensional tight-binding model with dispersion \refeq{1dtb} at half filling. 
The next-nearest-neighbor hopping is set to $T'=0$, for simplicity.

Fig.\ \ref{fig:alpha} shows results for the local Gilbert damping for a system with $L=10^{5}$ sites. 
Periodic boundary conditions are assumed such that $\alpha_{mm}(t) = \alpha_{0}(t)$ is $m$ independent. 
Within LRT, we find that $\alpha_{0}(t)$ exhibits an undamped oscillation, after some initial decay at shorter times. 
Fourier analysis yields a dominating frequency $\omega_{\rm vH} \approx 2.0$, which stems from the van Hove singularities at $\omega_{\rm vH}=\pm 2$ in the local density of states, as already explained in Ref.\ \cite{SP15}. 
Within ART the same oscillation frequency is found, but there is a strong damping of the oscillation such that $\alpha_{0}(t)$ converges to an essentially constant value $\alpha_{0} \approx 0.040 J^2$ on a time scale of roughly $t=50$. 

For the nearest-neighbor Gilbert damping $\alpha_{1}$, the same qualitative behavior is observed for both approaches, LRT and ART, as can be seen in Fig.\ \ref{fig:alphann}. 
However, the time-independent nearest-neighbor damping constant vanishes, $\alpha_{1} \approx 0$.
The same values for the local and the nearest-neighbor damping, $\alpha_{0} \approx 0.040 J^{2}$ and $\alpha_{1} \approx 0$, are obtained from LRT with a finite regularization parameter $\eta$. 
In Figs.\ \ref{fig:alpha} and \ref{fig:alphann}, results are shown for $\eta = 0.01$ (orange lines). 
However, within numerical accuracy, the damping constants, $\alpha_{0}$ and $\alpha_{1}$, do not depend on the choice for $\eta$, as long as $\eta$ is sufficiently small (but $\eta \gtrsim 1/L$, since the limit $\eta \searrow 0$ must be taken after the limit $L\to \infty$).
This independence of $\eta$ is expected, see the discussion in Refs.\ \onlinecite{SP15,EP24} and the discussion of the two-dimensional systems below.
It is remarkable, however, that both approaches, LRT and ART, yield the same local and nonlocal damping constants.

\subsection{Nonlocal spin friction}
\label{sec:nalpha}

We have also calculated $\alpha_{d}$ for larger distances $d = i_{m'}-i_{m}$.
Both approaches precisely reproduce the finding of Refs.\ \onlinecite{RON24,EP24}, namely
$\alpha_{d+2} = \alpha_{d} = \mbox{const} >0$ for even distances $d=0,2,4,...$ and  
$\alpha_{d+2} = \alpha_{d} = \mbox{const} =0$ for odd $d$. 
This distance (in-)dependence of $\alpha_{d}$ is a characteristic feature for the $D=1$ model at half filling.

\begin{figure}[t]
\includegraphics[width=0.9\linewidth]{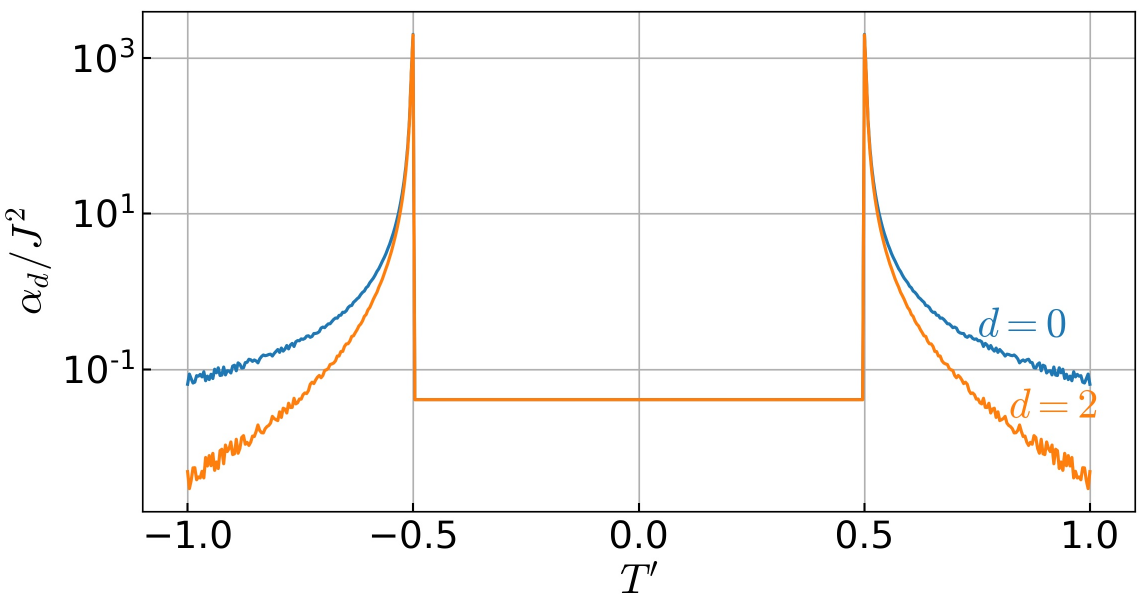}
\caption{
Local and next-nearest-neighbor damping parameters, $\alpha_{0}$ and $\alpha_{2}$, for the $D=1$ system at half filling as a function of the next-nearest-neighbor hopping $T'$. 
Calculations for $L = 10^6$.
}
\label{fig:tprime}
\end{figure}

Another, at first sight unintuitive numerical result is the $T'$ dependence of the local and the next-nearest-neighbor Gilbert damping. 
This is shown in Fig.\ \ref{fig:tprime}.
We find that $\alpha_{0}(T') = \alpha_{2}(T')=\mbox{const}$ for all $T'$ with $-T'_{\rm c} < T' < T'_{\rm c}$. 
The critical next-nearest-neighbor hopping is given by $T'_{\rm c} = T/2$.
Furthermore, $\alpha_{1}(T') = \alpha_{3}(T') = 0$ within the same $T'$ range (not shown).
For $| T' | > T'_{\rm c}$, the $T'$ dependence of $\alpha_{d}(T')$ is nontrivial. 
Again, this result is characteristic for $D=1$ and half filling.

\begin{figure}[t]
\includegraphics[width=0.9\linewidth]{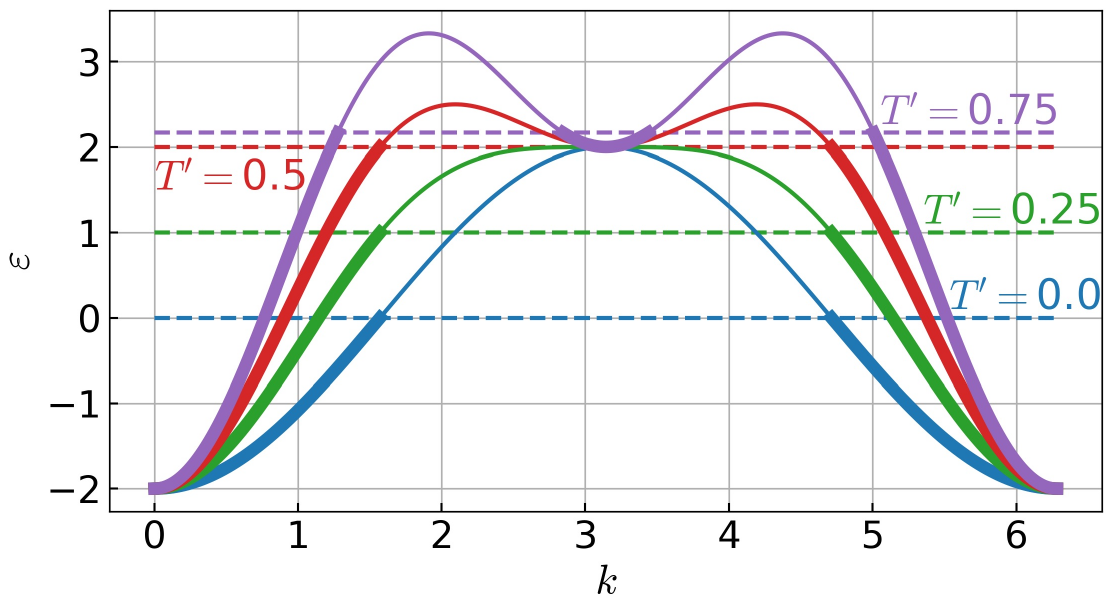}
\caption{
Dispersion (\ref{eq:1dtb}) for various $T'$. 
{\em Blue:} $T'=0.0$ (no offset).
{\em Green:} $T'=0.25$, with vertical offset $\Delta \varepsilon = 0.5$. 
{\em Red:} $T'=0.50$ ($\Delta \varepsilon = 1.0$). 
{\em Violet:} $T'=0.75$  ($\Delta \varepsilon = 1.5$). 
Horizontal dashed lines: respective chemical potentials $\mu$, corresponding to half filling.
Thick lines indicate occupied states. 
}
\label{fig:epsk}
\end{figure}

There is a simple proof for both, the peculiar $d$ and $T'$ dependencies of $\alpha_{d}(T')$, which is based on the density-of-states formula \refeq{dosform}.
We first note that $T'_{\rm c} = T/2$ is the critical value for a Lifschitz transition of the Fermi ``surface''. 
If $-T'_{\rm c} < T' < T'_{\rm c}$, the occupied $k$ points lie in the range $-k_{\rm F} < k < k_{\rm F}$ with $k_{\rm F} = \pi /2$ in the first Brillouin zone $[ -\pi, \pi ]$, i.e., $k_{\rm F}$ is independent of $T'$ in this $T'$ range.
However, for $|T'| > T'_{\rm c}$, the ``Fermi-surface volume'' splits into three disconnected parts.
This is illustrated with Fig.\ \ref{fig:epsk}.

For $D=1$, $-T'_{\rm c} < T' < T'_{\rm c}$, and for $L\to \infty$, we can express the spectral density $A_{d}(0) \equiv A_{i_{m} i_{m'}}(\omega = 0)$ with $d=i_{m'}-i_{m}$ in the form
\ba
A_{d}(0) 
&=&
\frac{1}{2\pi} \int_{-\pi}^{\pi} dk \, e^{ikd} \delta(\mu - \varepsilon(k))
\nonumber \\
&=&
\frac{1}{2\pi} \sum_{k_{\rm F}}  e^{ik_{\rm F} d}
\left(
\frac{d \varepsilon(k)}{dk} \Big|_{k=k_{\rm F}}
\right)^{-1}
\nonumber \\
&=&
\frac{1}{\pi} \cos \left( \frac{\pi}{2} d\right) \frac{1}{2 T} 
\: ,
\labeq{alphaa}
\ea
with Fermi wave vectors $k_{\rm F}=\pm \pi / 2$. 
With \refeq{dosform} this demonstrates that $\alpha_{d}(T')$ is independent of $T'$ and oscillates between $0$ and $\alpha_{0} = J^{2} / 8\pi T^{2} \approx 0.0398 J^{2}$  (with $T=1$) for odd and even $d$, respectively. 

\subsection{Single-spin dynamics in $D=1$}
\label{sec:sd1}

\begin{figure}[t]
\includegraphics[width=0.9\linewidth]{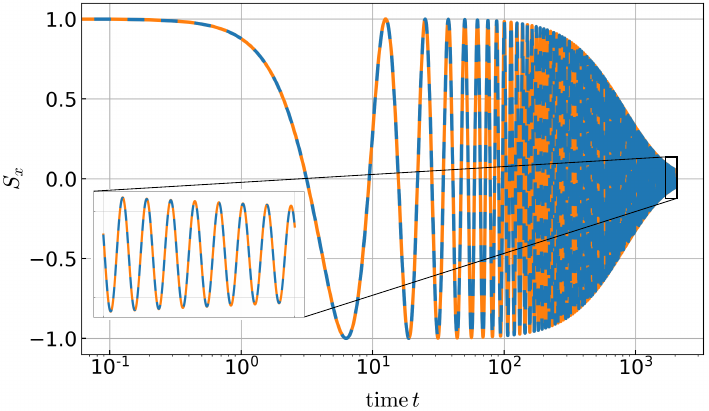}
\caption{
Time dependence of the $x$ component of a single impurity spin $\ff S$ coupled to a site of the $D=1$ tight-binding model via a local exchange interaction $J$. 
At time $t=0$, where $J$ is suddenly switched on, the spin points in the $x$ direction. 
For $t>0$, the spin dynamics is driven by a finite local magnetic field $\ff B = (0,0,B)$ in $z$ direction, which couples to $\ff S$.
{\em Dashed blue line:} LRT calculation with time-dependent damping.
{\em Solid orange line:} LRT calculation with constant damping. 
Parameters: $L = 4000$, $T'=0$, $J=0.3$, $B = 0.5$, half filling.
}
\label{fig:const}
\end{figure}

While LRT and ART yield the same Gilbert damping constant $\alpha_{mm'} = \lim_{t\to \infty} \alpha_{mm'}(t)$, the time dependence of the  damping is very different, as discussed above (Figs.\ \ref{fig:alpha}, \ref{fig:alphann}).
The resulting spin dynamics, however, turns out to be essentially independent of the approach used, at least in the regime, where the effective spin-only theory applies.
In addition, the time dependence of the damping is practically irrelevant for the spin dynamics. 

We start the related discussion with Fig.\ \ref{fig:const}, which shows the time evolution of the $x$ component of a single impurity spin coupled to a site of the one-dimensional model with $T'=0$. 
Initially, the impurity spin points in the $x$ direction, and the electron system is prepared in its ground state for $J=0$. 
At time $t=0$, we suddenly switch on the exchange coupling $J$ and, in addition, a local and time-independent magnetic field $\ff B=(0,0,B)$ in $z$ direction, which couples to the spin, i.e., $\hat{H} \mapsto \hat{H} - \ff B \ff S$ (see also the discussion at the end of \ref{sec:sd1}).
The purpose of the field is to drive the spin dynamics.
In fact, for $t>0$, the spin starts to precess around $\ff B$ with the Larmor frequency $\omega \approx B$, as seen in the oscillations of $S_{x}$ and of $S_{y}$ (not shown). 
On a time scale of a few thousand inverse hoppings the spin finally aligns to the field, i.e., $S_{z} \to 1$ (not shown) and $S_{x}, S_{y} \to 0$. 
Importantly, Fig.\ \ref{fig:const} shows the prediction of LRT, as obtained from the equation of motion
$\dot{\ff S}(t) = \alpha_{0}(t) \dot{\ff S}(t) \times \ff S(t) - \ff B \times \ff S(t)$ with time-dependent $\alpha_{0}(t)$ (dashed blue line) and for the same equation of motion but replacing $\alpha_{0}(t) \mapsto \alpha_{0}$, i.e., for time-independent $\alpha_{0}$ (solid orange line).

\begin{figure*}[t]
\includegraphics[width=0.99\linewidth]{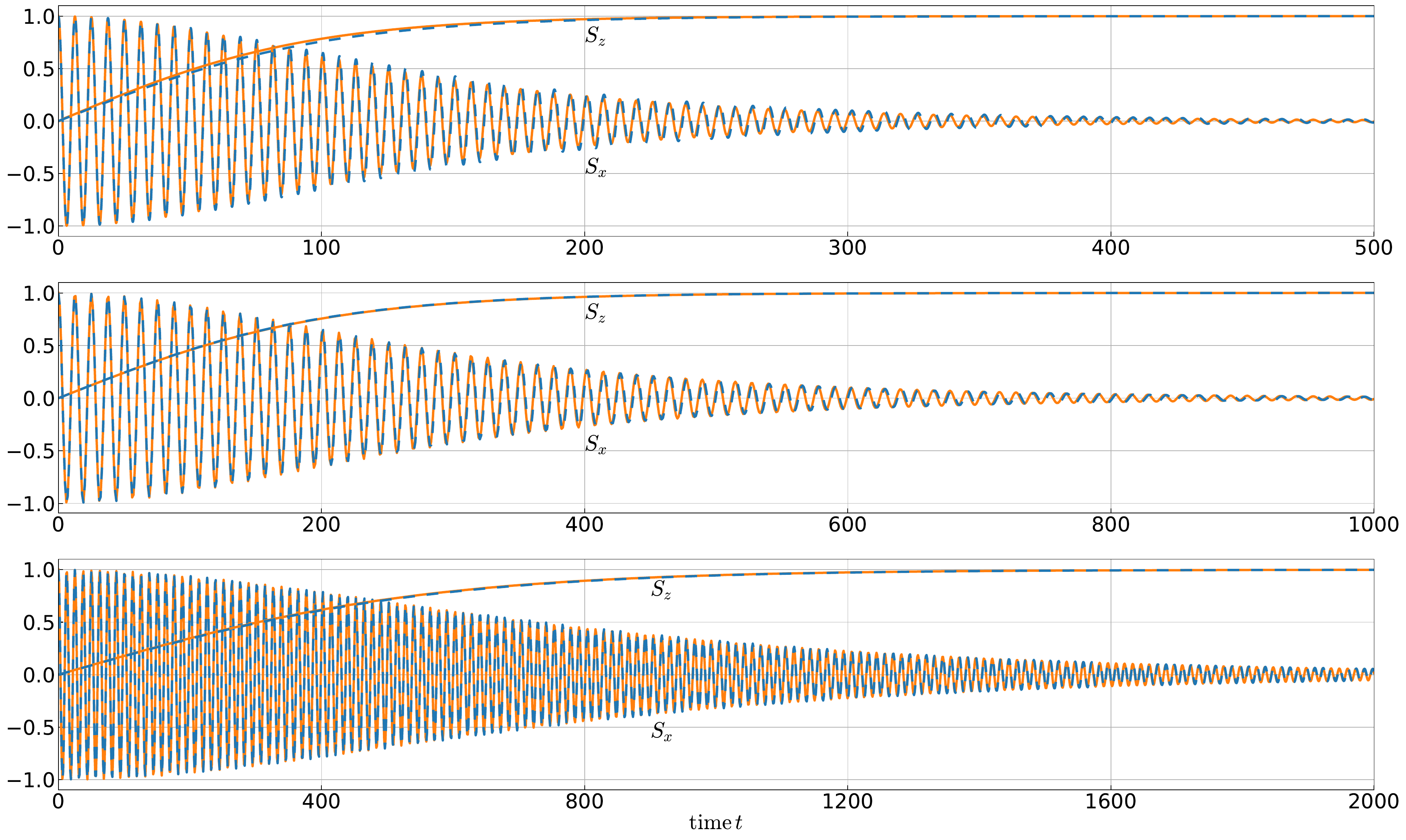}
\caption{
Time dependence of $S_{x}$ and of $S_{z}$ as obtained from the full theory (solid orange line), Eqs.\ (\ref{eq:eoms}) and (\ref{eq:eomr}), compared to the effective theory (dashed blue). 
{\em Top panel:}
Calculations with $J = 0.5$ and $B= 1.0$ for a chain of length $L=1000$ with periodic boundaries. $T'=0$.
{\em Middle panel:}
The same but for $J = 0.5$ and $B = 0.5$. 
System size: $L=2000$.
Note the different time scale.
{\em Bottom panel:}
The same but for $J = 0.3$ and $B = 0.5$. 
System size: $L=4000$. 
Note the different time scale.
}
\label{fig:dyn}
\end{figure*}

There are no significant differences, neither at early times nor for times close to the relaxation time.
Furthermore, inspection of the numerical data shows that the agreement becomes even better with decreasing field strength. 
The obvious interpretation is that the precession time scale $1/B$ set by the field strength $B$ is long as compared to the time scale for the oscillations of $\alpha_{d}$ that is set by the electronic band width, such that only the average over many $\alpha$ oscillations is actually relevant. 
This argument becomes increasingly pertinent with weaker (and thus more physical) field strengths $B$.

We have also calculated the spin dynamics within ART. 
The result perfectly agrees with that shown in Fig.\ \ref{fig:const}. 
After the previous discussion, this was only to be expected, since within ART $\alpha_{0}(t)$ oscillates with the same frequency as seen within LRT (cf.\ Fig.\ \ref{fig:alpha}) and converges to the constant $\alpha_{0}$ after a few oscillations anyway.
We conclude that the time dependence of the Gilbert damping is practically irrelevant for the resulting spin dynamics within both approaches, LRT and ART.
From this point on, we therefore use LRT with time-independent damping constants $\alpha_{mm'}$. 

The predictions of the effective spin-only theory agree very well with the results obtained from the full theory, i.e., from Eqs.\ (\ref{eq:eoms}) and (\ref{eq:eomr}). 
For a numerical evaluation of the full theory up to a propagation time $t_{\rm prop}$ and avoiding unwanted finite-size effects, i.e., interferences due to excitations propagating back to the impurity spin (in case of periodic boundary conditions), one must consider systems with 
$L \gtrsim v t_{\rm prop}$, where $v = d \varepsilon(k_{\rm F}) / dk$ is the Fermi velocity. 
At half filling and for $T'=0$, we have $v_{\rm F} = 2$.
This implies that a system with $L=1000$ sites is sufficiently large to see complete spin relaxation, if the relaxation time is $\tau = 500$.

As can be seen in Fig.\ \ref{fig:dyn}, this is the case, when choosing $J=0.5$ and $B=1.0$ (see top panel).
While $S_{x}$ (and $S_{y}$, not shown) oscillates with the Larmor frequency $\omega \approx B$, the $z$ component of the spin monotonously increases from $S_{z}=0$ for $t=0$ to $S_{z} \approx 1$ for $t \approx 300$.
The dynamical evolution is qualitatively the same for both, the full theory and the effective spin-only theory. 
The agreement is even quantitative, there is merely a small difference of a few per cent between full and effective theory visible in the $z$ component of $\ff S(t)$ around $t = 100$, and the relaxation is somewhat faster in the full theory. 

Physically relevant coupling strengths $J$ and magnetic fields $B$ (or effective ``Weiss'' fields produced via RKKY exchange in a multi-impurity-spin system) are generically much weaker. 
A weaker field $B$ implies a smaller precession frequency and thus a slower spin dynamics. 
Hence, this implies a more adiabatic motion and thus improves the short-retardation-time approximation. 
In fact, comparing the results at fixed $J=0.5$, for $B=1.0$ (Fig.\ \ref{fig:dyn}, top panel) with those obtained for $B=0.5$ (middle panel), one finds an even better agreement between full and effective theory for the smaller field strength.
Note that the relaxation time increases with decreasing $B$. 
We have thus extended the maximum propagation time to $t_{\rm prop}=1000$ and, accordingly, the system size to $L=2000$. 

Similarly, a weaker $J$, at fixed $B$, improves the weak-coupling approximation. 
In the bottom panel of Fig.\ \ref{fig:dyn}, for $J=0.3$ and $B=0.5$, the agreement between full and effective theory is perfect on the scale of the figure. 
The relaxation time increases once more.

The results discussed so far have been obtained by simultaneously switching on the coupling $J$ and the field $B$ at time $t=0$.
This setup for initiating the dynamics is the one that is conceptually consistent with the linear-response approach. 
Alternatively, we have tentatively initiated the dynamics by starting from the coupled system at finite $J$ and switching on the field $B$ only. 
Using this second setup and within the full theory, we did not find any significant differences in the spin dynamics. 
This is easily understood, since (i) for the considered coupling strengths there is only a very weak polarization of the conduction-electron local magnetic moment $\langle \ff s_{i_{0}} \rangle$ at $t=0$.
At $J=0.5$, for example, its magnitude amounts to $|\langle \ff s_{i_{0}} \rangle| \approx 0.064$, i.e., almost an order of magnitude smaller that the saturation value, and hence its feedback effect on the spin dynamics is weak. 
(ii) While in the first setup $\langle \ff s_{i_{0}} \rangle = 0$ at time $t=0$, the moment very quickly polarizes {\em in the course of time} to the same value $|\langle \ff s_{i_{0}} \rangle| \approx 0.064$ that is found at $t=0$ in the second setup. 
In fact, the polarization takes place on the fast electronic time scale and is fully completed already after $t=1/T=1$. 
We can thus state that the spin dynamics practically does not depend on the preparation of the initial state.

\subsection{Van Hove singularity}
\label{sec:vh}

Besides $J$, the strength of the (local) Gilbert damping also depends on the density of states at the Fermi energy, as Eq.\ (\ref{eq:dosform}) demonstrates. 
A divergence of the density of states at the Fermi energy then implies a divergent damping constant $\alpha_{0}$. 
The analysis of the effective equation of motion for a single spin \cite{Kik56,SP15,STZ24},
\be
\dot{\ff S}(t) 
=
\alpha_{0} \, \dot{\ff S}(t) \times \ff S(t)
\nonumber \\
-
\ff B \times \ff S(t)
\: , 
\ee
gives the exact analytical result
\be
  \tau \propto \frac{1 + \alpha_{0}^{2}}{\alpha_{0}} \frac{1}{B}
\ee
for $\tau$.  
The relaxation time diverges for $\alpha_{0} \to \infty$. 

We have checked this against the numerical evaluation of the full theory, Eqs.\ (\ref{eq:eoms}) and (\ref{eq:eomr}), for the half-filled system, $J=0.5$, $B=1.0$, and next-nearest-neighbor hopping $T'=-0.5$, where the density of states exhibits a strong $1/\sqrt{\omega}$ van Hove singularity at the Fermi energy (see Fig.\ \ref{fig:tprime}).
In addition, slightly smaller $T' \lesssim - 0.5$ have been considered. 
Contrary to the prediction of the effective theory, we find a finite relaxation time of the order of $\tau = \ca O(10^{2})$. 
This shows that the effective theory fails, if the (dimensionless) parameter, given by the product of $J$ with the density of states at the Fermi energy, is large.

\subsection{Two-spin dynamics in $D=1$}
\label{sec:sdtwo}

\begin{figure}[t]
\includegraphics[width=0.95\linewidth]{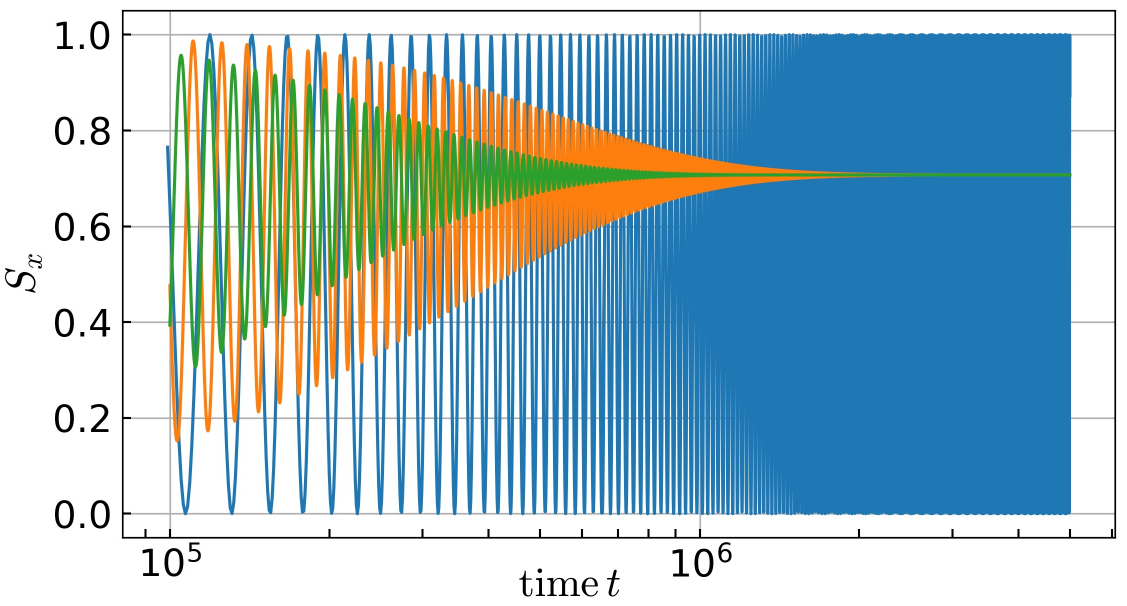}
\caption{
Time evolution of the $x$-component of $\ff S_{1}$ in a setup with two spins $\ff S_{1}$ and $\ff S_{2}$ at distance $d=2$ and for $J=0.1$.
Results for $T'=0$ (blue line), $T'=0.6$ (orange). 
The green line is the result of a calculation for $T'=0.6$, where $\alpha_{2}$ is set to zero.
Parameters: $L = 10^6$, $\eta = 10^{-5}$.
}
\label{fig:sdtprime}
\end{figure}

In case of two impurity spins at distance $d=i_{m'}-i_{m}$ and vanishing magnetic field, the effective equations of motion read: 
\ba
\dot{\ff S}_{1} = J_{d} \ff S_{2} \times \ff S_{1} + \alpha_{0} \ff S_{1} \times \dot{\ff S}_{1} + \alpha_{d} \ff S_{1} \times \dot{\ff S}_{2}
\:, 
\labeq{two1}
\\
\dot{\ff S}_{2} = J_{d} \ff S_{1} \times \ff S_{2} + \alpha_{0} \ff S_{2} \times \dot{\ff S}_{2} + \alpha_{d} \ff S_{2} \times \dot{\ff S}_{1}
\: .
\labeq{two2}
\ea
Here, we have disregarded the time dependence of $\alpha_{d}(t)$, as discussed above, but also of the RKKY coupling $J_{d}(t)$ [see \refeq{trkky}]. 
The latter shows an oscillatory time dependence with the same characteristic frequency $\omega_{\rm vH}=\pm 2$ that was found for $\alpha_{d}(t)$, but decays quickly on a time scale of a few tens of inverse hoppings. 
Using the same reasoning as for the damping, we can also ignore the $t$ dependence of $J_{d}(t)$, if $J$ is sufficiently small. 

As discussed above, for half filling and if $-T'_{\rm c} < T' < T'_{\rm c}$, the damping parameters $\alpha_{d}$ for all even $d$ are equal, while $\alpha_{d}=0$ for arbitrary odd $d$.
Hence, coupling the two impurity spins to neighboring lattice sites, the nonlocal damping vanishes, and, independent of the initial spin configuration, one finds a relaxation of the spin system to its ground state that is driven by the local damping parameter $\alpha_{0}$ only. 

For distance $d=2$, the situation is completely different (see also the discussion in Ref.\ \onlinecite{EP24}): 
Exploiting the fact that $\alpha_{0}=\alpha_{2}$ and adding the two equations (\ref{eq:two1}) and (\ref{eq:two2}), immediately implies that $\ff S_{\rm tot} = {\ff S}_{1} + \ff S_{2}$ is a constant of motion and, consequently, $\ff S_{1} \ff S_{2}$ is constant as well. 
We conclude that the spin system does not relax at all, if $d$ is even.

This is demonstrated with Fig.\ \ref{fig:sdtprime}, where the time dependence of the $x$ component of $\ff S_{1}$ is shown for a system with two impurity spins at distance $d=2$, initially prepared as mutually orthogonal, $\ff S_{1} \ff S_{2} =0$.
In fact, there is no relaxation for $T'=0$, and $S_{1x}$ exhibits an undamped oscillation.
This appears as counterintuitive. 

On the contrary, for $T' =0.6 > T'_{\rm c} =0.5$, the two-spin system relaxes to its (ferromagnetic) ground-state spin configuration.
The rather long relaxation time $\tau \sim 10^{6}$ is due to the chosen weak coupling strength $J=0.1$.
The relaxation is due to the fact that $\alpha_{0} = \alpha_{2}$ is no longer enforced by the arguments leading to \refeq{alphaa}:
At $T'=0.6$, we find $\alpha_{0} = 1.169$ and $\alpha_{2} = 0.581$.

Fig.\ \ref{fig:sdtprime} also shows the time evolution of $S_{1x}(t)$ for the same $T'=0.6$ but neglecting the nonlocal damping, i.e., the result of a calculation where we have {\em ad hoc} set $\alpha_{2}=0$. 
Intuitively, one would expect that switching off the nonlocal damping would lead to a longer relaxation time. 
However, as is seen in the figure, the opposite behavior is found, and $\tau$ in fact {\em decreases} if one sets $\alpha_{2}=0$.

\subsection{Local and nonlocal spin friction in $D=2$}
\label{sec:sd2}

\begin{figure}[t]
\includegraphics[width=0.9\linewidth]{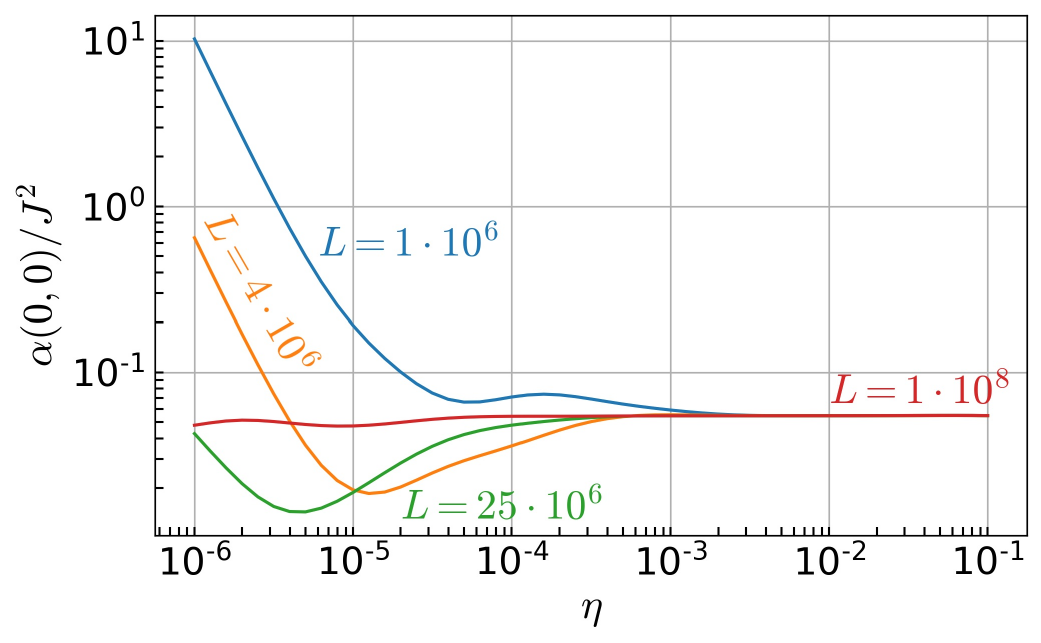}
\caption{
Local Gilbert-damping parameter $\alpha(0,0) / J^{2}$ as a function of the regularization parameter $\eta$ for different system sizes $L$. 
Calculations for the $D=2$ square lattice with next-nearest-neighbor hopping $T'=-0.3$.
}
\label{fig:etaloc}
\end{figure}

\begin{figure}[b]
\centering
\includegraphics[width=0.9\linewidth]{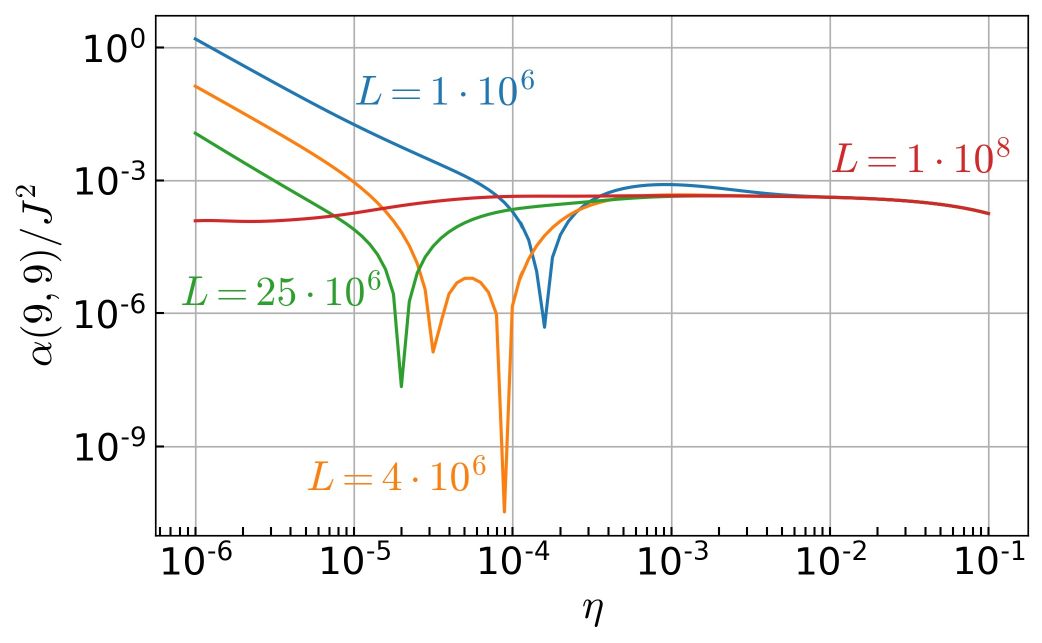}
\caption{
The same as Fig.\ \ref{fig:etaloc} but for the nonlocal Gilbert-damping parameter $\alpha(9,9) / J^{2}$ for sites linked by the distance vector $R=(9,9)$. 
}
\label{fig:etanonloc}
\end{figure}

The numerical computation of the local and the nonlocal elements of the (time-independent) Gilbert damping $\alpha_{mm'}$ on the $D=2$ square lattice proceeds along the lines described in Sec.\ \ref{sec:comp}.
As for $D=1$, a proper choice of the regularization parameter $\eta$ is decisive. 

Due to translational symmetry, the damping parameters $\alpha_{mm'}$ only depend on the distance vector $\ff R = (R_{x},R_{y})$ between the two sites $i_{m}$ and $i_{m'}$ in the square lattice.
Fig.\ \ref{fig:etaloc} shows the local damping parameter $\alpha(0,0)$, and Fig.\ \ref{fig:etanonloc} the nonlocal damping parameter $\alpha(9,9)$, i.e., for two spins $\ff S_{1}$ and $\ff S_{2}$ along the diagonal with distance vector $\ff R=(9,9)$, which is the maximum distance considered here.

Note that the limit $L\to \infty$ should be taken for a finite $\eta > 0$, and that the $\eta \searrow 0$ limit should be taken in the end. 
For $\ff R=(0,0)$ (Fig.\ \ref{fig:etaloc}) and for $\eta = 10^{-3}$, the thermodynamic limit is reached with $L \gtrsim 4\cdot 10^{6}$ in practice.
The same result for $\alpha(0,0)$ is obtained with larger parameters, e.g., with $\eta=10^{-2}$, where the thermodynamical limit is reached even earlier ($L=1\cdot 10^{6}$ is sufficient). 
Computations with, e.g., $L=10^{6}$ and $\eta = 10^{-2}$ are fully converged. 
A too small $\eta$ requires larger system sizes, otherwise one only resolves finite-size artifacts.
For a too large $\eta$, on the other hand, results start to get an unphysical $\eta$ dependence (even for sufficiently large $L$). 

Achieving convergence, $\lim_{\eta \searrow 0} \lim_{L\to\infty} (\cdots)$, is more difficult for longer distance vectors $\ff R$. 
At $\ff R=(9,9)$, see Fig.\ \ref{fig:etanonloc}, there is an $\eta$ independent plateau for sufficiently large $L$ (e.g., $L = 4\cdot 10^{6}$) in a smaller $\eta$ range, $\eta = 10^{-3} - 10^{-2}$ (note the logarithmic scale for $\alpha(\ff R)/J^{2}$).

In the rest of the paper, we consider classical spins exchange coupled to a half-filled conduction-electron system with nearest-neighbor hopping $T=1$ and finite next-nearest-neighbor hopping $T'$ on the $D=2$ square lattice. 
The tight-binding dispersion is given by \refeq{tb2}.
Considering a finite next-nearest-neighbor hopping is essential for $D=2$ and at half filling, since for $T'=0$ the density of states exhibits a logarithmic van Hove singularity at the corresponding chemical potential. 

\begin{figure}[t]
\centering
\includegraphics[width=0.9\linewidth]{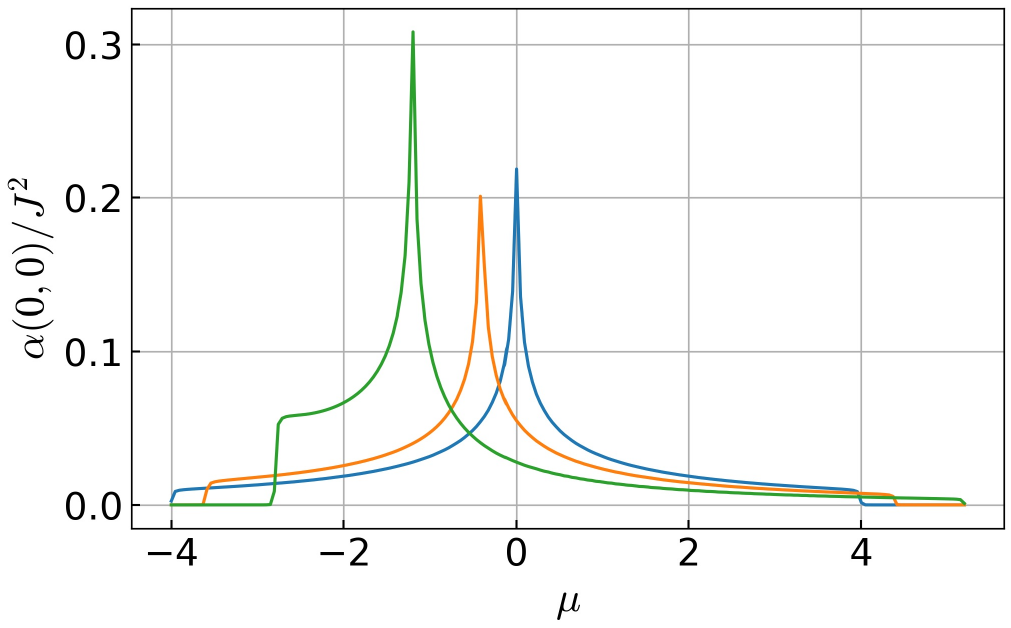}
\caption{
Local Gilbert damping $\alpha(0,0)$ for the $D=2$ tight-binding model as a function of the chemical potential $\mu$ for different values of the next-nearest-neighbor hopping. 
{\em Blue line:} $T^\prime = 0$.
{\em Orange:} $T^\prime = -0.1$. 
{\em Green}: $T^\prime = -0.3$.
Parameters: $L = 10^{6}$, $\eta = 10^{-2}$.
At $T'=-0.3$, half filling is achieved with $\mu \approx -0.66$.
}
\label{fig:alphap}
\end{figure}

This is also seen in Fig.\ \ref{fig:alphap}, which shows the $\mu$ dependence of the local damping $\alpha(0,0)$ for different $T'$. 
For $T'=0$, the chemical potential corresponding to half filling is $\mu = 0$, i.e., the Gilbert damping diverges and the effective theory would break down as discussed in Sec.\ \ref{sec:vh}. 
With decreasing $T'$, the location of the divergence shifts to lower chemical potentials. 
At $T'=-0.3$, the singularity is located at $\mu \approx -1.19$, 
while half filling is achieved with $\mu \approx -0.66$, where $\alpha(0,0) \approx 0.055$ is still small.

\subsection{Distance and directional dependence}
\label{sec:dist}

\begin{figure}[t]
\includegraphics[width=0.8\linewidth]{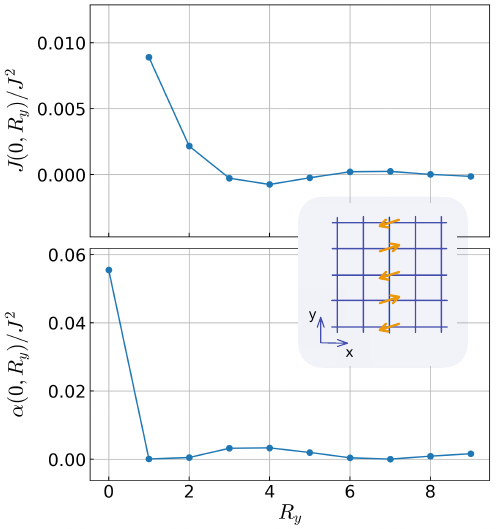}
\caption{
Distance dependence of the RKKY interaction $J(\ff R)$ (upper panel) and of the Gilbert damping $\alpha(\ff R)$ (lower panel), both normalized to unit local exchange $J$, along the $y$ axis of the square lattice, $\ff R=(0,R_{y})$. 
$T=1$, $T'=-0.3$, $L= 4 \cdot 10^{6}$, $\eta = 1 \cdot 10^{-3}$. 
The inset displays a ground-state spin configuration.
}
\label{fig:disty}
\end{figure}

In the one-dimensional case and for $T'$ with $-T'_{\rm c} < T' < T'_{\rm c}$, the counterintuitive relaxation dynamics of two spins at distance $d$ results from the extremely nonlocal spin friction given by $\alpha_{d}= \frac12 [1 + (-1)^{d}] \alpha_{0}$, as has been discussed in Sec.\ \ref{sec:nalpha} above. 

For higher-dimensional lattices, $D \ge 2$, the dependence of the Gilbert damping $\alpha(R)$ with $R=|\ff R|$ can be computed analytically, when assuming a free dispersion $\varepsilon(\ff k) = \frac{k^{2}}{2m}$. 
One easily finds: 
\be
  \alpha(R) \propto \frac{1}{R^{D-1}}  \qquad \mbox{for} \quad R \to \infty
\: .
\ee
This may be compared with the well-known distance dependence of the RKKY interaction 
\be
  J(R) \propto \frac{1}{R^{D}}  \qquad \mbox{for} \quad R \to \infty
\: ,
\ee
which decays more rapidly with $R$.
Analytical results for $\alpha(R)$ for all $R$ and including the proportionality constants (but still assuming a free dispersion) can be found in Ref.\ \onlinecite{RON24}. 

\begin{figure}[t]
\includegraphics[height=0.85\linewidth]{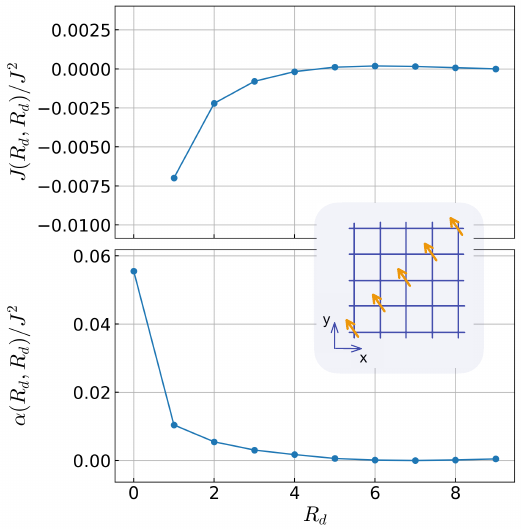}
\caption{
The same as Fig.\ \ref{fig:disty} but along the diagonal $\ff R = (R_{d}, R_{d})$ with $R_{x}=R_{y} = R_{d}$.
The inset displays a ground-state spin configuration.
\label{fig:distd}
}
\end{figure}

Here, for the tight-binding dispersion and for not too large distances, we show that the distance dependencies are less regular and that the directional dependencies are much more important.
Numerical results for the $D=2$ tight-binding system and a generic value for the next-nearest-neighbor hopping $T'=-0.3$ are shown in Figs.\ \ref{fig:disty} and \ref{fig:distd}.
Along the $y$ axis with $\ff R=(0,R_{y})$ (see Fig.\ \ref{fig:disty}), we find an oscillatory dependence of $J(0, R_{y})$ with decreasing amplitude as $R_{y}$ increases. 
The nonlocal Gilbert damping $\alpha(0, R_{y})$ is positive and shows slight oscillations with decreasing amplitude. 
Remarkably, all nonlocal elements $\alpha(0, R_{y})$ with $R_{y} \ge 1$ are smaller by more than one order of magnitude as compared to the local Gilbert damping $\alpha(0,0)$.

Along the diagonal of the square lattice (see Fig.\ \ref{fig:distd}), the RKKY coupling is negative (ferromagnetic) rather than oscillatory at short distances $R_{x}=R_{y} \le 4$. 
Likewise, the Gilbert damping exhibits a strong directional dependence. 
For nearest and next-nearest neighbors along the diagonal, i.e., for $\ff R=(1,1)$ and for $\ff R = (2,2)$, it is much stronger as compared to nearest- and next-nearest-neighbor positions along the $y$ axis, $\ff R=(0,1)$ and $\ff R=(0,2)$, for example.

\subsection{Effect of nonlocal Gilbert damping in $D=2$}
\label{sec:non2}

\begin{figure}[t]
\includegraphics[width=0.9\linewidth]{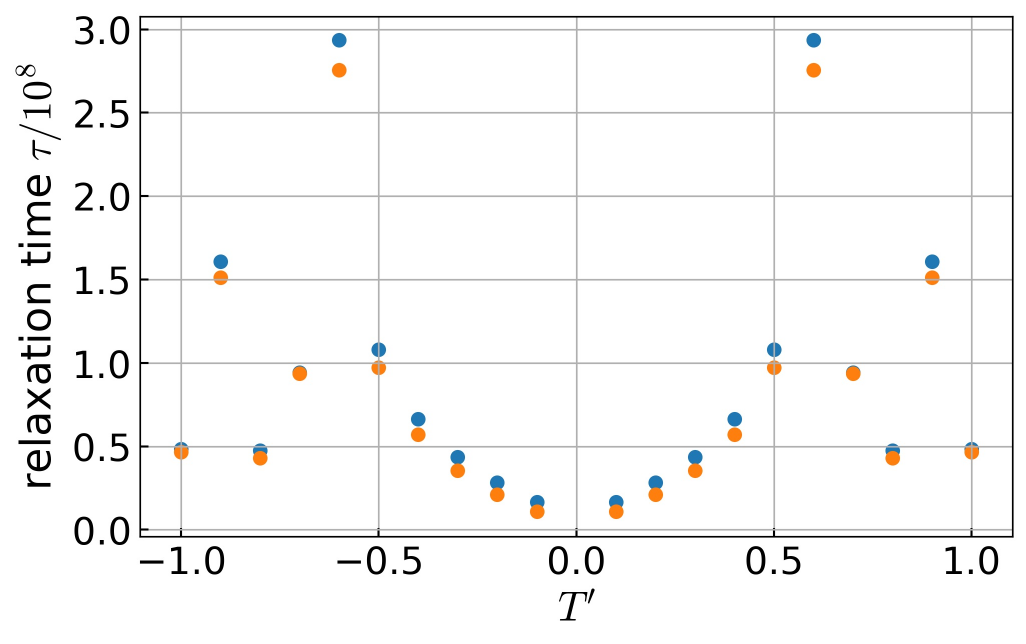}
\caption{
Relaxation time $\tau$ as function of the next-nearest-neighbor hopping $T'$ for a system with two classical spins locally exchange coupled with $J=0.1$ to next-nearest-neighbor sites of the square lattice, $\ff R=(1,1)$. 
In the initial state, the two impurity spins enclose the angle $\pi/2$.
{\em Blue:} computation including the nonlocal Gilbert damping. 
{\em Orange:} computation with nonlocal Gilbert damping set to zero. 
The conduction-electron system is at half filling.
Parameters: $L = 4 \cdot 10^6$, $\eta = 0.01$.
}
\label{fig:relax2}
\end{figure}

To study the effect of the nonlocal Gilbert damping on the relaxation dynamics of impurity spins in two dimensions we first consider a setup with two classical spins exchange coupled to next-nearest-neighbor sites, i.e., $\ff R=(1,1)$.
Figure \ref{fig:relax2} shows the relaxation time $\tau$ as a function of the next-nearest-neighbor hopping $T'$, starting from an initial state where the two classical spins are orthogonal. 

Of course, the definition of $\tau$ is somewhat arbitrary. 
As an operational criterion for (almost) full relaxation of $R$ impurity spins, we employ the condition
\be
  \frac{1}{R-1} \sum_{r=1}^{R-1} \left| \ff S_{r} \ff S_{r+1} \mp 1 \right| < \epsilon
  \; , 
\labeq{crit}  
\ee
with the $-$ sign in case of a ferromagnetic alignment in the ground state and with the $+$ sign for antiferromagnetic alignment. 
Furthermore, we choose $\epsilon = 0.001$.
We find that the criterion \refeq{crit} is completely fulfilled for a sufficiently long time evolution so that we can define the relaxation time $\tau$ as the latest time for which \refeq{crit} no longer applies.

As can be seen in Fig.\ \ref{fig:relax2}, $\tau$ monotonously increases with increasing $|T'|$ up to $|T'|=0.6$, while for $|T'| \ge 0.7$ the $\tau$ dependence becomes more complicated. 
This is partly related to the $T'$ dependence of the ground-state spin configuration, which is determined by the RKKY interaction. 
For $T'=0$ and $\ff R=(1,1)$, we have $J(\ff R) < 0$, i.e., ferromagnetic coupling.
A finite (positive or negative) $T'$ increases the tendency towards antiferromagnetic coupling.
In fact we find $J(\ff R) < 0$ (ferromagnetic) for the data points up to $T'=0.6$, and for $T'=0.8$, while $J(\ff R) > 0$ (antiferromagnetic) for $T'=0.7, 0.9, 1.0$.
This ``irregularity'' in the ground-state spin configuration and in the relaxation time is due to the nontrivial $T'$ dependence of both, the RKKY coupling and the Gilbert damping. 
In all cases the impurity-spin configuration in the fully relaxed state is identical with the ground-state configuration. 

The relaxation time scale crucially depends on $J$ and is clearly unreachable in practice by {\em full} quantum-classical theory at a local exchange coupling $J=0.1$.  
The numerical results of the effective spin-only theory (Fig.\ \ref{fig:relax2}) show that, roughly, the relaxation time is of the order of $\tau \sim 10^{8}$ (see blue symbols) and thus, at the same value for $J$, about two orders of magnitude larger than for the $D=1$ case (see Sec.\ \ref{sec:sdtwo}).

Let us note here that the spin dynamics is numerically very stable and does not change significantly when sharpening the tolerances for numerical errors in the solution of the system of differential equations. 
Furthermore, slight deviations in the initial state, e.g., a small change of the angle enclosed by the two spins, only leads to small changes of $\tau$. 

Interestingly, a clearly {\em shorter} relaxation time is obtained when disregarding the nonlocal Gilbert damping.
At $T'= \pm 0.3$, where the value for $\alpha(1,1)$ is about $18\%$ of the local damping (see Fig.\ \ref{fig:distd}), a decrease of $\tau$ by about $10\%$ is found, when {\em ad hoc} switching off the nonlocal damping. 
The effect becomes stronger with decreasing $|T'|$. 

Let us mention that for $T'=0$, the relaxation time is infinite in the effective theory due to the van Hove singularity at the Fermi energy of the half-filled conduction-electron system (see the discussion in Sec.\ \ref{sec:vh}).
The equations of motion \refeq{llg} are meaningless in this case, since $\alpha(\ff R)$ is divergent. 

The finding that including nonlocal Gilbert damping leads to {\em longer} relaxation times is reminiscent of the results found for the $D=1$ lattice. 
For the $D=1$ case, the effect could be explained analytically as the result of an emergent conserved quantity that is not caused by a symmetry of the Hamiltonian, see Sec.\ \ref{sec:sdtwo}. 
The main point is that an infinite relaxation time is obtained if $\alpha_{0} = \alpha_{d}$, which is exactly realized for impurity spins coupled to next-nearest-neighbor (distance $d=2$) sites on the $D=1$ lattice.
For the $D=2$ case, the ratio $\alpha(1,1) / \alpha(0,0)$ is clearly smaller than unity (see Fig.\ \ref{fig:distd} for $T'=-0.3$) and, therefore, 
there is a weak reminiscence of the effect only, but still the relaxation time {\em increases} due to nonlocal damping.
This is corroborated by the observation that the effect is practically absent if the impurity spins couple to nearest-neighbor sites along the $y$ axis of the square lattice, where we find $\alpha(0,1) / \alpha(0,0) \approx 0.001$ 
(see Fig.\ \ref{fig:disty} for $T'=-0.3$). 

\subsection{Relaxation of spin arrays}
\label{sec:array}

\begin{figure}[t]
\includegraphics[width=0.9\linewidth]{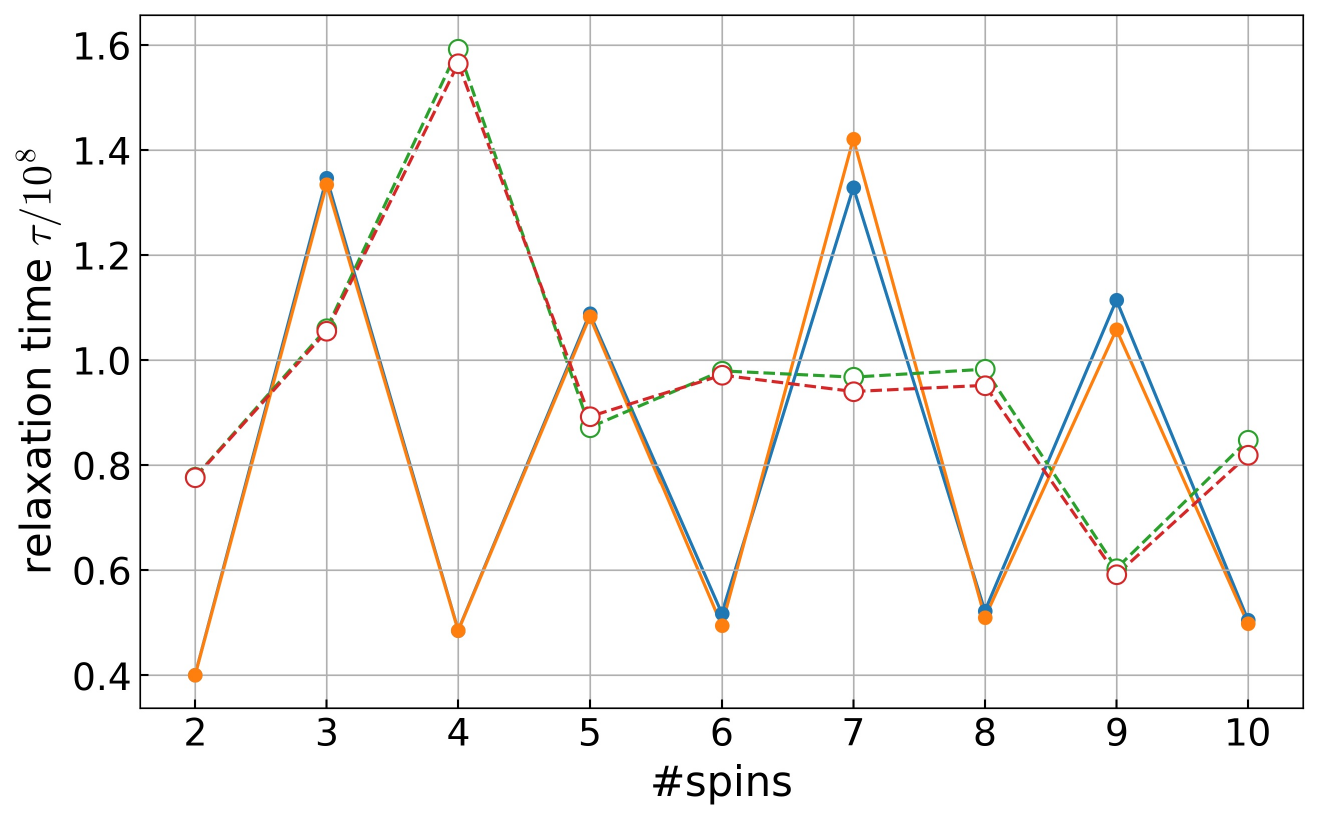}
\caption{
Relaxation time $\tau$ as function of the number of spins $R$, as determined via \refeq{crit} for $\epsilon=0.001$. 
Computations for a system of $R$ impurity spins ($R=2-10$) coupled to the sites $(0,0), (0,1), (0,2), ..., (0,R-1)$ of the square lattice, see inset of Fig.\ \ref{fig:disty}. 
{\em Filled circles:} data obtained for an initial spin configuration with $\ff S_{m} = O^{m-1} \ff S_{1}$ for $m=2,...,R$, where $O$ is the orthogonal $3\times 3$ matrix representing a $\pi / 2$ rotation around a fixed axis perpendicular to $\ff S_{1}$ ($\pi/2$ spin spiral).
{\em Open circles:} initial spin configuration given by $\ff S_{m} = (-1)^{m} \ff S_{2}$ for $m=3,...,R$ and $\ff S_{1} = O \ff S_{2}$ (antiferromagnetic configuration with first spin rotated by $\pi/2$.
$\ff S_{1}$ is kept fixed during the time evolution.
{\em Orange and red symbols:} relaxation time as obtained from computations with all nonlocal Gilbert-damping parameters set to zero.
{\em Blue and green symbols:} computations including the full nonlocal Gilbert damping.
Parameter: $L=4 \cdot 10^{6}$, $\eta=10^{-3}$, $T' = -  0.3$.
}
\label{fig:chainy}
\end{figure}

For the above-discussed setup with two impurity spins coupled to the square lattice the {\em local} Gilbert damping dominates the relaxation dynamics. 
Considering more impurity spins introduces additional complexity: 
The spatial structure of the RKKY interaction might lead to magnetic frustration, and the growing number of nonlocal damping terms might result in a qualitatively different relaxation dynamics.

\begin{figure}[b]
\centering
\includegraphics[width=0.9\linewidth]{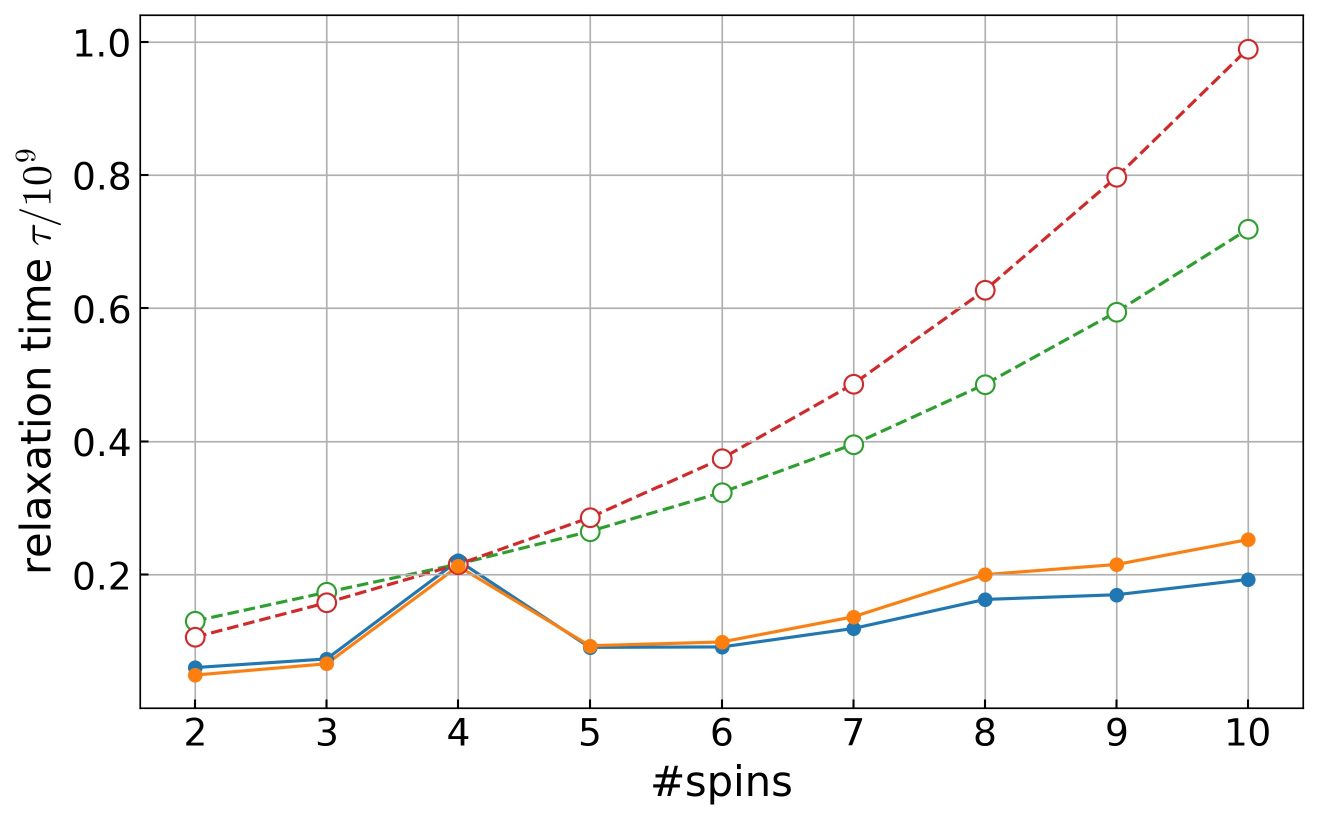}
\caption{
The same as Fig.\ \ref{fig:chainy} but for impurity spins coupled to the sites $(0,0), (1,1), (2,2), ..., (R-1,R-1)$ along a diagonal on the square lattice, see inset of Fig.\ \ref{fig:distd}. 
As in Fig.\ \ref{fig:chainy}, the first initial spin configuration (filled circles) is a $\pi/2$ spin spiral. 
Contrary to Fig.\ \ref{fig:chainy}, however, the second configuration (open circles) is given by a {\em ferromagnetic} spin configuration, except for the first spin $\ff S_{1}$, which is rotated by $\pi/2$ (and kept fixed during the time evolution).
}
\label{fig:chaind}
\end{figure}

Here, we study systems in two different chain geometries: 
(i) nearest-neighbor chains of $R$ impurity spins along, say, the $y$ direction, as visualized with the inset in Fig.\ \ref{fig:disty}, and (ii) 
next-nearest-neighbor chains of length $R$ along a diagonal of the square lattice, see the inset in Fig.\ \ref{fig:distd}.
The respective relaxation times $\tau$, defined via Eq.\ (\ref{eq:crit}), are shown as a function of the chain length $R$ in Figs.\ \ref{fig:chainy} and \ref{fig:chaind}.

We start with the discussion of the chain along the $y$ axis. 
Two different initial spin configurations have been considered: 
The filled circles in Fig.\ \ref{fig:chainy} refer to the first initial configuration: a $\pi/2$ spin spiral state at time $t=0$, i.e., each spin $\ff S_{m}$ is obtained from $\ff S_{m-1}$ by a $\pi/2$ rotation around an axis perperdicular to the first spin $\ff S_{1}$.

The computations have been performed for $T'=-0.3$, where we have a strongly positive RKKY coupling between nearest neighbors and a  smaller but still positive RKKY coupling between next-nearest neighbors, see Fig.\ \ref{fig:disty}. 
While this implies significant magnetic frustration, we find that the ground-state spin configuration of the chain is antiferromagnetic for all $R=2,...,10$.
This ground-state configuration is in fact reached on {\em roughly} the same time scale $\tau \sim 10^{8}$ (at $J=0.1$) for all chain lengths $R$.
We find that the relaxation time is clearly shorter for chains with even $R$ opposed to chains with an odd number of spins. 
This implies a rather regular oscillation of $\tau$ with $R$ and might be traced back to the choice of the initial state. 
Indeed, a less regular $R$ dependence of $\tau$ is seen for the second initial spin configuration, where we have assumed all spins
aligned antiferromagnetically, except for the first, which is rotated by $\pi/2$ around an axis perpendicular to the directions of the remaining spins and kept fixed during the subsequent dynamics. 

For both types of initial states there is no general {\em overall} increase or decrease of $\tau$ with increasing $R$, i.e., the relaxation time scale $\tau \sim 10^{8}$ roughly remains constant. 
This hints towards a dominantly local relaxation mechanism, where the relaxation time is mainly dominated by the approach of the local ground state, while reaching the overall collinear spin structure is not very crucial. 

This view is corroborated by the observation that the local Gilbert damping still dominates the relaxation dynamics: 
For both types of initial conditions we have performed computations with all nonlocal damping switched off, i.e., $\alpha(\ff R)=0$ for all $\ff R$ except for $\ff R=(0,0)$.
Results are displayed by orange and read circles in the figure.
While there is some effect of the nonlocal damping, it is generally small. 
This is remarkable since the number of nonlocal damping terms grows as $R(R-1)/2$ with $R$, i.e., quadratically for large $R$ opposed to the number $R$ of local damping terms.
We also note that switching off the nonlocal damping counterintuitively leads to shorter relaxation times $\tau$ for {\em some} $R$, as discussed above for $R=2$, but there seems to be no clear trend.

This is very much different for chains along the diagonal of the square lattice, see the results in Fig.\ \ref{fig:chaind}.
Here, the RKKY couplings $J(\ff R)$ for the first three distance vectors $\ff R=(1,1), (2,2), (3,3)$ are non-negligible and all negative (ferromagnetic), see Fig.\ \ref{fig:distd}.
Hence, there is no significant magnetic frustration. 
In fact, the ground-state spin configuration turns out as ferromagnetic for all $\ff R$ and is always reached as the final state of the relaxation dynamics.

For the diagonal spin chain we find that the relaxation time $\tau$ increases with the number of spins $R$ for both initial spin configurations, the global $\pi/2$ spin spiral (filled circles) and for an initial configuration (open circles) with only the first spin $\ff S_{1}$ rotated by $\pi/2$ out of the ferromagnetic ground-state configuration.
The growth of $\tau$ with $R$ is plausible in both cases, since there is a total impurity-spin difference, between the initial and the final configuration, that linearly grows with $R$. 
As the total spin of the full quantum-classical system (including the conduction electrons) is a constant of motion, this difference must be compensated in the course of time by a corresponding difference in the conduction-electron system. 
In other words, the necessary amount of spin dissipation grows with $R$ and leads to a (superlinearly) growing relaxation time. 
Note that for the second type of initial conditions, only the component of the total spin along the direction of $\ff S_{1}$ is constant.
This reasoning is also supported by the fact that overall the relaxation times are by about an order of magnitude larger (of the order of $\tau \sim 10^{9}$) for the spin chain along the diagonal compared to the chain along the $y$ axis ($\tau \sim 10^{8}$).

The data obtained for the initial $R=4$ spin spiral represents an exception and, to a lesser extent, those for $R=8$ as well, see the respective peaks of $\tau(R)$ in Fig.\ \ref{fig:chaind}. 
As already discussed above (Fig.\ \ref{fig:chainy}), this appears related to the period of the $\pi/2$ spin spiral, as for $R=4, 8, 12, ...$ the total impurity spin is zero and thus the necessary spin dissipation is at a maximum.
We have checked this by analogous computations for a $2\pi /3$ spin spiral. 
As expected the corresponding relaxation times peak at $R=3, 6, 9$ with strongly decreasing peak heights as $R$ increases.

Finally, switching off all nonlocal Gilbert damping constants, $\alpha(\ff R)=0$ for $\ff R \ne 0$, has a substantial impact on the relaxation time for the spin chain on the diagonal. 
Very consistently, the relaxation time slightly {\em decreases} for both types of initial states, if $R=1,2,3$ -- this is the counterintuitive effect observed and discussed earlier. 
For $R=5,6, ....$, on the other hand, we find an increasing relaxation time. 
We attribute this trend reversal to the fact that the overall larger nonlocal damping (see Figs.\ \ref{fig:disty} and \ref{fig:distd}) in combination with the growing number of nonlocal terms as $R$ increases, now substantially adds to the local damping mechanism. 

\subsection{Weakly interacting systems}
\label{sec:int}

The derivation (see Appendix \ref{sec:appdos}) of Eq.\ (\ref{eq:dosform}) is based on the application of Wick's theorem and thus cannot be used to compute the Gilbert damping in the case of a correlated electron system. 
Therefore, we compute $\alpha_{mm'}$ via the frequency derivative of the magnetic susceptibility at $\omega=0$, i.e., via Eq.\ (\ref{eq:altrep}).
Addressing weakly interacting systems, the random-phase approximation (RPA) \cite{LW60,Mor85,RHT+18,LKP23} is employed here.
We start from Eq.\ (\ref{eq:chinum}) for the non-interacting retarded magnetic susceptibility, denoted here as $\chi_{mm'}^{(0)}(t)$, to get $\chi^{(0)}(\ff k,\omega)$ via Fourier transformation from real to $\ff k$ space and from time to frequency representation. 
The interacting RPA susceptibility is then obtained as
\be
\chi(\ff k, \omega) = \frac{\chi^{(0)}(\ff k,\omega)}{1+U\chi^{(0)}(\ff k, \omega)}
\ee
for the $D=2$ Hubbard model on the square lattice,
\be
\hat{H}_{\text{el}}
= 
\sum_{\langle ii'\rangle} \sum_{\sigma=\uparrow, \downarrow} T_{ii'} c^\dagger_{i\sigma} c_{i^\prime\sigma} 
+ U \sum_{i} n_{i\uparrow} n_{i\downarrow}
\: ,
\labeq{hamhub}
\ee
where $U$ is the strength of the on-site Hubbard interaction. 

On the level of one-particle excitations, the RPA corresponds to the Hartree-Fock approach.
We consider the paramagnetic phase of the Hubbard model at weak $U$ and finite next-nearest-neighbor hopping $T'\ne 0$.
As has been found within Hartree-Fock theory \cite{HV98b}, the paramagnetic state of the system becomes unstable towards an antiferromagnetic state at a finite critical interaction $U_{\rm c} = U_{\rm c}(T')$. 

Our numerical calculations have been performed for lattices with $L=512 \times 512$ sites, periodic boundary conditions, a large inverse temperature $\beta=500$, and using a small regularization parameter $\eta=0.015$.
We have checked that the results do not significantly depend on these choices and are representative for the zero-temperature and the thermodynamical limit. 
In particular, we recover the known results for $U_{c}$ \cite{HV98b}. 
At $T'=-0.3$, for instance, we find $U_{c} \approx 2.53$, which is straightforwardly obtained as the singularity of $\chi(\ff k, \omega)$ at $\ff k=(\pi,\pi)$ and $\omega=0$.

\begin{figure}[t]
\includegraphics[width=0.8\linewidth]{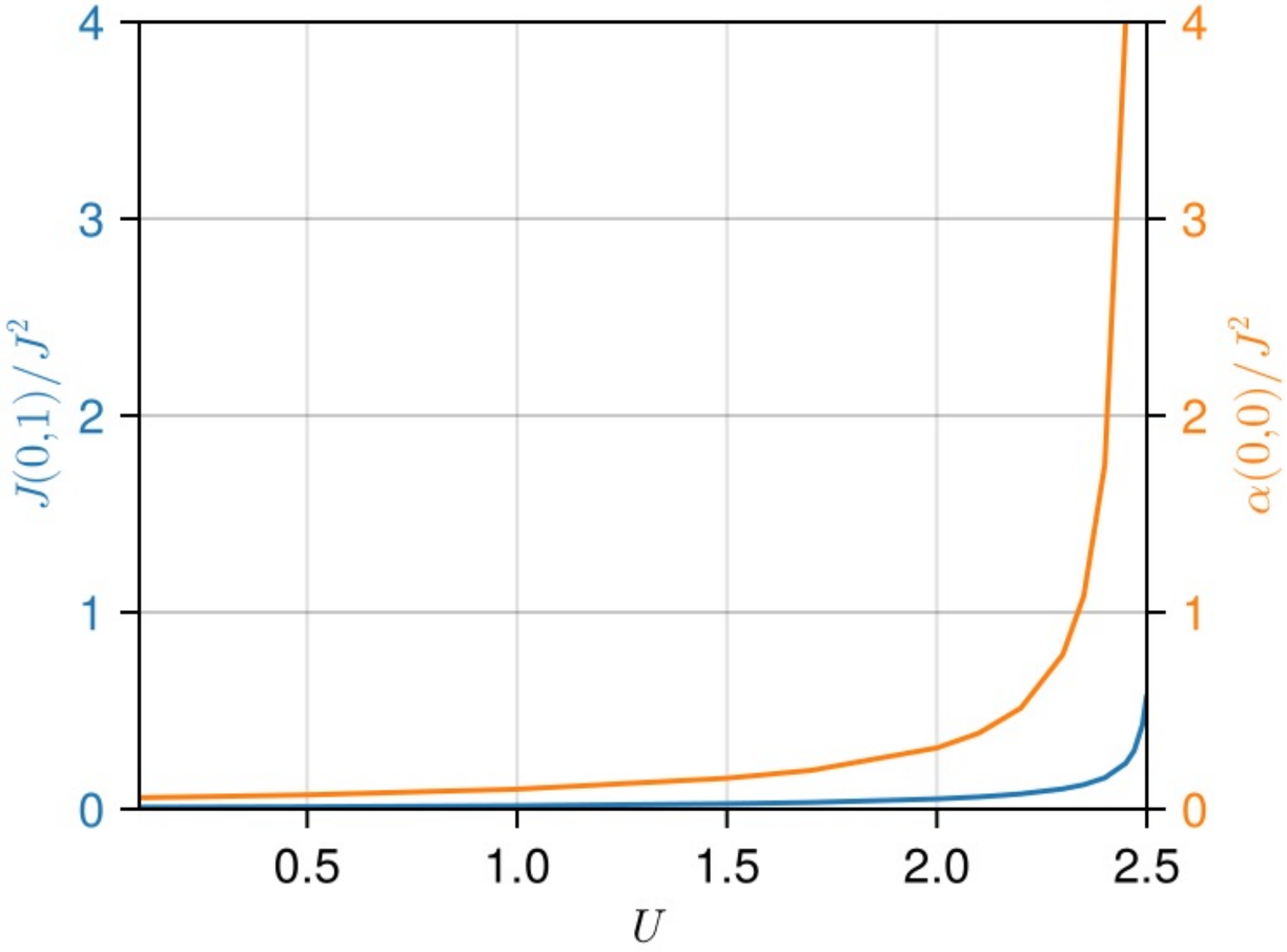}
\caption{
$U$ dependence of the RKKY interaction $J(\ff R) / J^{2}$ between nearest neighbors $\ff R=(0,1)$ (blue) and the local Gilbert damping $\alpha(\ff R)$ with $\ff R=(0,0)$, as obtained from RPA calculations for the half-filled Hubbard model with $T'=-0.3$ on the square lattice.
Parameters: $L=512 \times 512$, $\eta=0.015$, $\beta =500$.
}
\label{fig:udep}
\end{figure}

In reciprocal space, the frequency derivative of the susceptibility can be computed analytically: 
\be
\frac{d}{d\omega} \chi(\ff k, \omega) = \frac{d\chi^{(0)}(\ff k,\omega)/d\omega}{[1+U\chi^{(0)}(\ff k, \omega)]^{2}}
\: . 
\ee
Via Fourier transformation of $\chi(\ff k,0)$ and of $d \chi(\ff k, 0) / d \omega$ we find the distance-dependent RKKY interaction $J(\ff R)$ and the Gilbert damping $\alpha(\ff R)$ in real space.
Fig.\ \ref{fig:udep} shows the nearest-neighbor RKKY coupling $J(0,1) / J^{2}$ as well as the local Gilbert damping $\alpha(0,0)/J^{2}$ as functions of $U$ for $T'=-0.3$. 
In reciprocal space, on approaching the phase transition via $U \to U_{\rm c} \approx 2.53$, we have 
$\chi(\ff k, 0) \sim (U_{\rm c} - U)^{-1}$ at $\ff k=(\pi, \pi)$ while $d \chi(\ff k, 0) / d \omega \sim  (U_{\rm c} - U)^{-2}$.
The stronger divergence of $d \chi / d \omega$ as compared to $\chi$ leads to the stronger enhancement of $\alpha(\ff R)$, as compared to $J(\ff R)$, that can be seen in the figure.

Generally, the increase of the local Gilbert damping with increasing $U$ for the square lattice is consistent with the results of a previous study performed for a single impurity spin coupled to the $D=1$ Hubbard model \cite{SRP16a}, where a decreasing relaxation time with increasing $U$ has been found in the weak-interaction regime. 

\begin{figure}[t]
\includegraphics[width=0.95\linewidth]{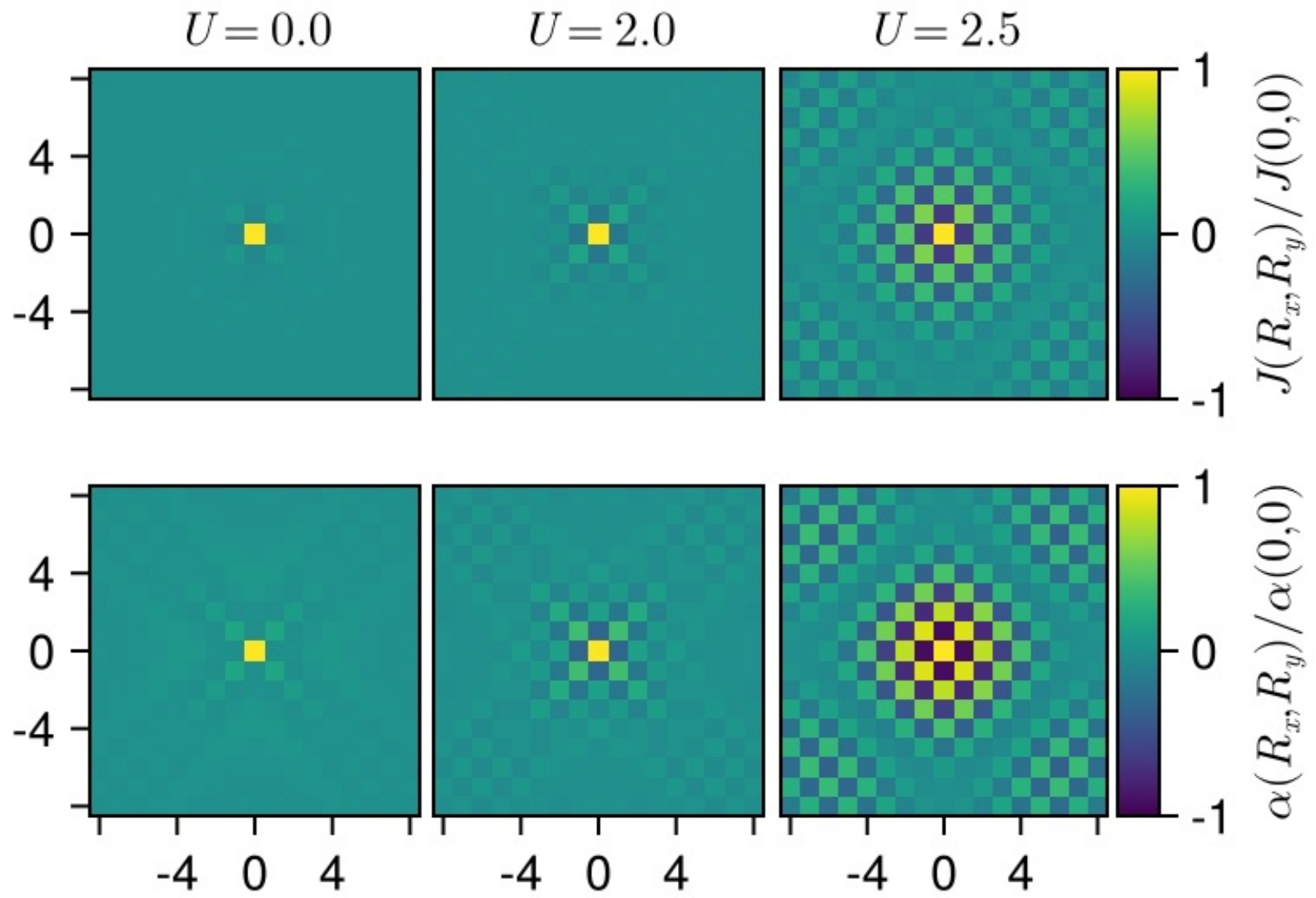}
\caption{
Distance and directional dependence of the RKKY interaction $J(\ff R)/J(\ff 0)$ and the Gilbert damping $\alpha(\ff R)/\alpha(\ff 0)$, normalized to the local values (color scale) for $R_{x}, R_{y} = -8, -7, ..., 7,8$. 
RPA calculations for the half-filled Hubbard model at $T'=-0.3$ and for $U=0$, $U=2$, and $U=2.5$, as indicated. 
Parameters: $L=512 \times 512$, $\eta=0.015$, $\beta = 500$.
}
\label{fig:rpa}
\end{figure}

More importantly, however, we find that the Hubbard interaction causes a strongly {\em nonlocal} Gilbert damping. 
This is demonstrated by the results shown in Fig.\ \ref{fig:rpa}, where the full spatial dependence of the RKKY interaction $J(\ff R)$ and of the Gilbert damping $\alpha(\ff R)$ is plotted, after normalization to the respective local element.
For $U=0$ (panels on left), one recovers the above-discussed directional dependence of the Gilbert damping (bottom panel). 
As compared to the RKKY interaction (top), this is much more pronounced. 
With increasing $U$, see the middle panels for $U=2.0$, the directional dependence of $\alpha(\ff R)$ is strongly enhanced due to the strong increase of the nonlocal damping parameters, relative to $\alpha(0,0)$. 

At $U=2.5$ (upper right panel), the RKKY coupling exhibits a checkerboard pattern with $J(\ff R)$ strongly oscillating between positive (antiferromagnetic) and negative (ferromagnetic) values.
Comparing with $U=2.0$, we see that this pattern extends spatially and is expected to eventually cover the entire lattice upon approaching the phase transition at $U_{c} \approx 2.53$. 
This is indicated by the closed nodal line of RKKY-coupling zeros extending spatially with increasing $U$.

Similarly, in the same spatial region enclosed by this line, the Gilbert damping at $U=2.5$ (lower right panel) is found to oscillate.
Eventually, for $U\to U_{c}$, the spatial structure of the Gilbert damping is expected to be given by $\alpha(R_{x}, R_{y}) / \alpha(0,0) \to  (-1)^{R_{x}+R_{y}}$. 
This is very reminiscent of the distance dependence of the Gilbert damping on the $D=1$ chain,
$\alpha_{d} / \alpha_{0} = \frac12 [1 + (-1)^{d}]$, which has been discussed in Sec.\ \ref{sec:nalpha} for the noninteracting case.

For a system with, e.g., two impurity spins coupled to next-nearest-neighbor sites with $\ff R=(1,1)$, and if one trusted the effective theory right at the phase transition, the spatial structure of the (divergent) Gilbert damping would imply that the total spin $\ff S_{1}+\ff S_{2}$ and the enclosed angle $\arccos (\ff S_{1} \ff S_{2})$ are constants of motion and thus that the system never relaxes to its ground-state spin configuration.
However, it remains an open question, how the relaxation dynamics would look like in $D=2$ and for $U \to U_{c}$, since the absolute (unnormalized) Gilbert damping is divergent (Fig.\ \ref{fig:udep}) and, hence, the effective spin-only theory is likely to break down at the critical point.

\section{Conclusions}
\label{sec:con}

Local magnetic moments, modelled as classical spins and locally exchange coupled to an extended metallic system of conduction electrons at zero temperature undergo a relaxation dynamics. 
To reduce the computational complexity, it is tempting to describe this dynamics within an effective spin-only theory. 
Actually, this is necessary to access the emergent relaxation times scale, which can be orders of magnitude longer than the fast femtosecond time scale of the conduction-electron dynamics. 

This reduction to a spin-only theory is possible by invoking two mutually interrelated approximations: 
(i) The exchange-coupling strength $J$ is small as compared to, e.g., the nearest-neighbor hopping matrix element $T$. 
(ii) The electron dynamics follows the spin dynamics almost adiabatically, i.e., the retardation time of the electron system is short compared to the spin-dynamics time scale. 
Within the framework of linear-response theory (LRT), one focusses on step (i) first and adopts (ii) in a subsequent step, while in the framework of adiabatic-response theory (ART), one starts from (ii) and later on adopts the weak-$J$ approximation (i).
Both approaches, LRT and ART, yield a system of nonlinear differential equations of motion involving the impurity spin $\ff S_{m}$ only, as well as time-dependent and nonlocal parameters, namely the RKKY interaction strengths $J_{mm'}(t)$ and the Gilbert-damping parameters $\alpha_{mm'}(t)$, and possibly external magnetic fields coupling to the spins. 
The parameters $J_{mm'}(t)$ and $\alpha_{mm'}(t)$ are properties of the conduction-electron system only and closely related to its retarded magnetic susceptibility. 

The reliability of the effective spin-only theory has been tested against the numerical solution of the full coupled quantum-classical electron-spin dynamics for a single spin, driven by an external field $\ff B$, coupled to a $D=1$ dimensional tight-binding electron system. 
The expressions for $J_{mm'}(t)$ and $\alpha_{mm'}(t)$, as obtained within LRT and ART, exhibit a quite different time dependence. 
It has been found, however, that the static RKKY interaction and the static Gilbert damping $J_{mm'}=\lim_{t\to \infty} J_{mm'}(t)$ and $\alpha_{mm'}=\lim_{t\to \infty} \alpha_{mm'}(t)$ perfectly agree. 
Furthermore, the strong temporal oscillations of $J_{mm'}(t)$ and $\alpha_{mm'}(t)$ seen within LRT are irrelevant for the spin dynamics as they take place on the fast electron time scale only. 
In fact, if $J$ is sufficiently weak and if the spin-dynamics time scale $\sim 1/B$ is sufficiently slow, almost perfect agreement is found when comparing the LRT and ART results against each other {\em and} against the results of the full theory. 

To achieve computationally manageable relaxation times (within the full theory), this comparison was done for values of $J$ and $B$ of the order of $T$. 
Realistic values for $J$ are very specific to the concrete system considered but will typically be an order of magnitude smaller than $T$:
Assuming a nearest-neighbor RKKY exchange of the order of $J_{\rm RKKY} \sim 0.01 T$ (with $J_{\rm RKKY} \sim 1$meV) or smaller, see, e.g., Ref.\ \cite{ZWL+10}, we may deduce from $J_{\rm RKKY} = J^{2} \chi \sim J^{2} / T$ a typical value $J=0.1 T$ for the local exchange. 
Realistic values for $B$ are even smaller. 
Hence, for computations addressing real systems, one may expect that the reliability of the effective theory is even better.

In the absence of an external driving field $\ff B$, the time scale on which the spin dynamics takes place is governed by the RKKY coupling. 
Since $J_{mm'} \propto J^{2}$, one actually has to control the exchange-coupling strength $J$ only. 
We have derived a simple expression for the Gilbert damping in the wide-band limit, $\alpha_{mm'} = \frac{\pi}{2} J^{2} \rho_{mm'}^{2}$, which involves the (nonlocal) densities of states $\rho_{mm'}$ at the Fermi energy. 
This shows that the dimensionless parameter $J \rho_{mm}$ is a suitable control parameter for the reliability of the effective theory. 
In fact, if the local density of states $\rho_{mm}$ diverges, as in the case of a van Hove singularity, the effective theory is found to break down.

The $D=1$ case turns out to be critical in the sense that the Gilbert damping $\alpha_{d}$ does not decay as a function of the distance $d =i_{m} - i_{m'}$ but rather oscillates between the local value $\alpha_{0}$ and zero for even and odd $d$, respectively. 
This causes an anomalous spin dynamics where relaxation is prohibited by an emergent conserved quantity that does not derive from a symmetry of the full Hamiltonian. 
This anomaly has already been discussed in Ref.\ \cite{EP24} for $T'=0$. 
Here, we could demonstrate that it occurs for absolute value of the next-nearest neighbor hoppings $T'$ smaller than a critical value $T'_{\rm c} = T/2$ and were able to relate the critical value to a Lifschitz transition of the Fermi ``surface''. 
Our analysis also explains why the relaxation time counterintuitively {\em decreases} when disregarding the nonlocal terms $m\ne m'$ in the matrix of Gilbert-damping parameters $\alpha_{mm'}$.

This is actually a recurring theme in the discussion of relaxation dynamics in $D=2$ dimension:
For two, initially orthogonal impurity spins coupled to next-nearest-neighbor sites on the square lattice and for all values of $T'$ ($|T'| < T$), we again find a decrease of the relaxation time when {\em ad hoc} switching off the nonlocal Gilbert damping.
As compared to $D=1$, however, the effect is much smaller such that even disregarding the nonlocality of the Gibert damping might be a reasonable approximation, depending on the value for $T'$.

For the $D=2$ square lattice and distances $R=|\ff R|$ up to $R \sim 10$ lattice constants, we did not see a universal simple power law $\alpha(R) \propto R^{-(D-1)}$, which is expected for large $R$ (assuming a free, $\propto k^{2}$ dispersion).
More importantly, the nonlocal Gilbert damping exhibits a strong directional dependence. 
This reflects itself, e.g., in the relaxation dynamics of one-dimensional spin chains coupled to the square lattice.
For chains with 2-10 impurity spins coupled to nearest-neigboring sites along the $x$ or, equivalently, the $y$ direction, the relaxation time $\tau$ turns out to oscillate with the length of the chain without a significant overall increase of $\tau$, independent of the initial spin configuration. 
Contrary, for chains of spins coupled to next-nearest-neigboring sites along a diagonal direction, there is an overall superlinear increase of $\tau$ with the chain length. 
The same behavior is found for an initial $\pi/2$ spin-spiral configuration, as compared to an initial state, where all spins are in their ground-state (ferromagnetic) configuration, except for an edge spin rotated by $\pi/2$ that is kept fixed during the time evolution. 
The strong increase of $\tau$ with the chain length can be explained with the necessarily increasing amount of spin dissipation.

There are several main lines of relevant future research:
(i) Our studies can straightforwardly be extended to more realistic models for the electronic structure and also combined with {\em ab initio} band-structure calculations. 
In particular, it would be interesting to consider multi-orbital systems and to see the effect of band degeneracy on the nonlocality of the Gilbert damping. 

(ii) The present study had focussed on systems with a gapless metallic electronic structure and weak local exchange $J$, where Gilbert damping is the dominant leading-order effect. 
For gapped systems, on the other hand, the spin dynamics is affected by the geometrical spin torque in addition \cite{SP17,BN20,MP22,LKP23}.
Studying semi-metals, where one would expect a drastically reduced damping and a still well-defined geometrical spin torque \cite{MP22} are worth studying. 
Furthermore, somewhat relaxing the almost adiabatic time evolution, by truncating at the next higher order in the expansion in the typical retardation time $\tau_{\rm ret}$, brings in additional effects such as nutation \cite{Fra08,BH12,SRP16b,LLP22}.

(iii) Modelling the impurity local magnetic moments as classical spins can only be a first step towards a theory of spin friction and dissipative dynamics of {\em quantum} spins. 
Clearly, this is a hard correlation problem as it essentially requires a conceptually and computationally feasible treatment of the long-time dynamics in a multi-impurity Kondo model.

More promising is (iv) to study the effect of electron correlations on the relaxation dynamics of classical impurity spins. 
This type of problem can be addressed within the presented LRT and ART frameworks. 
It essentially requires a computation of the zero-temperature (or the finite-temperature equilibrium) retarded magnetic susceptibility close to $\omega=0$. 
In a first attempt to address a weakly correlated conduction-electron system, we here have applied the RPA to the half-filled Hubbard model at finite next-nearest-neighbor hopping $T'$. 
This is already interesting, since with increasing $U$ the paramagnetic phase of the electron system becomes unstable towards a $\ff k = (\pi,\pi)$ antiferromagnetic phase. 
Close to the phase transition, the Gilbert damping matrix $\alpha_{mm'}$ is found to be overall strongly enhanced and strongly nonlocal.
The numerical data suggest that, right at the phase transition, the distance dependence is simply given by $\alpha(\ff R) \propto e^{i\ff k \ff R}$ with $\ff k=(\pi,\pi)$, which would cause spin dynamics without any dissipation. 
However, right at the phase transition the effective spin-only theory itself is expected to break down. 
Important next steps would therefore be to replace the RPA by a more advanced technique, to study the system for $U$ {\em above} the critical interaction, and to address the strong-coupling regime. 

\acknowledgments
This work was supported by the Deutsche Forschungsgemeinschaft (DFG, German Research Foundation) through the research unit QUAST, FOR 5249 (project P8), Project ID No. 449872909, and through the Cluster of Excellence “Advanced Imaging of Matter”- EXC 2056 - Project ID No. 390715994.

\appendix

\section{Representation of the nonlocal Gilbert damping in terms of the nonlocal density of states}
\label{sec:appdos}

To derive \refeq{dosform} for the Gilbert damping matrix, we start from the definition of the retarded susceptibility given by \refeq{chi}. 
The computations are done at finite temperature $1/\beta$, and the zero-temperature limit $\beta \to \infty$ is taken at the end.
Since $\langle [ s^{\alpha}_{i_{m}}(t) , s^{\alpha'}_{i_{m'}}(0) ] \rangle$ is a free expectation value, Wick's theorem applies and can be used to decompose the two-particle correlation into products of one-particle correlations. 
A straightforward calculation yields:
\ba
\chi_{mm'}^{\alpha\alpha'}(t) 
& = &
\delta^{\alpha\alpha'} \Theta(t) e^{-\eta t} 
\nonumber \\
& \times &\mbox{Im}
\left(
\langle c_{i_{m} \sigma}(t) c^{\dagger}_{i_{m'}\sigma}(0) \rangle
\langle c^{\dagger}_{i_{m'}\sigma}(0) c_{i_{m} \sigma}(-t) \rangle
\right)
\: .
\nonumber \\
\ea
The one-particle correlation functions can be obtained from the (nonlocal) one-particle density of states, \refeq{spden}, as
\ba
\langle c_{i \sigma}(t) c^{\dagger}_{i' \sigma}(0) \rangle 
&=& 
\int dx \, f(-x) A_{ii'}(x) e^{-ixt}
\: , 
\nonumber \\
\langle c^{\dagger}_{i'\sigma}(0) c_{i \sigma}(-t) \rangle
&=&
\int dy \, f(y) A_{ii'}(y) e^{iyt}
\: .
\labeq{specth}
\ea
Here, $f(x) = 1/(e^{\beta x} +1)$ denotes the Fermi function. 
Using \refeq{specth} we find for the frequency-dependent susceptibility
\be
\chi_{ii'}^{\alpha\alpha'}(\omega) 
= 
\delta^{\alpha\alpha'} \!\!  \iint dx dy \, f(-x) f(y)
A_{ii'}(x)
A_{ii'}(y)
\rho(x,y)
\labeq{chi1}
\ee
with 
\ba
\rho(x,y) &=& \int dt \, e^{i\omega t} \Theta(t) e^{-\eta t} \mbox{Im}(e^{-ixt} e^{iyt})
\nonumber \\
&=&
\frac12 \left( 
\frac{1}{\omega - x + y + i\eta} - \frac{1}{\omega + x - y + i\eta}
\right)
\: .
\nonumber \\
\ea
Note that $A_{ii'}(x)$ is real.
Inserting this result in \refeq{chi1} and using the identity $f(-x) f(y) - f(x) f(-y) = f(y) - f(x)$, we arrive at
\ba
\chi_{ii'}^{\alpha\alpha'}(\omega) 
= 
\frac{\delta^{\alpha\alpha'}}{2} \iint dx dy \, (f(y) - f(x))
\frac{A_{ii'}(x) A_{ii'}(y)}{\omega - x + y + i \eta}
\labeq{chi2}
\: .
\nonumber \\
\ea
For the second representation of the Gilbert damping in \refeq{altrep} we only need the imaginary part of $\chi_{ii'}^{\alpha\alpha'}(\omega)$.
Using the identity $(-1/\pi) \mbox{Im} (x + i \eta)^{-1} = \delta(x)$, taking the $\omega$ derivative of $\mbox{Im} \, \chi_{ii'}^{\alpha\alpha'}(\omega)$, and evaluating the result at $\omega=0$, yields
\ba
\alpha_{ii'} 
&=& 
\frac{\pi}{2} J^{2} \iint dx dy \, \delta'(y-x) 
\nonumber \\
& \times &
(f(y) - f(x)) A_{ii'}(x) A_{ii'}(y)
\: .
\ea
The prime at the $\delta$ function denotes the derivative with respect to the argument.
After exchanging the integration variables $x \leftrightarrow y$ for the summand involving $f(x)$ and using $\delta'(-x) = - \delta(x)$, this simplifies to
\ba
\alpha_{ii'} 
=
\pi J^{2} \iint dx dy \, \delta'(y-x) f(y) A_{ii'}(x) A_{ii'}(y)
\: .
\nonumber \\
\ea
To carry out the integration over $y$, we use integration by parts. This leaves us with 
\ba
\alpha_{ii'} 
&=&
- \pi J^{2} \int dx \, f'(x) A_{ii'}(x)^{2} 
\nonumber \\
&-& \pi J^{2} \int dx \, f(x) A_{ii'}(x) A'_{ii'}(x)
\: .
\ea
Consider the second term. 
We have $A_{ii'}(x) A'_{ii'}(x) = \frac12 (d/dx) A_{ii'}(x)^{2}$, so that we can use integration by parts once more. 
For a tight-binding density of states there is no residual boundary term.
Hence, we finally get the result
\be
\alpha_{ii'} 
=
- \frac{\pi}{2} J^{2} \int dx \, f'(x) A_{ii'}(x)^{2} 
\: .
\labeq{alphados}
\ee
Since $f'(x) \to - \delta(x)$ in the zero-temperature limit, this gives \refeq{dosform}.


\begin{thebibliography}{80}%
\makeatletter
\providecommand \@ifxundefined [1]{%
 \@ifx{#1\undefined}
}%
\providecommand \@ifnum [1]{%
 \ifnum #1\expandafter \@firstoftwo
 \else \expandafter \@secondoftwo
 \fi
}%
\providecommand \@ifx [1]{%
 \ifx #1\expandafter \@firstoftwo
 \else \expandafter \@secondoftwo
 \fi
}%
\providecommand \natexlab [1]{#1}%
\providecommand \enquote  [1]{``#1''}%
\providecommand \bibnamefont  [1]{#1}%
\providecommand \bibfnamefont [1]{#1}%
\providecommand \citenamefont [1]{#1}%
\providecommand \href@noop [0]{\@secondoftwo}%
\providecommand \href [0]{\begingroup \@sanitize@url \@href}%
\providecommand \@href[1]{\@@startlink{#1}\@@href}%
\providecommand \@@href[1]{\endgroup#1\@@endlink}%
\providecommand \@sanitize@url [0]{\catcode `\\12\catcode `\$12\catcode
  `\&12\catcode `\#12\catcode `\^12\catcode `\_12\catcode `\%12\relax}%
\providecommand \@@startlink[1]{}%
\providecommand \@@endlink[0]{}%
\providecommand \url  [0]{\begingroup\@sanitize@url \@url }%
\providecommand \@url [1]{\endgroup\@href {#1}{\urlprefix }}%
\providecommand \urlprefix  [0]{URL }%
\providecommand \Eprint [0]{\href }%
\providecommand \doibase [0]{https://doi.org/}%
\providecommand \selectlanguage [0]{\@gobble}%
\providecommand \bibinfo  [0]{\@secondoftwo}%
\providecommand \bibfield  [0]{\@secondoftwo}%
\providecommand \translation [1]{[#1]}%
\providecommand \BibitemOpen [0]{}%
\providecommand \bibitemStop [0]{}%
\providecommand \bibitemNoStop [0]{.\EOS\space}%
\providecommand \EOS [0]{\spacefactor3000\relax}%
\providecommand \BibitemShut  [1]{\csname bibitem#1\endcsname}%
\let\auto@bib@innerbib\@empty
\bibitem [{\citenamefont {Chappert}\ \emph {et~al.}(2007)\citenamefont
  {Chappert}, \citenamefont {Fert},\ and\ \citenamefont {Van~Dau}}]{CFVD07}%
  \BibitemOpen
  \bibfield  {author} {\bibinfo {author} {\bibfnamefont {C.}~\bibnamefont
  {Chappert}}, \bibinfo {author} {\bibfnamefont {A.}~\bibnamefont {Fert}},\
  and\ \bibinfo {author} {\bibfnamefont {F.~N.}\ \bibnamefont {Van~Dau}},\
  }\bibfield  {title} {\bibinfo {title} {The emergence of spin electronics in
  data storage},\ }\href {https://doi.org/10.1038/nmat2024} {\bibfield
  {journal} {\bibinfo  {journal} {Nat. Mater.}\ }\textbf {\bibinfo {volume}
  {6}},\ \bibinfo {pages} {813} (\bibinfo {year} {2007})}\BibitemShut {NoStop}%
\bibitem [{\citenamefont {Chumak}\ \emph {et~al.}(2015)\citenamefont {Chumak},
  \citenamefont {Vasyuchka}, \citenamefont {Serga},\ and\ \citenamefont
  {Hillebrands}}]{CVSH15}%
  \BibitemOpen
  \bibfield  {author} {\bibinfo {author} {\bibfnamefont {A.~V.}\ \bibnamefont
  {Chumak}}, \bibinfo {author} {\bibfnamefont {V.~I.}\ \bibnamefont
  {Vasyuchka}}, \bibinfo {author} {\bibfnamefont {A.~A.}\ \bibnamefont
  {Serga}},\ and\ \bibinfo {author} {\bibfnamefont {B.}~\bibnamefont
  {Hillebrands}},\ }\bibfield  {title} {\bibinfo {title} {The next wave},\
  }\href {https://doi.org/10.1038/nphys3367} {\bibfield  {journal} {\bibinfo
  {journal} {Nature Physics}\ }\textbf {\bibinfo {volume} {11}},\ \bibinfo
  {pages} {437} (\bibinfo {year} {2015})}\BibitemShut {NoStop}%
\bibitem [{\citenamefont {Gomonay}\ and\ \citenamefont {Loktev}(2014)}]{GL14}%
  \BibitemOpen
  \bibfield  {author} {\bibinfo {author} {\bibfnamefont {E.~V.}\ \bibnamefont
  {Gomonay}}\ and\ \bibinfo {author} {\bibfnamefont {V.~M.}\ \bibnamefont
  {Loktev}},\ }\bibfield  {title} {\bibinfo {title} {Spintronics of
  antiferromagnetic systems},\ }\href {https://doi.org/10.1063/1.4862467}
  {\bibfield  {journal} {\bibinfo  {journal} {Low Temp. Phys.}\ }\textbf
  {\bibinfo {volume} {40}},\ \bibinfo {pages} {17} (\bibinfo {year}
  {2014})}\BibitemShut {NoStop}%
\bibitem [{\citenamefont {Baltz}\ \emph {et~al.}(2018)\citenamefont {Baltz},
  \citenamefont {Manchon}, \citenamefont {Tsoi}, \citenamefont {Moriyama},
  \citenamefont {Ono},\ and\ \citenamefont {Tserkovnyak}}]{BMT+18}%
  \BibitemOpen
  \bibfield  {author} {\bibinfo {author} {\bibfnamefont {V.}~\bibnamefont
  {Baltz}}, \bibinfo {author} {\bibfnamefont {A.}~\bibnamefont {Manchon}},
  \bibinfo {author} {\bibfnamefont {M.}~\bibnamefont {Tsoi}}, \bibinfo {author}
  {\bibfnamefont {T.}~\bibnamefont {Moriyama}}, \bibinfo {author}
  {\bibfnamefont {T.}~\bibnamefont {Ono}},\ and\ \bibinfo {author}
  {\bibfnamefont {Y.}~\bibnamefont {Tserkovnyak}},\ }\bibfield  {title}
  {\bibinfo {title} {Antiferromagnetic spintronics},\ }\href
  {https://doi.org/10.1103/RevModPhys.90.015005} {\bibfield  {journal}
  {\bibinfo  {journal} {Rev. Mod. Phys.}\ }\textbf {\bibinfo {volume} {90}},\
  \bibinfo {pages} {015005} (\bibinfo {year} {2018})}\BibitemShut {NoStop}%
\bibitem [{\citenamefont {Tatara}\ \emph {et~al.}(2008)\citenamefont {Tatara},
  \citenamefont {Kohno},\ and\ \citenamefont {Shibata}}]{TKS08}%
  \BibitemOpen
  \bibfield  {author} {\bibinfo {author} {\bibfnamefont {G.}~\bibnamefont
  {Tatara}}, \bibinfo {author} {\bibfnamefont {H.}~\bibnamefont {Kohno}},\ and\
  \bibinfo {author} {\bibfnamefont {J.}~\bibnamefont {Shibata}},\ }\bibfield
  {title} {\bibinfo {title} {Microscopic approach to current-driven domain wall
  dynamics},\ }\href
  {https://doi.org/https://doi.org/10.1016/j.physrep.2008.07.003} {\bibfield
  {journal} {\bibinfo  {journal} {Physics Reports}\ }\textbf {\bibinfo {volume}
  {468}},\ \bibinfo {pages} {213} (\bibinfo {year} {2008})}\BibitemShut
  {NoStop}%
\bibitem [{\citenamefont {Skubic}\ \emph {et~al.}(2008)\citenamefont {Skubic},
  \citenamefont {Hellsvik}, \citenamefont {Nordstr\"om},\ and\ \citenamefont
  {Eriksson}}]{SHNE08}%
  \BibitemOpen
  \bibfield  {author} {\bibinfo {author} {\bibfnamefont {B.}~\bibnamefont
  {Skubic}}, \bibinfo {author} {\bibfnamefont {J.}~\bibnamefont {Hellsvik}},
  \bibinfo {author} {\bibfnamefont {L.}~\bibnamefont {Nordstr\"om}},\ and\
  \bibinfo {author} {\bibfnamefont {O.}~\bibnamefont {Eriksson}},\ }\bibfield
  {title} {\bibinfo {title} {A method for atomistic spin dynamics simulations:
  implementation and examples},\ }\href
  {https://doi.org/10.1088/0953-8984/20/31/315203} {\bibfield  {journal}
  {\bibinfo  {journal} {J. Phys.: Condens. Matter}\ }\textbf {\bibinfo {volume}
  {20}},\ \bibinfo {pages} {315203} (\bibinfo {year} {2008})}\BibitemShut
  {NoStop}%
\bibitem [{\citenamefont {Bertotti}\ \emph {et~al.}(2009)\citenamefont
  {Bertotti}, \citenamefont {Mayergoyz},\ and\ \citenamefont
  {Serpico}}]{BMS09}%
  \BibitemOpen
  \bibfield  {author} {\bibinfo {author} {\bibfnamefont {G.}~\bibnamefont
  {Bertotti}}, \bibinfo {author} {\bibfnamefont {I.~D.}\ \bibnamefont
  {Mayergoyz}},\ and\ \bibinfo {author} {\bibfnamefont {C.}~\bibnamefont
  {Serpico}},\ }\href@noop {} {\emph {\bibinfo {title} {Nonlinear Magnetization
  Dynamics in Nanosystemes}}}\ (\bibinfo  {publisher} {Elsevier},\ \bibinfo
  {address} {Amsterdam},\ \bibinfo {year} {2009})\BibitemShut {NoStop}%
\bibitem [{\citenamefont {Evans}\ \emph {et~al.}(2014)\citenamefont {Evans},
  \citenamefont {Fan}, \citenamefont {Chureemart}, \citenamefont {Ostler},
  \citenamefont {Ellis},\ and\ \citenamefont {Chantrell}}]{EFC+14}%
  \BibitemOpen
  \bibfield  {author} {\bibinfo {author} {\bibfnamefont {R.~F.~L.}\
  \bibnamefont {Evans}}, \bibinfo {author} {\bibfnamefont {W.~J.}\ \bibnamefont
  {Fan}}, \bibinfo {author} {\bibfnamefont {P.}~\bibnamefont {Chureemart}},
  \bibinfo {author} {\bibfnamefont {T.~A.}\ \bibnamefont {Ostler}}, \bibinfo
  {author} {\bibfnamefont {M.~O.~A.}\ \bibnamefont {Ellis}},\ and\ \bibinfo
  {author} {\bibfnamefont {R.~W.}\ \bibnamefont {Chantrell}},\ }\bibfield
  {title} {\bibinfo {title} {Atomistic spin model simulations of magnetic
  nanomaterials},\ }\href {http://stacks.iop.org/0953-8984/26/i=10/a=103202}
  {\bibfield  {journal} {\bibinfo  {journal} {J. Phys.: Condens. Matter}\
  }\textbf {\bibinfo {volume} {26}},\ \bibinfo {pages} {103202} (\bibinfo
  {year} {2014})}\BibitemShut {NoStop}%
\bibitem [{vz()}]{vz}%
  \BibitemOpen
  \href@noop {} {}\bibinfo {note} {S. V. Vonsovsky, Zh. \'Eksp. Teor. Fiz.
  {\bfseries 16}, 981 (1946); C. Zener, Phys. Rev. {\bfseries 81}, 440 (1951);
  S. V. Vonsovsky and E. A. Turov, Zh. \'Eksp. Teor. Fiz. {\bfseries 24}, 419
  (1953).}\BibitemShut {Stop}%
\bibitem [{\citenamefont {Elze}(2012)}]{Elz12}%
  \BibitemOpen
  \bibfield  {author} {\bibinfo {author} {\bibfnamefont {H.-T.}\ \bibnamefont
  {Elze}},\ }\bibfield  {title} {\bibinfo {title} {Linear dynamics of
  quantum-classical hybrids},\ }\href
  {https://doi.org/10.1103/PhysRevA.85.052109} {\bibfield  {journal} {\bibinfo
  {journal} {Phys. Rev. A}\ }\textbf {\bibinfo {volume} {85}},\ \bibinfo
  {pages} {052109} (\bibinfo {year} {2012})}\BibitemShut {NoStop}%
\bibitem [{\citenamefont {Sayad}\ and\ \citenamefont {Potthoff}(2015)}]{SP15}%
  \BibitemOpen
  \bibfield  {author} {\bibinfo {author} {\bibfnamefont {M.}~\bibnamefont
  {Sayad}}\ and\ \bibinfo {author} {\bibfnamefont {M.}~\bibnamefont
  {Potthoff}},\ }\bibfield  {title} {\bibinfo {title} {Spin dynamics and
  relaxation in the classical-spin {{Kondo}}-impurity model beyond the
  {Landau-Lifschitz-{Gilbert}} equation},\ }\href
  {http://stacks.iop.org/1367-2630/17/i=11/a=113058} {\bibfield  {journal}
  {\bibinfo  {journal} {New J. Phys.}\ }\textbf {\bibinfo {volume} {17}},\
  \bibinfo {pages} {113058} (\bibinfo {year} {2015})}\BibitemShut {NoStop}%
\bibitem [{\citenamefont {Elbracht}\ and\ \citenamefont
  {Potthoff}(2020)}]{EP20}%
  \BibitemOpen
  \bibfield  {author} {\bibinfo {author} {\bibfnamefont {M.}~\bibnamefont
  {Elbracht}}\ and\ \bibinfo {author} {\bibfnamefont {M.}~\bibnamefont
  {Potthoff}},\ }\bibfield  {title} {\bibinfo {title} {Accessing long
  timescales in the relaxation dynamics of spins coupled to a
  conduction-electron system using absorbing boundary conditions},\ }\href
  {https://doi.org/10.1103/PhysRevB.102.115434} {\bibfield  {journal} {\bibinfo
   {journal} {Phys. Rev. B}\ }\textbf {\bibinfo {volume} {102}},\ \bibinfo
  {pages} {115434} (\bibinfo {year} {2020})}\BibitemShut {NoStop}%
\bibitem [{\citenamefont {Elbracht}\ and\ \citenamefont
  {Potthoff}(2021)}]{EP21}%
  \BibitemOpen
  \bibfield  {author} {\bibinfo {author} {\bibfnamefont {M.}~\bibnamefont
  {Elbracht}}\ and\ \bibinfo {author} {\bibfnamefont {M.}~\bibnamefont
  {Potthoff}},\ }\bibfield  {title} {\bibinfo {title} {Long-time relaxation
  dynamics of a spin coupled to a {Chern} insulator},\ }\href
  {https://doi.org/10.1103/PhysRevB.103.024301} {\bibfield  {journal} {\bibinfo
   {journal} {Phys. Rev. B}\ }\textbf {\bibinfo {volume} {103}},\ \bibinfo
  {pages} {024301} (\bibinfo {year} {2021})}\BibitemShut {NoStop}%
\bibitem [{\citenamefont {Kirilyuk}\ \emph {et~al.}(2010)\citenamefont
  {Kirilyuk}, \citenamefont {Kimel},\ and\ \citenamefont {Rasing}}]{KKR10}%
  \BibitemOpen
  \bibfield  {author} {\bibinfo {author} {\bibfnamefont {A.}~\bibnamefont
  {Kirilyuk}}, \bibinfo {author} {\bibfnamefont {A.~V.}\ \bibnamefont
  {Kimel}},\ and\ \bibinfo {author} {\bibfnamefont {T.}~\bibnamefont
  {Rasing}},\ }\bibfield  {title} {\bibinfo {title} {Ultrafast optical
  manipulation of magnetic order},\ }\href
  {https://doi.org/10.1103/RevModPhys.82.2731} {\bibfield  {journal} {\bibinfo
  {journal} {Rev. Mod. Phys.}\ }\textbf {\bibinfo {volume} {82}},\ \bibinfo
  {pages} {2731} (\bibinfo {year} {2010})}\BibitemShut {NoStop}%
\bibitem [{\citenamefont {Wei\ss{}e}(2009)}]{Wei09}%
  \BibitemOpen
  \bibfield  {author} {\bibinfo {author} {\bibfnamefont {A.}~\bibnamefont
  {Wei\ss{}e}},\ }\bibfield  {title} {\bibinfo {title} {Green-function-based
  {Monte Carlo} method for classical fields coupled to fermions},\ }\href
  {https://doi.org/10.1103/PhysRevLett.102.150604} {\bibfield  {journal}
  {\bibinfo  {journal} {Phys. Rev. Lett.}\ }\textbf {\bibinfo {volume} {102}},\
  \bibinfo {pages} {150604} (\bibinfo {year} {2009})}\BibitemShut {NoStop}%
\bibitem [{\citenamefont {Capelle}\ and\ \citenamefont {Gyorffy}(2003)}]{CG03}%
  \BibitemOpen
  \bibfield  {author} {\bibinfo {author} {\bibfnamefont {K.}~\bibnamefont
  {Capelle}}\ and\ \bibinfo {author} {\bibfnamefont {B.~L.}\ \bibnamefont
  {Gyorffy}},\ }\bibfield  {title} {\bibinfo {title} {Exploring dynamical
  magnetism with time-dependent density-functional theory: From spin
  fluctuations to {Gilbert} damping},\ }\href
  {http://stacks.iop.org/0295-5075/61/i=3/a=354} {\bibfield  {journal}
  {\bibinfo  {journal} {Europhys. Lett.}\ }\textbf {\bibinfo {volume} {61}},\
  \bibinfo {pages} {354} (\bibinfo {year} {2003})}\BibitemShut {NoStop}%
\bibitem [{\citenamefont {Onoda}\ and\ \citenamefont {Nagaosa}(2006)}]{ON06}%
  \BibitemOpen
  \bibfield  {author} {\bibinfo {author} {\bibfnamefont {M.}~\bibnamefont
  {Onoda}}\ and\ \bibinfo {author} {\bibfnamefont {N.}~\bibnamefont
  {Nagaosa}},\ }\bibfield  {title} {\bibinfo {title} {Dynamics of localized
  spins coupled to the conduction electrons with charge and spin currents},\
  }\href {https://doi.org/10.1103/PhysRevLett.96.066603} {\bibfield  {journal}
  {\bibinfo  {journal} {Phys. Rev. Lett.}\ }\textbf {\bibinfo {volume} {96}},\
  \bibinfo {pages} {066603} (\bibinfo {year} {2006})}\BibitemShut {NoStop}%
\bibitem [{\citenamefont {Bhattacharjee}\ \emph {et~al.}(2012)\citenamefont
  {Bhattacharjee}, \citenamefont {Nordstr\"om},\ and\ \citenamefont
  {Fransson}}]{BNF12}%
  \BibitemOpen
  \bibfield  {author} {\bibinfo {author} {\bibfnamefont {S.}~\bibnamefont
  {Bhattacharjee}}, \bibinfo {author} {\bibfnamefont {L.}~\bibnamefont
  {Nordstr\"om}},\ and\ \bibinfo {author} {\bibfnamefont {J.}~\bibnamefont
  {Fransson}},\ }\bibfield  {title} {\bibinfo {title} {Atomistic spin dynamic
  method with both damping and moment of inertia effects included from first
  principles},\ }\href {https://doi.org/10.1103/PhysRevLett.108.057204}
  {\bibfield  {journal} {\bibinfo  {journal} {Phys. Rev. Lett.}\ }\textbf
  {\bibinfo {volume} {108}},\ \bibinfo {pages} {057204} (\bibinfo {year}
  {2012})}\BibitemShut {NoStop}%
\bibitem [{\citenamefont {Umetsu}\ \emph {et~al.}(2012)\citenamefont {Umetsu},
  \citenamefont {Miura},\ and\ \citenamefont {Sakuma}}]{UMS12}%
  \BibitemOpen
  \bibfield  {author} {\bibinfo {author} {\bibfnamefont {N.}~\bibnamefont
  {Umetsu}}, \bibinfo {author} {\bibfnamefont {D.}~\bibnamefont {Miura}},\ and\
  \bibinfo {author} {\bibfnamefont {A.}~\bibnamefont {Sakuma}},\ }\bibfield
  {title} {\bibinfo {title} {Microscopic theory for {Gilbert} damping in
  materials with inhomogeneous spin dynamics},\ }\href
  {https://doi.org/http://dx.doi.org/10.1063/1.3675999} {\bibfield  {journal}
  {\bibinfo  {journal} {J. Appl. Phys.}\ }\textbf {\bibinfo {volume} {111}},\
  \bibinfo {eid} {07D117} (\bibinfo {year} {2012})}\BibitemShut {NoStop}%
\bibitem [{llg()}]{llg}%
  \BibitemOpen
  \href@noop {} {}\bibinfo {note} {L.~D. Landau and E.~M. Lifshitz, Physik.
  Zeits. Sowjetunion \textbf{8}, 153 (1935); T. {Gilbert}, Phys. Rev.
  \textbf{100}, 1243 (1955); T. {Gilbert}, Magnetics, IEEE Transactions on
  \textbf{40}, 3443 (2004).}\BibitemShut {Stop}%
\bibitem [{\citenamefont {Ruderman}\ and\ \citenamefont {Kittel}(1954)}]{RK54}%
  \BibitemOpen
  \bibfield  {author} {\bibinfo {author} {\bibfnamefont {M.~A.}\ \bibnamefont
  {Ruderman}}\ and\ \bibinfo {author} {\bibfnamefont {C.}~\bibnamefont
  {Kittel}},\ }\bibfield  {title} {\bibinfo {title} {Indirect exchange coupling
  of nuclear magnetic moments by conduction electrons},\ }\href
  {https://doi.org/10.1103/PhysRev.96.99} {\bibfield  {journal} {\bibinfo
  {journal} {Phys. Rev.}\ }\textbf {\bibinfo {volume} {96}},\ \bibinfo {pages}
  {99} (\bibinfo {year} {1954})}\BibitemShut {NoStop}%
\bibitem [{\citenamefont {Kasuya}(1956)}]{Kas56}%
  \BibitemOpen
  \bibfield  {author} {\bibinfo {author} {\bibfnamefont {T.}~\bibnamefont
  {Kasuya}},\ }\bibfield  {title} {\bibinfo {title} {A theory of metallic
  ferro- and antiferromagnetism on {Zener's} model},\ }\href
  {https://doi.org/10.1143/PTP.16.45} {\bibfield  {journal} {\bibinfo
  {journal} {Prog. Theor. Phys.}\ }\textbf {\bibinfo {volume} {16}},\ \bibinfo
  {pages} {45} (\bibinfo {year} {1956})}\BibitemShut {NoStop}%
\bibitem [{\citenamefont {Yosida}(1957)}]{Yos57}%
  \BibitemOpen
  \bibfield  {author} {\bibinfo {author} {\bibfnamefont {K.}~\bibnamefont
  {Yosida}},\ }\bibfield  {title} {\bibinfo {title} {Magnetic properties of
  {Cu-Mn} alloys},\ }\href {https://doi.org/10.1103/PhysRev.106.893} {\bibfield
   {journal} {\bibinfo  {journal} {Phys. Rev.}\ }\textbf {\bibinfo {volume}
  {106}},\ \bibinfo {pages} {893} (\bibinfo {year} {1957})}\BibitemShut
  {NoStop}%
\bibitem [{\citenamefont {Bajpai}\ and\ \citenamefont
  {Nikoli\ifmmode~\acute{c}\else \'{c}\fi{}}(2019)}]{BN19}%
  \BibitemOpen
  \bibfield  {author} {\bibinfo {author} {\bibfnamefont {U.}~\bibnamefont
  {Bajpai}}\ and\ \bibinfo {author} {\bibfnamefont {B.~K.}\ \bibnamefont
  {Nikoli\ifmmode~\acute{c}\else \'{c}\fi{}}},\ }\bibfield  {title} {\bibinfo
  {title} {Time-retarded damping and magnetic inertia in the
  {Landau-Lifshitz-Gilbert} equation self-consistently coupled to electronic
  time-dependent nonequilibrium {Green} functions},\ }\href
  {https://doi.org/10.1103/PhysRevB.99.134409} {\bibfield  {journal} {\bibinfo
  {journal} {Phys. Rev. B}\ }\textbf {\bibinfo {volume} {99}},\ \bibinfo
  {pages} {134409} (\bibinfo {year} {2019})}\BibitemShut {NoStop}%
\bibitem [{\citenamefont {Reyes-Osorio}\ and\ \citenamefont
  {Nikoli\ifmmode~\acute{c}\else \'{c}\fi{}}(2024)}]{RON24}%
  \BibitemOpen
  \bibfield  {author} {\bibinfo {author} {\bibfnamefont {F.}~\bibnamefont
  {Reyes-Osorio}}\ and\ \bibinfo {author} {\bibfnamefont {B.~K.}\ \bibnamefont
  {Nikoli\ifmmode~\acute{c}\else \'{c}\fi{}}},\ }\bibfield  {title} {\bibinfo
  {title} {{Gilbert} damping in metallic ferromagnets from {Schwinger-Keldysh}
  field theory: Intrinsically nonlocal, nonuniform, and made anisotropic by
  spin-orbit coupling},\ }\href {https://doi.org/10.1103/PhysRevB.109.024413}
  {\bibfield  {journal} {\bibinfo  {journal} {Phys. Rev. B}\ }\textbf {\bibinfo
  {volume} {109}},\ \bibinfo {pages} {024413} (\bibinfo {year}
  {2024})}\BibitemShut {NoStop}%
\bibitem [{\citenamefont {Thonig}\ \emph {et~al.}(2015)\citenamefont {Thonig},
  \citenamefont {Henk},\ and\ \citenamefont {Eriksson}}]{THE15}%
  \BibitemOpen
  \bibfield  {author} {\bibinfo {author} {\bibfnamefont {D.}~\bibnamefont
  {Thonig}}, \bibinfo {author} {\bibfnamefont {J.}~\bibnamefont {Henk}},\ and\
  \bibinfo {author} {\bibfnamefont {O.}~\bibnamefont {Eriksson}},\ }\bibfield
  {title} {\bibinfo {title} {{Gilbert}-like damping caused by time retardation
  in atomistic magnetization dynamics},\ }\href
  {https://doi.org/10.1103/PhysRevB.92.104403} {\bibfield  {journal} {\bibinfo
  {journal} {Phys. Rev. B}\ }\textbf {\bibinfo {volume} {92}},\ \bibinfo
  {pages} {104403} (\bibinfo {year} {2015})}\BibitemShut {NoStop}%
\bibitem [{\citenamefont {Kune\v{s}}\ and\ \citenamefont
  {Kambersk\'y}(2002)}]{KK02}%
  \BibitemOpen
  \bibfield  {author} {\bibinfo {author} {\bibfnamefont {J.}~\bibnamefont
  {Kune\v{s}}}\ and\ \bibinfo {author} {\bibfnamefont {V.}~\bibnamefont
  {Kambersk\'y}},\ }\bibfield  {title} {\bibinfo {title} {First-principles
  investigation of the damping of fast magnetization precession in
  ferromagnetic $3d$ metals},\ }\href
  {https://doi.org/10.1103/PhysRevB.65.212411} {\bibfield  {journal} {\bibinfo
  {journal} {Phys. Rev. B}\ }\textbf {\bibinfo {volume} {65}},\ \bibinfo
  {pages} {212411} (\bibinfo {year} {2002})}\BibitemShut {NoStop}%
\bibitem [{\citenamefont {Kambersk\'y}(2007)}]{Kam07}%
  \BibitemOpen
  \bibfield  {author} {\bibinfo {author} {\bibfnamefont {V.}~\bibnamefont
  {Kambersk\'y}},\ }\bibfield  {title} {\bibinfo {title} {Spin-orbital
  {Gilbert} damping in common magnetic metals},\ }\href
  {https://doi.org/10.1103/PhysRevB.76.134416} {\bibfield  {journal} {\bibinfo
  {journal} {Phys. Rev. B}\ }\textbf {\bibinfo {volume} {76}},\ \bibinfo
  {pages} {134416} (\bibinfo {year} {2007})}\BibitemShut {NoStop}%
\bibitem [{\citenamefont {Gilmore}\ \emph {et~al.}(2007)\citenamefont
  {Gilmore}, \citenamefont {Idzerda},\ and\ \citenamefont {Stiles}}]{GIS07}%
  \BibitemOpen
  \bibfield  {author} {\bibinfo {author} {\bibfnamefont {K.}~\bibnamefont
  {Gilmore}}, \bibinfo {author} {\bibfnamefont {Y.~U.}\ \bibnamefont
  {Idzerda}},\ and\ \bibinfo {author} {\bibfnamefont {M.~D.}\ \bibnamefont
  {Stiles}},\ }\bibfield  {title} {\bibinfo {title} {Identification of the
  dominant precession-damping mechanism in {Fe, Co, and Ni} by first-principles
  calculations},\ }\href {https://doi.org/10.1103/PhysRevLett.99.027204}
  {\bibfield  {journal} {\bibinfo  {journal} {Phys. Rev. Lett.}\ }\textbf
  {\bibinfo {volume} {99}},\ \bibinfo {pages} {027204} (\bibinfo {year}
  {2007})}\BibitemShut {NoStop}%
\bibitem [{\citenamefont {Brataas}\ \emph {et~al.}(2008)\citenamefont
  {Brataas}, \citenamefont {Tserkovnyak},\ and\ \citenamefont {Bauer}}]{BTB08}%
  \BibitemOpen
  \bibfield  {author} {\bibinfo {author} {\bibfnamefont {A.}~\bibnamefont
  {Brataas}}, \bibinfo {author} {\bibfnamefont {Y.}~\bibnamefont
  {Tserkovnyak}},\ and\ \bibinfo {author} {\bibfnamefont {G.~E.~W.}\
  \bibnamefont {Bauer}},\ }\bibfield  {title} {\bibinfo {title} {Scattering
  theory of {Gilbert} damping},\ }\href
  {https://doi.org/10.1103/PhysRevLett.101.037207} {\bibfield  {journal}
  {\bibinfo  {journal} {Phys. Rev. Lett.}\ }\textbf {\bibinfo {volume} {101}},\
  \bibinfo {pages} {037207} (\bibinfo {year} {2008})}\BibitemShut {NoStop}%
\bibitem [{\citenamefont {Hickey}\ and\ \citenamefont {Moodera}(2009)}]{HM09}%
  \BibitemOpen
  \bibfield  {author} {\bibinfo {author} {\bibfnamefont {M.~C.}\ \bibnamefont
  {Hickey}}\ and\ \bibinfo {author} {\bibfnamefont {J.~S.}\ \bibnamefont
  {Moodera}},\ }\bibfield  {title} {\bibinfo {title} {Origin of intrinsic
  {Gilbert} damping},\ }\href {https://doi.org/10.1103/PhysRevLett.102.137601}
  {\bibfield  {journal} {\bibinfo  {journal} {Phys. Rev. Lett.}\ }\textbf
  {\bibinfo {volume} {102}},\ \bibinfo {pages} {137601} (\bibinfo {year}
  {2009})}\BibitemShut {NoStop}%
\bibitem [{\citenamefont {Garate}\ and\ \citenamefont
  {MacDonald}(2009)}]{GMcD09}%
  \BibitemOpen
  \bibfield  {author} {\bibinfo {author} {\bibfnamefont {I.}~\bibnamefont
  {Garate}}\ and\ \bibinfo {author} {\bibfnamefont {A.}~\bibnamefont
  {MacDonald}},\ }\bibfield  {title} {\bibinfo {title} {{Gilbert} damping in
  conducting ferromagnets. {I. Kohn-Sham} theory and atomic-scale
  inhomogeneity},\ }\href {https://doi.org/10.1103/PhysRevB.79.064403}
  {\bibfield  {journal} {\bibinfo  {journal} {Phys. Rev. B}\ }\textbf {\bibinfo
  {volume} {79}},\ \bibinfo {pages} {064403} (\bibinfo {year}
  {2009})}\BibitemShut {NoStop}%
\bibitem [{\citenamefont {Starikov}\ \emph {et~al.}(2010)\citenamefont
  {Starikov}, \citenamefont {Kelly}, \citenamefont {Brataas}, \citenamefont
  {Tserkovnyak},\ and\ \citenamefont {Bauer}}]{SKB+10}%
  \BibitemOpen
  \bibfield  {author} {\bibinfo {author} {\bibfnamefont {A.~A.}\ \bibnamefont
  {Starikov}}, \bibinfo {author} {\bibfnamefont {P.~J.}\ \bibnamefont {Kelly}},
  \bibinfo {author} {\bibfnamefont {A.}~\bibnamefont {Brataas}}, \bibinfo
  {author} {\bibfnamefont {Y.}~\bibnamefont {Tserkovnyak}},\ and\ \bibinfo
  {author} {\bibfnamefont {G.~E.~W.}\ \bibnamefont {Bauer}},\ }\bibfield
  {title} {\bibinfo {title} {Unified first-principles study of {Gilbert}
  damping, spin-flip diffusion, and resistivity in transition metal alloys},\
  }\href {https://doi.org/10.1103/PhysRevLett.105.236601} {\bibfield  {journal}
  {\bibinfo  {journal} {Phys. Rev. Lett.}\ }\textbf {\bibinfo {volume} {105}},\
  \bibinfo {pages} {236601} (\bibinfo {year} {2010})}\BibitemShut {NoStop}%
\bibitem [{\citenamefont {Sakuma}(2012)}]{Sak12}%
  \BibitemOpen
  \bibfield  {author} {\bibinfo {author} {\bibfnamefont {A.}~\bibnamefont
  {Sakuma}},\ }\bibfield  {title} {\bibinfo {title} {First-principles study on
  the {Gilbert} damping constants of transition metal alloys, {Fe-Ni and Fe-Pt}
  systems},\ }\href {https://doi.org/10.1143/JPSJ.81.084701} {\bibfield
  {journal} {\bibinfo  {journal} {J. Phys. Soc. Jpn.}\ }\textbf {\bibinfo
  {volume} {81}},\ \bibinfo {pages} {084701} (\bibinfo {year}
  {2012})}\BibitemShut {NoStop}%
\bibitem [{\citenamefont {Ebert}\ \emph {et~al.}(2011)\citenamefont {Ebert},
  \citenamefont {Mankovsky}, \citenamefont {K\"odderitzsch},\ and\
  \citenamefont {Kelly}}]{EMKK11}%
  \BibitemOpen
  \bibfield  {author} {\bibinfo {author} {\bibfnamefont {H.}~\bibnamefont
  {Ebert}}, \bibinfo {author} {\bibfnamefont {S.}~\bibnamefont {Mankovsky}},
  \bibinfo {author} {\bibfnamefont {D.}~\bibnamefont {K\"odderitzsch}},\ and\
  \bibinfo {author} {\bibfnamefont {P.~J.}\ \bibnamefont {Kelly}},\ }\bibfield
  {title} {\bibinfo {title} {Ab initio calculation of the {Gilbert} damping
  parameter via the linear response formalism},\ }\href
  {https://doi.org/10.1103/PhysRevLett.107.066603} {\bibfield  {journal}
  {\bibinfo  {journal} {Phys. Rev. Lett.}\ }\textbf {\bibinfo {volume} {107}},\
  \bibinfo {pages} {066603} (\bibinfo {year} {2011})}\BibitemShut {NoStop}%
\bibitem [{\citenamefont {F\"ahnle}\ and\ \citenamefont {Illg}(2011)}]{FI11}%
  \BibitemOpen
  \bibfield  {author} {\bibinfo {author} {\bibfnamefont {M.}~\bibnamefont
  {F\"ahnle}}\ and\ \bibinfo {author} {\bibfnamefont {C.}~\bibnamefont
  {Illg}},\ }\bibfield  {title} {\bibinfo {title} {Electron theory of fast and
  ultrafast dissipative magnetization dynamics},\ }\href
  {http://stacks.iop.org/0953-8984/23/i=49/a=493201} {\bibfield  {journal}
  {\bibinfo  {journal} {J. Phys.: Condens. Matter}\ }\textbf {\bibinfo {volume}
  {23}},\ \bibinfo {pages} {493201} (\bibinfo {year} {2011})}\BibitemShut
  {NoStop}%
\bibitem [{\citenamefont {Mankovsky}\ \emph {et~al.}(2013)\citenamefont
  {Mankovsky}, \citenamefont {K\"odderitzsch}, \citenamefont {Woltersdorf},\
  and\ \citenamefont {Ebert}}]{MKWE13}%
  \BibitemOpen
  \bibfield  {author} {\bibinfo {author} {\bibfnamefont {S.}~\bibnamefont
  {Mankovsky}}, \bibinfo {author} {\bibfnamefont {D.}~\bibnamefont
  {K\"odderitzsch}}, \bibinfo {author} {\bibfnamefont {G.}~\bibnamefont
  {Woltersdorf}},\ and\ \bibinfo {author} {\bibfnamefont {H.}~\bibnamefont
  {Ebert}},\ }\bibfield  {title} {\bibinfo {title} {First-principles
  calculation of the {Gilbert} damping parameter via the linear response
  formalism with application to magnetic transition metals and alloys},\ }\href
  {https://doi.org/10.1103/PhysRevB.87.014430} {\bibfield  {journal} {\bibinfo
  {journal} {Phys. Rev. B}\ }\textbf {\bibinfo {volume} {87}},\ \bibinfo
  {pages} {014430} (\bibinfo {year} {2013})}\BibitemShut {NoStop}%
\bibitem [{\citenamefont {Mondal}\ \emph {et~al.}(2016)\citenamefont {Mondal},
  \citenamefont {Berritta},\ and\ \citenamefont {Oppeneer}}]{MBO16}%
  \BibitemOpen
  \bibfield  {author} {\bibinfo {author} {\bibfnamefont {R.}~\bibnamefont
  {Mondal}}, \bibinfo {author} {\bibfnamefont {M.}~\bibnamefont {Berritta}},\
  and\ \bibinfo {author} {\bibfnamefont {P.~M.}\ \bibnamefont {Oppeneer}},\
  }\bibfield  {title} {\bibinfo {title} {Relativistic theory of spin relaxation
  mechanisms in the {Landau-Lifshitz-Gilbert} equation of spin dynamics},\
  }\href {https://doi.org/10.1103/PhysRevB.94.144419} {\bibfield  {journal}
  {\bibinfo  {journal} {Phys. Rev. B}\ }\textbf {\bibinfo {volume} {94}},\
  \bibinfo {pages} {144419} (\bibinfo {year} {2016})}\BibitemShut {NoStop}%
\bibitem [{\citenamefont {Mondal}\ \emph {et~al.}(2017)\citenamefont {Mondal},
  \citenamefont {Berritta}, \citenamefont {Nandy},\ and\ \citenamefont
  {Oppeneer}}]{MBNO17}%
  \BibitemOpen
  \bibfield  {author} {\bibinfo {author} {\bibfnamefont {R.}~\bibnamefont
  {Mondal}}, \bibinfo {author} {\bibfnamefont {M.}~\bibnamefont {Berritta}},
  \bibinfo {author} {\bibfnamefont {A.~K.}\ \bibnamefont {Nandy}},\ and\
  \bibinfo {author} {\bibfnamefont {P.~M.}\ \bibnamefont {Oppeneer}},\
  }\bibfield  {title} {\bibinfo {title} {Relativistic theory of magnetic
  inertia in ultrafast spin dynamics},\ }\href
  {https://doi.org/10.1103/PhysRevB.96.024425} {\bibfield  {journal} {\bibinfo
  {journal} {Phys. Rev. B}\ }\textbf {\bibinfo {volume} {96}},\ \bibinfo
  {pages} {024425} (\bibinfo {year} {2017})}\BibitemShut {NoStop}%
\bibitem [{\citenamefont {Guimaraes}\ \emph {et~al.}(2019)\citenamefont
  {Guimaraes}, \citenamefont {Suckert}, \citenamefont {Chico}, \citenamefont
  {Bouaziz}, \citenamefont {dos Santos~Dias},\ and\ \citenamefont
  {Lounis}}]{GSC+19}%
  \BibitemOpen
  \bibfield  {author} {\bibinfo {author} {\bibfnamefont {F.~S.~M.}\
  \bibnamefont {Guimaraes}}, \bibinfo {author} {\bibfnamefont {J.~R.}\
  \bibnamefont {Suckert}}, \bibinfo {author} {\bibfnamefont {J.}~\bibnamefont
  {Chico}}, \bibinfo {author} {\bibfnamefont {J.}~\bibnamefont {Bouaziz}},
  \bibinfo {author} {\bibfnamefont {M.}~\bibnamefont {dos Santos~Dias}},\ and\
  \bibinfo {author} {\bibfnamefont {S.}~\bibnamefont {Lounis}},\ }\bibfield
  {title} {\bibinfo {title} {Comparative study of methodologies to compute the
  intrinsic {Gilbert} damping: interrelations, validity and physical
  consequences},\ }\href {https://doi.org/10.1088/1361-648X/ab1239} {\bibfield
  {journal} {\bibinfo  {journal} {Journal of Physics: Condensed Matter}\
  }\textbf {\bibinfo {volume} {31}},\ \bibinfo {pages} {255802} (\bibinfo
  {year} {2019})}\BibitemShut {NoStop}%
\bibitem [{\citenamefont {Ado}\ \emph {et~al.}(2020)\citenamefont {Ado},
  \citenamefont {Ostrovsky},\ and\ \citenamefont {Titov}}]{AOT20}%
  \BibitemOpen
  \bibfield  {author} {\bibinfo {author} {\bibfnamefont {I.~A.}\ \bibnamefont
  {Ado}}, \bibinfo {author} {\bibfnamefont {P.~M.}\ \bibnamefont {Ostrovsky}},\
  and\ \bibinfo {author} {\bibfnamefont {M.}~\bibnamefont {Titov}},\ }\bibfield
   {title} {\bibinfo {title} {Anisotropy of spin-transfer torques and {Gilbert}
  damping induced by {Rashba} coupling},\ }\href
  {https://doi.org/10.1103/PhysRevB.101.085405} {\bibfield  {journal} {\bibinfo
   {journal} {Phys. Rev. B}\ }\textbf {\bibinfo {volume} {101}},\ \bibinfo
  {pages} {085405} (\bibinfo {year} {2020})}\BibitemShut {NoStop}%
\bibitem [{\citenamefont {Tserkovnyak}\ and\ \citenamefont
  {Mecklenburg}(2008)}]{TM08}%
  \BibitemOpen
  \bibfield  {author} {\bibinfo {author} {\bibfnamefont {Y.}~\bibnamefont
  {Tserkovnyak}}\ and\ \bibinfo {author} {\bibfnamefont {M.}~\bibnamefont
  {Mecklenburg}},\ }\bibfield  {title} {\bibinfo {title} {Electron transport
  driven by nonequilibrium magnetic textures},\ }\href
  {https://doi.org/10.1103/PhysRevB.77.134407} {\bibfield  {journal} {\bibinfo
  {journal} {Phys. Rev. B}\ }\textbf {\bibinfo {volume} {77}},\ \bibinfo
  {pages} {134407} (\bibinfo {year} {2008})}\BibitemShut {NoStop}%
\bibitem [{\citenamefont {Hankiewicz}\ \emph {et~al.}(2008)\citenamefont
  {Hankiewicz}, \citenamefont {Vignale},\ and\ \citenamefont
  {Tserkovnyak}}]{HVT08}%
  \BibitemOpen
  \bibfield  {author} {\bibinfo {author} {\bibfnamefont {E.~M.}\ \bibnamefont
  {Hankiewicz}}, \bibinfo {author} {\bibfnamefont {G.}~\bibnamefont
  {Vignale}},\ and\ \bibinfo {author} {\bibfnamefont {Y.}~\bibnamefont
  {Tserkovnyak}},\ }\bibfield  {title} {\bibinfo {title} {Inhomogeneous
  {Gilbert} damping from impurities and electron-electron interactions},\
  }\href {https://doi.org/10.1103/PhysRevB.78.020404} {\bibfield  {journal}
  {\bibinfo  {journal} {Phys. Rev. B}\ }\textbf {\bibinfo {volume} {78}},\
  \bibinfo {pages} {020404} (\bibinfo {year} {2008})}\BibitemShut {NoStop}%
\bibitem [{\citenamefont {Zhang}\ and\ \citenamefont {Zhang}(2009)}]{ZZ09}%
  \BibitemOpen
  \bibfield  {author} {\bibinfo {author} {\bibfnamefont {S.}~\bibnamefont
  {Zhang}}\ and\ \bibinfo {author} {\bibfnamefont {S.~S.-L.}\ \bibnamefont
  {Zhang}},\ }\bibfield  {title} {\bibinfo {title} {Generalization of the
  {Landau-Lifshitz-Gilbert} equation for conducting ferromagnets},\ }\href
  {https://doi.org/10.1103/PhysRevLett.102.086601} {\bibfield  {journal}
  {\bibinfo  {journal} {Phys. Rev. Lett.}\ }\textbf {\bibinfo {volume} {102}},\
  \bibinfo {pages} {086601} (\bibinfo {year} {2009})}\BibitemShut {NoStop}%
\bibitem [{\citenamefont {Kim}\ \emph {et~al.}(2012)\citenamefont {Kim},
  \citenamefont {Moon}, \citenamefont {Lee},\ and\ \citenamefont
  {Lee}}]{KMLL12}%
  \BibitemOpen
  \bibfield  {author} {\bibinfo {author} {\bibfnamefont {K.-W.}\ \bibnamefont
  {Kim}}, \bibinfo {author} {\bibfnamefont {J.-H.}\ \bibnamefont {Moon}},
  \bibinfo {author} {\bibfnamefont {K.-J.}\ \bibnamefont {Lee}},\ and\ \bibinfo
  {author} {\bibfnamefont {H.-W.}\ \bibnamefont {Lee}},\ }\bibfield  {title}
  {\bibinfo {title} {Prediction of giant spin motive force due to {Rashba}
  spin-orbit coupling},\ }\href
  {https://doi.org/10.1103/PhysRevLett.108.217202} {\bibfield  {journal}
  {\bibinfo  {journal} {Phys. Rev. Lett.}\ }\textbf {\bibinfo {volume} {108}},\
  \bibinfo {pages} {217202} (\bibinfo {year} {2012})}\BibitemShut {NoStop}%
\bibitem [{\citenamefont {Yuan}\ \emph {et~al.}(2016)\citenamefont {Yuan},
  \citenamefont {Yuan}, \citenamefont {Xia},\ and\ \citenamefont
  {Wang}}]{YYXW16}%
  \BibitemOpen
  \bibfield  {author} {\bibinfo {author} {\bibfnamefont {H.~Y.}\ \bibnamefont
  {Yuan}}, \bibinfo {author} {\bibfnamefont {Z.}~\bibnamefont {Yuan}}, \bibinfo
  {author} {\bibfnamefont {K.}~\bibnamefont {Xia}},\ and\ \bibinfo {author}
  {\bibfnamefont {X.~R.}\ \bibnamefont {Wang}},\ }\bibfield  {title} {\bibinfo
  {title} {Influence of nonlocal damping on the field-driven domain wall
  motion},\ }\href {https://doi.org/10.1103/PhysRevB.94.064415} {\bibfield
  {journal} {\bibinfo  {journal} {Phys. Rev. B}\ }\textbf {\bibinfo {volume}
  {94}},\ \bibinfo {pages} {064415} (\bibinfo {year} {2016})}\BibitemShut
  {NoStop}%
\bibitem [{\citenamefont {Mankovsky}\ \emph {et~al.}(2018)\citenamefont
  {Mankovsky}, \citenamefont {Wimmer},\ and\ \citenamefont {Ebert}}]{MWE18}%
  \BibitemOpen
  \bibfield  {author} {\bibinfo {author} {\bibfnamefont {S.}~\bibnamefont
  {Mankovsky}}, \bibinfo {author} {\bibfnamefont {S.}~\bibnamefont {Wimmer}},\
  and\ \bibinfo {author} {\bibfnamefont {H.}~\bibnamefont {Ebert}},\ }\bibfield
   {title} {\bibinfo {title} {{Gilbert} damping in noncollinear magnetic
  systems},\ }\href {https://doi.org/10.1103/PhysRevB.98.104406} {\bibfield
  {journal} {\bibinfo  {journal} {Phys. Rev. B}\ }\textbf {\bibinfo {volume}
  {98}},\ \bibinfo {pages} {104406} (\bibinfo {year} {2018})}\BibitemShut
  {NoStop}%
\bibitem [{\citenamefont {Verba}\ \emph {et~al.}(2018)\citenamefont {Verba},
  \citenamefont {Tiberkevich},\ and\ \citenamefont {Slavin}}]{VTS18}%
  \BibitemOpen
  \bibfield  {author} {\bibinfo {author} {\bibfnamefont {R.}~\bibnamefont
  {Verba}}, \bibinfo {author} {\bibfnamefont {V.}~\bibnamefont {Tiberkevich}},\
  and\ \bibinfo {author} {\bibfnamefont {A.}~\bibnamefont {Slavin}},\
  }\bibfield  {title} {\bibinfo {title} {Damping of linear spin-wave modes in
  magnetic nanostructures: Local, nonlocal, and coordinate-dependent damping},\
  }\href {https://doi.org/10.1103/PhysRevB.98.104408} {\bibfield  {journal}
  {\bibinfo  {journal} {Phys. Rev. B}\ }\textbf {\bibinfo {volume} {98}},\
  \bibinfo {pages} {104408} (\bibinfo {year} {2018})}\BibitemShut {NoStop}%
\bibitem [{\citenamefont {Elbracht}\ and\ \citenamefont
  {Potthoff}(2024)}]{EP24}%
  \BibitemOpen
  \bibfield  {author} {\bibinfo {author} {\bibfnamefont {M.}~\bibnamefont
  {Elbracht}}\ and\ \bibinfo {author} {\bibfnamefont {M.}~\bibnamefont
  {Potthoff}},\ }\bibfield  {title} {\bibinfo {title} {Prerelaxation in
  quantum, classical, and quantum-classical two-impurity models},\ }\href
  {https://doi.org/10.1103/PhysRevResearch.6.033275} {\bibfield  {journal}
  {\bibinfo  {journal} {Phys. Rev. Res.}\ }\textbf {\bibinfo {volume} {6}},\
  \bibinfo {pages} {033275} (\bibinfo {year} {2024})}\BibitemShut {NoStop}%
\bibitem [{\citenamefont {Campisi}\ \emph {et~al.}(2012)\citenamefont
  {Campisi}, \citenamefont {Denisov},\ and\ \citenamefont {H\"anggi}}]{CDH12}%
  \BibitemOpen
  \bibfield  {author} {\bibinfo {author} {\bibfnamefont {M.}~\bibnamefont
  {Campisi}}, \bibinfo {author} {\bibfnamefont {S.}~\bibnamefont {Denisov}},\
  and\ \bibinfo {author} {\bibfnamefont {P.}~\bibnamefont {H\"anggi}},\
  }\bibfield  {title} {\bibinfo {title} {Geometric magnetism in open quantum
  systems},\ }\href {https://doi.org/10.1103/PhysRevA.86.032114} {\bibfield
  {journal} {\bibinfo  {journal} {Phys. Rev. A}\ }\textbf {\bibinfo {volume}
  {86}},\ \bibinfo {pages} {032114} (\bibinfo {year} {2012})}\BibitemShut
  {NoStop}%
\bibitem [{\citenamefont {Sakuma}(2006)}]{Sak06}%
  \BibitemOpen
  \bibfield  {author} {\bibinfo {author} {\bibfnamefont {A.}~\bibnamefont
  {Sakuma}},\ }\bibfield  {title} {\bibinfo {title} {Microscopic description of
  {Landau-Lifshitz-Gilbert} type equation based on the s-d model},\ }\href@noop
  {} {\bibfield  {journal} {\bibinfo  {journal} {eprint
  arXiv:cond-mat/0602075}\ } (\bibinfo {year} {2006})},\ \Eprint
  {https://arxiv.org/abs/cond-mat/0602075} {cond-mat/0602075} \BibitemShut
  {NoStop}%
\bibitem [{\citenamefont {Katsura}\ \emph {et~al.}(2006)\citenamefont
  {Katsura}, \citenamefont {Balatsky}, \citenamefont {Nussinov},\ and\
  \citenamefont {Nagaosa}}]{KBNN06}%
  \BibitemOpen
  \bibfield  {author} {\bibinfo {author} {\bibfnamefont {H.}~\bibnamefont
  {Katsura}}, \bibinfo {author} {\bibfnamefont {A.~V.}\ \bibnamefont
  {Balatsky}}, \bibinfo {author} {\bibfnamefont {Z.}~\bibnamefont {Nussinov}},\
  and\ \bibinfo {author} {\bibfnamefont {N.}~\bibnamefont {Nagaosa}},\
  }\bibfield  {title} {\bibinfo {title} {Voltage dependence of
  landau-lifshitz-gilbert damping of spin in a current-driven tunnel
  junction},\ }\href {https://doi.org/10.1103/PhysRevB.73.212501} {\bibfield
  {journal} {\bibinfo  {journal} {Phys. Rev. B}\ }\textbf {\bibinfo {volume}
  {73}},\ \bibinfo {pages} {212501} (\bibinfo {year} {2006})}\BibitemShut
  {NoStop}%
\bibitem [{\citenamefont {N\'u\~nez}\ and\ \citenamefont {Duine}(2008)}]{ND08}%
  \BibitemOpen
  \bibfield  {author} {\bibinfo {author} {\bibfnamefont {A.~S.}\ \bibnamefont
  {N\'u\~nez}}\ and\ \bibinfo {author} {\bibfnamefont {R.~A.}\ \bibnamefont
  {Duine}},\ }\bibfield  {title} {\bibinfo {title} {Effective temperature and
  gilbert damping of a current-driven localized spin},\ }\href
  {https://doi.org/10.1103/PhysRevB.77.054401} {\bibfield  {journal} {\bibinfo
  {journal} {Phys. Rev. B}\ }\textbf {\bibinfo {volume} {77}},\ \bibinfo
  {pages} {054401} (\bibinfo {year} {2008})}\BibitemShut {NoStop}%
\bibitem [{\citenamefont {Turek}\ \emph {et~al.}(2015)\citenamefont {Turek},
  \citenamefont {Kudrnovsk\'y},\ and\ \citenamefont {Drchal}}]{TKD15}%
  \BibitemOpen
  \bibfield  {author} {\bibinfo {author} {\bibfnamefont {I.}~\bibnamefont
  {Turek}}, \bibinfo {author} {\bibfnamefont {J.}~\bibnamefont
  {Kudrnovsk\'y}},\ and\ \bibinfo {author} {\bibfnamefont {V.}~\bibnamefont
  {Drchal}},\ }\bibfield  {title} {\bibinfo {title} {Nonlocal torque operators
  in ab initio theory of the gilbert damping in random ferromagnetic alloys},\
  }\href {https://doi.org/10.1103/PhysRevB.92.214407} {\bibfield  {journal}
  {\bibinfo  {journal} {Phys. Rev. B}\ }\textbf {\bibinfo {volume} {92}},\
  \bibinfo {pages} {214407} (\bibinfo {year} {2015})}\BibitemShut {NoStop}%
\bibitem [{\citenamefont {Sakuma}(2013)}]{Sak13}%
  \BibitemOpen
  \bibfield  {author} {\bibinfo {author} {\bibfnamefont {A.}~\bibnamefont
  {Sakuma}},\ }\bibfield  {title} {\bibinfo {title} {Microscopic theory of
  gilbert damping for transition metal systems},\ }\href
  {https://doi.org/10.3379/msjmag.1310R001} {\bibfield  {journal} {\bibinfo
  {journal} {Journal of the Magnetics Society of Japan}\ }\textbf {\bibinfo
  {volume} {37}},\ \bibinfo {pages} {343} (\bibinfo {year} {2013})}\BibitemShut
  {NoStop}%
\bibitem [{\citenamefont {Smorka}\ \emph {et~al.}(2024)\citenamefont {Smorka},
  \citenamefont {Thoss},\ and\ \citenamefont {\ifmmode~\check{Z}\else
  \v{Z}\fi{}onda}}]{STZ24}%
  \BibitemOpen
  \bibfield  {author} {\bibinfo {author} {\bibfnamefont {R.}~\bibnamefont
  {Smorka}}, \bibinfo {author} {\bibfnamefont {M.}~\bibnamefont {Thoss}},\ and\
  \bibinfo {author} {\bibfnamefont {M.}~\bibnamefont {\ifmmode~\check{Z}\else
  \v{Z}\fi{}onda}},\ }\bibfield  {title} {\bibinfo {title} {Dynamics of spin
  relaxation in nonequilibrium magnetic nanojunctions},\ }\href
  {https://doi.org/10.1088/1367-2630/ad1fa9} {\bibfield  {journal} {\bibinfo
  {journal} {New J. Phys.}\ }\textbf {\bibinfo {volume} {26}},\ \bibinfo
  {pages} {013056} (\bibinfo {year} {2024})}\BibitemShut {NoStop}%
\bibitem [{\citenamefont {Jarzynski}(1997)}]{Jar97}%
  \BibitemOpen
  \bibfield  {author} {\bibinfo {author} {\bibfnamefont {C.}~\bibnamefont
  {Jarzynski}},\ }\bibfield  {title} {\bibinfo {title} {Nonequilibrium equality
  for free energy differences},\ }\href
  {https://doi.org/10.1103/PhysRevLett.78.2690} {\bibfield  {journal} {\bibinfo
   {journal} {Phys. Rev. Lett.}\ }\textbf {\bibinfo {volume} {78}},\ \bibinfo
  {pages} {2690} (\bibinfo {year} {1997})}\BibitemShut {NoStop}%
\bibitem [{\citenamefont {Talkner}\ \emph {et~al.}(2007)\citenamefont
  {Talkner}, \citenamefont {Lutz},\ and\ \citenamefont {H\"anggi}}]{TLH07}%
  \BibitemOpen
  \bibfield  {author} {\bibinfo {author} {\bibfnamefont {P.}~\bibnamefont
  {Talkner}}, \bibinfo {author} {\bibfnamefont {E.}~\bibnamefont {Lutz}},\ and\
  \bibinfo {author} {\bibfnamefont {P.}~\bibnamefont {H\"anggi}},\ }\bibfield
  {title} {\bibinfo {title} {Fluctuation theorems: Work is not an observable},\
  }\href {https://doi.org/10.1103/PhysRevE.75.050102} {\bibfield  {journal}
  {\bibinfo  {journal} {Phys. Rev. E}\ }\textbf {\bibinfo {volume} {75}},\
  \bibinfo {pages} {050102} (\bibinfo {year} {2007})}\BibitemShut {NoStop}%
\bibitem [{\citenamefont {Campisi}\ \emph {et~al.}(2011)\citenamefont
  {Campisi}, \citenamefont {H\"anggi},\ and\ \citenamefont {Talkner}}]{CHT11}%
  \BibitemOpen
  \bibfield  {author} {\bibinfo {author} {\bibfnamefont {M.}~\bibnamefont
  {Campisi}}, \bibinfo {author} {\bibfnamefont {P.}~\bibnamefont {H\"anggi}},\
  and\ \bibinfo {author} {\bibfnamefont {P.}~\bibnamefont {Talkner}},\
  }\bibfield  {title} {\bibinfo {title} {Colloquium: Quantum fluctuation
  relations: Foundations and applications},\ }\href
  {https://doi.org/10.1103/RevModPhys.83.771} {\bibfield  {journal} {\bibinfo
  {journal} {Rev. Mod. Phys.}\ }\textbf {\bibinfo {volume} {83}},\ \bibinfo
  {pages} {771} (\bibinfo {year} {2011})}\BibitemShut {NoStop}%
\bibitem [{\citenamefont {Michel}\ and\ \citenamefont {Potthoff}(2022)}]{MP22}%
  \BibitemOpen
  \bibfield  {author} {\bibinfo {author} {\bibfnamefont {S.}~\bibnamefont
  {Michel}}\ and\ \bibinfo {author} {\bibfnamefont {M.}~\bibnamefont
  {Potthoff}},\ }\bibfield  {title} {\bibinfo {title} {Spin {Berry} curvature
  of the {Haldane} model},\ }\href
  {https://doi.org/10.1103/PhysRevB.106.235423} {\bibfield  {journal} {\bibinfo
   {journal} {Phys. Rev. B}\ }\textbf {\bibinfo {volume} {106}},\ \bibinfo
  {pages} {235423} (\bibinfo {year} {2022})}\BibitemShut {NoStop}%
\bibitem [{\citenamefont {Haldane}(1988)}]{Hal88}%
  \BibitemOpen
  \bibfield  {author} {\bibinfo {author} {\bibfnamefont {F.~D.~M.}\
  \bibnamefont {Haldane}},\ }\bibfield  {title} {\bibinfo {title} {Model for a
  quantum {Hall} effect without {Landau} levels: Condensed-matter realization
  of the "parity anomaly"},\ }\href
  {https://doi.org/10.1103/PhysRevLett.61.2015} {\bibfield  {journal} {\bibinfo
   {journal} {Phys. Rev. Lett.}\ }\textbf {\bibinfo {volume} {61}},\ \bibinfo
  {pages} {2015} (\bibinfo {year} {1988})}\BibitemShut {NoStop}%
\bibitem [{\citenamefont {Stahl}\ and\ \citenamefont {Potthoff}(2017)}]{SP17}%
  \BibitemOpen
  \bibfield  {author} {\bibinfo {author} {\bibfnamefont {C.}~\bibnamefont
  {Stahl}}\ and\ \bibinfo {author} {\bibfnamefont {M.}~\bibnamefont
  {Potthoff}},\ }\bibfield  {title} {\bibinfo {title} {Anomalous spin
  precession under a geometrical torque},\ }\href
  {https://doi.org/10.1103/PhysRevLett.119.227203} {\bibfield  {journal}
  {\bibinfo  {journal} {Phys. Rev. Lett.}\ }\textbf {\bibinfo {volume} {119}},\
  \bibinfo {pages} {227203} (\bibinfo {year} {2017})}\BibitemShut {NoStop}%
\bibitem [{\citenamefont {Lenzing}\ \emph {et~al.}(2022)\citenamefont
  {Lenzing}, \citenamefont {Lichtenstein},\ and\ \citenamefont
  {Potthoff}}]{LLP22}%
  \BibitemOpen
  \bibfield  {author} {\bibinfo {author} {\bibfnamefont {N.}~\bibnamefont
  {Lenzing}}, \bibinfo {author} {\bibfnamefont {A.~I.}\ \bibnamefont
  {Lichtenstein}},\ and\ \bibinfo {author} {\bibfnamefont {M.}~\bibnamefont
  {Potthoff}},\ }\bibfield  {title} {\bibinfo {title} {Emergent non-{Abelian}
  gauge theory in coupled spin-electron dynamics},\ }\href
  {https://doi.org/10.1103/PhysRevB.106.094433} {\bibfield  {journal} {\bibinfo
   {journal} {Phys. Rev. B}\ }\textbf {\bibinfo {volume} {106}},\ \bibinfo
  {pages} {094433} (\bibinfo {year} {2022})}\BibitemShut {NoStop}%
\bibitem [{\citenamefont {Lenzing}\ \emph {et~al.}(2023)\citenamefont
  {Lenzing}, \citenamefont {Kr\"uger},\ and\ \citenamefont {Potthoff}}]{LKP23}%
  \BibitemOpen
  \bibfield  {author} {\bibinfo {author} {\bibfnamefont {N.}~\bibnamefont
  {Lenzing}}, \bibinfo {author} {\bibfnamefont {D.}~\bibnamefont {Kr\"uger}},\
  and\ \bibinfo {author} {\bibfnamefont {M.}~\bibnamefont {Potthoff}},\
  }\bibfield  {title} {\bibinfo {title} {Geometrical torque on magnetic moments
  coupled to a correlated antiferromagnet},\ }\href
  {https://doi.org/10.1103/PhysRevResearch.5.L032012} {\bibfield  {journal}
  {\bibinfo  {journal} {Phys. Rev. Res.}\ }\textbf {\bibinfo {volume} {5}},\
  \bibinfo {pages} {L032012} (\bibinfo {year} {2023})}\BibitemShut {NoStop}%
\bibitem [{\citenamefont {Berry}\ and\ \citenamefont {Robbins}(1993)}]{BR93b}%
  \BibitemOpen
  \bibfield  {author} {\bibinfo {author} {\bibfnamefont {M.}~\bibnamefont
  {Berry}}\ and\ \bibinfo {author} {\bibfnamefont {J.}~\bibnamefont
  {Robbins}},\ }\bibfield  {title} {\bibinfo {title} {Chaotic classical and
  half-classical adiabatic reactions: geometric magnetism and deterministic
  friction},\ }\href {https://doi.org/10.1098/rspa.1993.0127} {\bibfield
  {journal} {\bibinfo  {journal} {Proc. R. Soc. London A}\ }\textbf {\bibinfo
  {volume} {442}},\ \bibinfo {pages} {659} (\bibinfo {year}
  {1993})}\BibitemShut {NoStop}%
\bibitem [{\citenamefont {Berry}(1984)}]{Ber84}%
  \BibitemOpen
  \bibfield  {author} {\bibinfo {author} {\bibfnamefont {M.~V.}\ \bibnamefont
  {Berry}},\ }\bibfield  {title} {\bibinfo {title} {Quantal phase factors
  accompanying adiabatic changes},\ }\href
  {https://doi.org/10.1098/rspa.1984.0023} {\bibfield  {journal} {\bibinfo
  {journal} {Proc. R. Soc. London A}\ }\textbf {\bibinfo {volume} {392}},\
  \bibinfo {pages} {45} (\bibinfo {year} {1984})}\BibitemShut {NoStop}%
\bibitem [{\citenamefont {Verner}(2010)}]{Ver10}%
  \BibitemOpen
  \bibfield  {author} {\bibinfo {author} {\bibfnamefont {J.~H.}\ \bibnamefont
  {Verner}},\ }\bibfield  {title} {\bibinfo {title} {Numerically optimal
  {Runge-Kutta} pairs with interpolants},\ }\href
  {https://doi.org/10.1007/s11075-009-9290-3} {\bibfield  {journal} {\bibinfo
  {journal} {Numer. Algor.}\ }\textbf {\bibinfo {volume} {53}},\ \bibinfo
  {pages} {383} (\bibinfo {year} {2010})}\BibitemShut {NoStop}%
\bibitem [{\citenamefont {Rackauckas}\ and\ \citenamefont {Nie}(2017)}]{RN17}%
  \BibitemOpen
  \bibfield  {author} {\bibinfo {author} {\bibfnamefont {C.}~\bibnamefont
  {Rackauckas}}\ and\ \bibinfo {author} {\bibfnamefont {Q.}~\bibnamefont
  {Nie}},\ }\bibfield  {title} {\bibinfo {title} {Differentialequations.jl - a
  performant and feature-rich ecosystem for solving differential equations in
  {Julia}},\ }\href {https://doi.org/10.5334/jors.151} {\bibfield  {journal}
  {\bibinfo  {journal} {The Journal of Open Research Software}\ }\textbf
  {\bibinfo {volume} {5}} (\bibinfo {year} {2017})}\BibitemShut {NoStop}%
\bibitem [{\citenamefont {Steinebach}(2023)}]{Ste23}%
  \BibitemOpen
  \bibfield  {author} {\bibinfo {author} {\bibfnamefont {G.}~\bibnamefont
  {Steinebach}},\ }\bibfield  {title} {\bibinfo {title} {Construction of
  {Rosenbrock-Wanner} method and numerical benchmarks within the {Julia}
  differential equations package},\ }\href
  {https://doi.org/10.1007/s10543-023-00967-x} {\bibfield  {journal} {\bibinfo
  {journal} {BIT Numer. Math}\ }\textbf {\bibinfo {volume} {63}},\ \bibinfo
  {pages} {27} (\bibinfo {year} {2023})}\BibitemShut {NoStop}%
\bibitem [{\citenamefont {Kikuchi}(1956)}]{Kik56}%
  \BibitemOpen
  \bibfield  {author} {\bibinfo {author} {\bibfnamefont {R.}~\bibnamefont
  {Kikuchi}},\ }\bibfield  {title} {\bibinfo {title} {On the minimum of
  magnetization reversal time},\ }\href {https://doi.org/10.1063/1.1722262}
  {\bibfield  {journal} {\bibinfo  {journal} {J. Appl. Phys.}\ }\textbf
  {\bibinfo {volume} {27}},\ \bibinfo {pages} {1352} (\bibinfo {year}
  {1956})}\BibitemShut {NoStop}%
\bibitem [{\citenamefont {Luttinger}\ and\ \citenamefont {Ward}(1960)}]{LW60}%
  \BibitemOpen
  \bibfield  {author} {\bibinfo {author} {\bibfnamefont {J.~M.}\ \bibnamefont
  {Luttinger}}\ and\ \bibinfo {author} {\bibfnamefont {J.~C.}\ \bibnamefont
  {Ward}},\ }\bibfield  {title} {\bibinfo {title} {Ground-state energy of a
  many-fermion system. ii},\ }\href {https://doi.org/10.1103/PhysRev.118.1417}
  {\bibfield  {journal} {\bibinfo  {journal} {Phys. Rev.}\ }\textbf {\bibinfo
  {volume} {118}},\ \bibinfo {pages} {1417} (\bibinfo {year}
  {1960})}\BibitemShut {NoStop}%
\bibitem [{\citenamefont {Moriya}(1985)}]{Mor85}%
  \BibitemOpen
  \bibfield  {author} {\bibinfo {author} {\bibfnamefont {T.}~\bibnamefont
  {Moriya}},\ }\href@noop {} {\emph {\bibinfo {title} {Spin Fluctuations in
  Itinerant Electron Magnetism}}},\ \bibinfo {series} {Springer Series in
  Solid-State Sciences}, Vol.~\bibinfo {volume} {56}\ (\bibinfo  {publisher}
  {Springer},\ \bibinfo {address} {Berlin},\ \bibinfo {year}
  {1985})\BibitemShut {NoStop}%
\bibitem [{\citenamefont {Rohringer}\ \emph {et~al.}(2018)\citenamefont
  {Rohringer}, \citenamefont {Hafermann}, \citenamefont {Toschi}, \citenamefont
  {Katanin}, \citenamefont {Antipov}, \citenamefont {Katsnelson}, \citenamefont
  {Lichtenstein}, \citenamefont {Rubtsov},\ and\ \citenamefont
  {Held}}]{RHT+18}%
  \BibitemOpen
  \bibfield  {author} {\bibinfo {author} {\bibfnamefont {G.}~\bibnamefont
  {Rohringer}}, \bibinfo {author} {\bibfnamefont {H.}~\bibnamefont
  {Hafermann}}, \bibinfo {author} {\bibfnamefont {A.}~\bibnamefont {Toschi}},
  \bibinfo {author} {\bibfnamefont {A.~A.}\ \bibnamefont {Katanin}}, \bibinfo
  {author} {\bibfnamefont {A.~E.}\ \bibnamefont {Antipov}}, \bibinfo {author}
  {\bibfnamefont {M.~I.}\ \bibnamefont {Katsnelson}}, \bibinfo {author}
  {\bibfnamefont {A.~I.}\ \bibnamefont {Lichtenstein}}, \bibinfo {author}
  {\bibfnamefont {A.~N.}\ \bibnamefont {Rubtsov}},\ and\ \bibinfo {author}
  {\bibfnamefont {K.}~\bibnamefont {Held}},\ }\bibfield  {title} {\bibinfo
  {title} {Diagrammatic routes to nonlocal correlations beyond dynamical mean
  field theory},\ }\href {https://doi.org/10.1103/RevModPhys.90.025003}
  {\bibfield  {journal} {\bibinfo  {journal} {Rev. Mod. Phys.}\ }\textbf
  {\bibinfo {volume} {90}},\ \bibinfo {pages} {025003} (\bibinfo {year}
  {2018})}\BibitemShut {NoStop}%
\bibitem [{\citenamefont {Hofstetter}\ and\ \citenamefont
  {Vollhardt}(1998)}]{HV98b}%
  \BibitemOpen
  \bibfield  {author} {\bibinfo {author} {\bibfnamefont {W.}~\bibnamefont
  {Hofstetter}}\ and\ \bibinfo {author} {\bibfnamefont {D.}~\bibnamefont
  {Vollhardt}},\ }\bibfield  {title} {\bibinfo {title} {Frustration of
  antiferromagnetism in the t-t'-hubbard model at weak coupling},\ }\href
  {https://doi.org/https://doi.org/10.1002/andp.19985100105} {\bibfield
  {journal} {\bibinfo  {journal} {Ann. Physik}\ }\textbf {\bibinfo {volume}
  {510}},\ \bibinfo {pages} {48} (\bibinfo {year} {1998})}\BibitemShut
  {NoStop}%
\bibitem [{\citenamefont {Sayad}\ \emph
  {et~al.}(2016{\natexlab{a}})\citenamefont {Sayad}, \citenamefont {Rausch},\
  and\ \citenamefont {Potthoff}}]{SRP16a}%
  \BibitemOpen
  \bibfield  {author} {\bibinfo {author} {\bibfnamefont {M.}~\bibnamefont
  {Sayad}}, \bibinfo {author} {\bibfnamefont {R.}~\bibnamefont {Rausch}},\ and\
  \bibinfo {author} {\bibfnamefont {M.}~\bibnamefont {Potthoff}},\ }\bibfield
  {title} {\bibinfo {title} {Relaxation of a classical spin coupled to a
  strongly correlated electron system},\ }\href
  {https://doi.org/10.1103/PhysRevLett.117.127201} {\bibfield  {journal}
  {\bibinfo  {journal} {Phys. Rev. Lett.}\ }\textbf {\bibinfo {volume} {117}},\
  \bibinfo {pages} {127201} (\bibinfo {year} {2016}{\natexlab{a}})}\BibitemShut
  {NoStop}%
\bibitem [{\citenamefont {Zhou}\ \emph {et~al.}(2010)\citenamefont {Zhou},
  \citenamefont {Wiebe}, \citenamefont {Lounis}, \citenamefont {Vedmedenko},
  \citenamefont {Meier}, \citenamefont {Bl{\"u}gel}, \citenamefont
  {Dederichs},\ and\ \citenamefont {Wiesendanger}}]{ZWL+10}%
  \BibitemOpen
  \bibfield  {author} {\bibinfo {author} {\bibfnamefont {L.}~\bibnamefont
  {Zhou}}, \bibinfo {author} {\bibfnamefont {J.}~\bibnamefont {Wiebe}},
  \bibinfo {author} {\bibfnamefont {S.}~\bibnamefont {Lounis}}, \bibinfo
  {author} {\bibfnamefont {E.}~\bibnamefont {Vedmedenko}}, \bibinfo {author}
  {\bibfnamefont {F.}~\bibnamefont {Meier}}, \bibinfo {author} {\bibfnamefont
  {S.}~\bibnamefont {Bl{\"u}gel}}, \bibinfo {author} {\bibfnamefont {P.~H.}\
  \bibnamefont {Dederichs}},\ and\ \bibinfo {author} {\bibfnamefont
  {R.}~\bibnamefont {Wiesendanger}},\ }\bibfield  {title} {\bibinfo {title}
  {Strength and directionality of surface Ruderman--Kittel--Kasuya--Yosida
  interaction mapped on the atomic scale},\ }\href
  {https://doi.org/10.1038/nphys1514} {\bibfield  {journal} {\bibinfo
  {journal} {Nat. Physics}\ }\textbf {\bibinfo {volume} {6}},\ \bibinfo {pages}
  {187} (\bibinfo {year} {2010})}\BibitemShut {NoStop}%
\bibitem [{\citenamefont {Bajpai}\ and\ \citenamefont
  {Nikoli\ifmmode~\acute{c}\else \'{c}\fi{}}(2020)}]{BN20}%
  \BibitemOpen
  \bibfield  {author} {\bibinfo {author} {\bibfnamefont {U.}~\bibnamefont
  {Bajpai}}\ and\ \bibinfo {author} {\bibfnamefont {B.~K.}\ \bibnamefont
  {Nikoli\ifmmode~\acute{c}\else \'{c}\fi{}}},\ }\bibfield  {title} {\bibinfo
  {title} {Spintronics meets nonadiabatic molecular dynamics: Geometric spin
  torque and damping on dynamical classical magnetic texture due to an
  electronic open quantum system},\ }\href
  {https://doi.org/10.1103/PhysRevLett.125.187202} {\bibfield  {journal}
  {\bibinfo  {journal} {Phys. Rev. Lett.}\ }\textbf {\bibinfo {volume} {125}},\
  \bibinfo {pages} {187202} (\bibinfo {year} {2020})}\BibitemShut {NoStop}%
\bibitem [{\citenamefont {Fransson}(2008)}]{Fra08}%
  \BibitemOpen
  \bibfield  {author} {\bibinfo {author} {\bibfnamefont {J.}~\bibnamefont
  {Fransson}},\ }\bibfield  {title} {\bibinfo {title} {Detection of spin
  reversal and nutations through current measurements},\ }\href
  {https://doi.org/10.1088/0957-4484/19/28/285714} {\bibfield  {journal}
  {\bibinfo  {journal} {Nanotechnology}\ }\textbf {\bibinfo {volume} {19}},\
  \bibinfo {pages} {285714} (\bibinfo {year} {2008})}\BibitemShut {NoStop}%
\bibitem [{\citenamefont {B\"ottcher}\ and\ \citenamefont {Henk}(2012)}]{BH12}%
  \BibitemOpen
  \bibfield  {author} {\bibinfo {author} {\bibfnamefont {D.}~\bibnamefont
  {B\"ottcher}}\ and\ \bibinfo {author} {\bibfnamefont {J.}~\bibnamefont
  {Henk}},\ }\bibfield  {title} {\bibinfo {title} {Significance of nutation in
  magnetization dynamics of nanostructures},\ }\href
  {https://doi.org/10.1103/PhysRevB.86.020404} {\bibfield  {journal} {\bibinfo
  {journal} {Phys. Rev. B}\ }\textbf {\bibinfo {volume} {86}},\ \bibinfo
  {pages} {020404} (\bibinfo {year} {2012})}\BibitemShut {NoStop}%
\bibitem [{\citenamefont {Sayad}\ \emph
  {et~al.}(2016{\natexlab{b}})\citenamefont {Sayad}, \citenamefont {Rausch},\
  and\ \citenamefont {Potthoff}}]{SRP16b}%
  \BibitemOpen
  \bibfield  {author} {\bibinfo {author} {\bibfnamefont {M.}~\bibnamefont
  {Sayad}}, \bibinfo {author} {\bibfnamefont {R.}~\bibnamefont {Rausch}},\ and\
  \bibinfo {author} {\bibfnamefont {M.}~\bibnamefont {Potthoff}},\ }\bibfield
  {title} {\bibinfo {title} {Inertia effects in the real-time dynamics of a
  quantum spin coupled to a fermi sea},\ }\href
  {https://doi.org/10.1209/0295-5075/116/17001} {\bibfield  {journal} {\bibinfo
   {journal} {Europhys. Lett.}\ }\textbf {\bibinfo {volume} {116}},\ \bibinfo
  {pages} {17001} (\bibinfo {year} {2016}{\natexlab{b}})}\BibitemShut {NoStop}%
\end{thebibliography}

%

\end{document}